\documentclass[usenatbib]{basi}
\usepackage[T1]{fontenc}
\usepackage[british]{babel}
\usepackage[varg]{txfonts}

\usepackage{rotating}
\usepackage{dcolumn}

\newcommand{\aap}{A\&A}
\newcommand{\apj}{ApJ}
\newcommand{\apjl}{ApJ}
\newcommand{\apjs}{ApJS}
\newcommand{\mnras}{MNRAS}
\newcommand{\nat}{Nature}
\newcommand{\aj}{AJ}
\newcommand{\araa}{ARA\&A}
\newcommand{\etal}{et al.}
\newcommand{\MgII}{Mg{\sevenrm II}}

\font\sevenrm=cmr7 scaled 1000

\newcommand{\hbeta}{H{$\beta$}}
\newcommand{\halpha}{H{$\alpha$}}
\newcommand{\lya}{Ly$\alpha$}
\newcommand{\CIV}{C{\sevenrm IV}}

\newcommand{\HeII}{He{\sevenrm II}}
\newcommand{\FeII}{Fe{\sevenrm II}}

\newcommand{\CIII}{C{\sevenrm III]}}

\newcommand{\AlIII}{Al{\sevenrm III}}

\newcommand{\SiIII}{Si{\sevenrm III]}}

\def\MgII{Mg\,{\sc ii}}

\def \OIII {[O\,{\sc iii}]}

\newcommand{\OIIIab}{[O{\sevenrm\,III}]\,$\lambda\lambda$4959,5007}

\newcommand{\SIIab}{[S{\sevenrm\,II}]\,$\lambda\lambda$6717,6731}

\begin{document}
\title[QUASAR MASSES]{The Mass of Quasars\thanks{In this review I use the
terms ``quasar'' and ``Active Galactic Nucleus (AGN)'' interchangeably to
refer to all active supermassive black holes, although traditionally quasars
are loosed defined as the luminous ($L_{\rm bol}\gtrsim 10^{12}\,L_\odot$)
subset of AGNs. By default I use quasars and AGNs to refer to unobscured
(Type 1), broad-line objects unless otherwise specified. }}
\author[SHEN]%
       {Yue Shen$^{1,2}$\thanks{email: \texttt{yshen@obs.carnegiescience.edu}}\\
       $^1$Carnegie Observatories, 813 Santa Barbara Street, Pasadena,
CA 91101, USA\\
       $^2$Hubble Fellow}

\pubyear{2013}
\volume{00}
\pagerange{\pageref{firstpage}--\pageref{lastpage}}

\date{Received --- ; accepted ---}

\maketitle
\label{firstpage}

\begin{abstract}
I review the current status of quasar black hole (BH) mass estimations.
Spectroscopic methods have been developed to estimate BH mass in broad line
quasars to an accuracy of $\sim 0.5$ dex. Despite their popularity,
significant issues and confusion remain regarding these mass estimators. I
provide an in-depth discussion on the merits and caveats of the single-epoch
(SE) virial BH mass estimators, and a detailed derivation of the statistical
biases of these SE mass estimates resulting from their errors. I show that
error-induced sample biases on the order of a factor of several are likely
present in the SE mass estimates for flux-limited, statistical quasar
samples, and the distribution of SE masses in finite luminosity bins can be
narrower than the nominal uncertainty of these mass estimates. I then discuss
the latest applications of SE virial masses in quasar studies, including the
early growth of supermassive black holes, quasar demography in the
mass-luminosity plane, and the evolution of the BH-host scaling relations,
with specific emphases on selection effects and sample biases in the SE
masses. I conclude that there is a pressing need to understand and deal with
the errors in these BH mass estimates, and to improve these BH weighing
methods with substantially more and better reverberation mapping data.
\end{abstract}


\section{Introduction}\label{s:intro}

Shortly after the discovery of quasars at great cosmological distances
\citep{Schmidt_1963}, it was realized that the energy required to power these
luminous and compact sources must be of gravitational origin rather than from
nuclear reaction
\citep[e.g.,][]{Hoyle_Fowler_1963,Salpeter_1964,Zeldovich_Novikov_1964,Lynden-Bell_1969}.
The standard picture now is that mass is accreted onto a supermassive black
hole (SMBH) at the center of the galaxy, and the gravitational energy is
released during this accretion process to power quasar activity. If the SMBH
grows mostly via this accretion process, its mass growth rate is simply:
$\dot{M}_{\rm BH}=\lambda L_{\rm Edd}(1-\epsilon)/(\epsilon c^2)$, where
$L_{\rm Edd}=lM_{\rm BH}=1.26\times 10^{38}(M_{\rm BH}/M_\odot)\,{\rm
erg\,s^{-1}}$ is the Eddington luminosity of the BH, $\lambda=L_{\rm
bol}/L_{\rm Edd}$ is the Eddington ratio, and $\epsilon$ is the radiative
efficiency, i.e., the fraction of accreted rest mass energy converted into
radiation. If both $\lambda$ and $\epsilon$ are non-evolving, the BH mass
increases by one $e$-fold on a characteristic timescale $t_{e}\equiv \epsilon
c^2/[(1-\epsilon)\lambda l]\approx
4.5\times10^8\frac{\epsilon}{\lambda(1-\epsilon)}\,$yr, also known as the
Salpeter time or $e$-folding time. If quasars do not radiate beyond the
Eddington limit $\lambda=1$, the observed luminosity provides a lower-limit
on their BH mass \citep[e.g.,][]{Zeldovich_Novikov_1964}. The discovery of
luminous quasars (with $L_{\rm bol}\gtrsim 10^{47}\,{\rm erg\,s^{-1}}$) at
$z>6$ \citep[e.g.,][]{Fan_etal_2001,Mortlock_etal_2011} then suggests that
SMBHs with $M_{\rm BH}>10^9\,M_\odot$ are already formed in the first billion
year after the Big Bang.

In the past two decades or so, there has been tremendous progress in the
demographic studies of SMBHs in the nuclei of nearby galaxies \citep[for
recent reviews, see, e.g.,][]{Kormendy_Richstone_1995,
Ferrarese_Ford_2005,Kormendy_Ho_2013}. It has come to the consensus that
SMBHs with masses of $\sim 10^5-10^{10}\,M_\odot$ are almost ubiquitous at
the center of massive galaxies with a significant spheroidal (bulge)
component, and also exist in at least some low-mass galaxies. More
remarkably, the mass of the nuclear BH is tightly correlated with the
properties of the bulge in the local samples
\citep[e.g.,][]{Gebhardt_etal_2000a,Ferrarese_Merritt_2000,Graham_etal_2001,Tremaine_etal_2002,
Marconi_Hunt_2003,Aller_Richstone_2007,Gultekin_etal_2009}, allowing an
estimate of the local SMBH mass function by convolutions with galaxy bulge
distribution functions. These BH-bulge scaling relations promoted the notion
of BH-galaxy co-evolution, during which the energy release from the accreting
SMBH self-regulates its growth, and impacts the formation and evolution of
the bulge via feedback processes
\citep[e.g.,][]{Silk_Rees_1998,King_2003,Di_Matteo_etal_2005}. Such feedback
from active SMBHs (i.e., AGN feedback) has also been invoked in most
present-day theoretical modeling of galaxy formation, to bring better
agreement with the observed statistics of massive galaxies. However, the
significance of AGN feedback and BH-host co-evolution is still under some
debate and is an active area of research.

An elegant argument tying the local relic SMBH population to the past active
population is the So{\l}tan argument \citep{Soltan_1982}: if SMBHs grow
mainly through a luminous (or obscured) quasar phase, then the accreted
luminosity density of quasars to $z = 0$, $\rho_{\rm \bullet,acc}$, should
equal the local relic BH mass density $\rho_{\bullet}$:
\begin{equation}\label{eqn:soltan}
\rho_{\rm \bullet,acc}=\int_0^{\infty}\frac{dt}{dz}dz\int_0^\infty
\frac{(1-\epsilon)L}{\epsilon c^2}\Phi(L,z)dL\approx
\rho_{\bullet}\ ,
\end{equation}
where $\Phi(L,z)$ is the bolometric luminosity function (LF) per $L$
interval. Given the observed quasar luminosity function, a reasonably good
match between $\rho_{\rm \bullet,acc}$ and $\rho_{\rm \bullet}$ can be
achieved if the average radiative efficiency $\epsilon\sim 0.1$
\citep[e.g.,][also see Salucci et~al.\ 1999; Fabian 1999; Elvis et~al.
2002]{Yu_Tremaine_2002,Shankar_etal_2004,Marconi_etal_2004}, consistent with
the mean $\epsilon$ value constrained from individual quasars with spectral
fitting \citep[e.g.,][]{Davis_Laor_2011}. The So{\l}tan argument and its
variants have been used extensively in recent years to model the growth of
SMBHs with constraints from the demographies of local BH relics and the past
AGN population \citep[for a recent review, see][]{Shankar_2009}. These
exercises are mainly facilitated by the advent of modern large-scale,
multiwavelength sky surveys, which have provided large and homogeneous data
sets many folds more than what was available twenty years ago, as well as
measurements of the abundance and clustering properties of quasars with
unprecedented precision.

The growth of SMBHs is among the key science topics in modern galaxy
formation studies \citep[for a relatively complete summary of recent progress
on this topic, see, e.g.,][and references therein]{Alexander_Hickox_2012}. As
one of the few fundamental quantities describing a BH, the mass of quasars is
of paramount importance to essentially all quasar-related science: the
evolution and phenomenology of quasars, accretion physics, the relations and
interplays between SMBHs and their host galaxies.

In this review I discuss the current status of quasar BH mass estimations and
how these developments can further our understandings of the physics and
evolution of SMBHs. I presume the reader has a basic understanding of AGNs
and I will skip elaborations on the usual AGN terminologies, which can be
found in AGN textbooks \citep[e.g.,][]{Peterson_1997,Krolik_1999}. This
review is mostly pragmatic without going into the detailed and sometimes
poorly understood physics behind observations; some further readings can be
found in the quoted references.

There are several recent reviews on measuring active and inactive BH masses
\citep[e.g.,][]{Peterson_2010,Czerny_2010,
Vestergaard_etal_2011,Marziani_Sulentic_2012}, which summarized some general
concepts and practical procedures in measuring SMBH masses. While some of the
common materials are also covered in the current review for completeness, the
scope and focus of this review are different: after an introduction on BH
mass measurements in \S\ref{s:method}, I describe in detail the caveats and
statistical biases of the most frequently used BH mass estimators in
\S\ref{s:bias}, in light of recent work invoking statistical quasar samples;
several applications of these BH mass estimates to quasar studies are
discussed in \S\ref{s:app}, and I conclude this review in \S\ref{s:future}
with a discussion on future perspectives of improving BH weighing methods. A
flat $\Lambda$CDM cosmology is adopted throughout this review, with
$\Omega_\Lambda=0.7$, $\Omega_0=0.3$ and $H_0=70\,{\rm km\,s^{-1}Mpc^{-1}}$.

\section{Methods to Measure Quasar BH Masses}\label{s:method}

\subsection{Virial BH Masses: From Reverberation Mapping to Single-Epoch
Methods}\label{s:virial_mass}

{\em Reverberation mapping}\quad The broad emission line regions in AGNs are
powered by photoionizations from the central source
\citep[e.g.,][]{Peterson_1997}, an assumption now widely accepted based on
observations of correlated broad line and continuum variations. In fact, this
photoionization assumption led to the first suggestions
\citep[e.g.,][]{Bahcall_etal_1972} of lagged broad line responses to
continuum variations, where the lag reflects the light travel time from the
ionizing source to the broad line region (BLR). In the 1970s there were
already reported lag measurements between broad emission line and continuum
variations in several local Seyfert nuclei \citep[e.g.,][]{Lyutyj_1974}.
Despite the low data quality that may impact the reliability of the
detection, these studies were among the first attempts to directly measure
BLR sizes. This idea was later developed in greater detail by
\citet[][]{Blandford_McKee_1982}, who suggested that by mapping the response
function of the broad emission line to continuum variations one can in
principle reconstruct the structure and kinematics of the BLR, a technique
they coined ``reverberation mapping'' (RM). Today RM has become a practical
and powerful tool to study BLRs \citep[see reviews by,
e.g.,][]{Peterson_1993,Netzer_Peterson_1997,Horne_etal_2004}, whose spatial
extent ($\sim$ sub-pc) is too small to be resolved by current
instrumentation. There are now several dozens of AGNs and quasars (most are
at $z<0.3$) with average lag measurements
\citep[e.g.,][]{Kaspi_etal_2000,Peterson_etal_2004,Bentz_etal_2009b},
although only a handful of them have decent velocity-resolved delay maps
\citep[e.g.,][]{Denney_etal_2009a,Bentz_etal_2010,Grier_etal_2013} to utilize
the full power of RM.

RM lag measurements provide an estimate of the typical size of the BLR. If we
further assume that the BLR is virialized and the motion of the emitting
clouds is dominated by the gravitational field of the central BH, then the
mass of the BH is determined by \citep[e.g.,][]{Ho_1999,Wandel_etal_1999}:
\begin{equation}\label{eq:rm_mass}
M_{\rm RM}=\frac{V_{\rm vir}^2R}{G}=f\frac{W^2R}{G}\ ,
\end{equation}
where $V_{\rm vir}$ is the virial velocity and $R$ is the BLR size. In
practice we use the width of the broad line, $W$, as an indicator of the
virial velocity, assuming that the broad line is Doppler broadened by the
virial motion of the emitting gas. The product $W^2R/G$ is called the virial
product. There are two commonly used line width definitions, the
``full-width-at-half-maximum'' (FWHM), and the line dispersion $\sigma_{\rm
line}$ \citep[i.e., the second moment of the line, ][]{Peterson_etal_2004}.
The pros and cons of both definitions will be discussed later. In computing
the RM BH masses, both line widths are measured from the rms spectra from the
monitoring period, thus only the variable part of the line contributes to the
line width calculation.

To account for our ignorance of the structure and geometry of the BLR which
determine the relation between the virial velocity and the line-of-sight
(LOS) velocity inferred from $W$, we have introduced a virial coefficient (or
geometrical factor), $f$\footnote{Some studies define the virial coefficient
differently, i.e., $V_{\rm vir}\equiv fW$
\citep[e.g.,][]{McLure_Jarvis_2002,Decarli_etal_2008a}.}, in Eqn.\
(\ref{eq:rm_mass}). This is a big simplification, because the BLR structure
and viewing angle determine the entire line profile, and the line width,
being only one characteristic of the line profile, can not fully describe the
underlying kinematic structure. Similarly it is an approximation to describe
the BLR with a single radius $R$. Nevertheless, given the difficulties and
ambiguities of modeling the line profile directly, Eqn.\ (\ref{eq:rm_mass})
involving line widths and $f$ is used almost universally. For BLR clouds in
randomly orientated orbits, an often quoted value is $f\approx 3/4$ (3) if
$W={\rm FWHM}\ (\sigma_{\rm line})$ \citep{Netzer_1990}, although such a $f$
value is derived under some simplifications and approximations and is not a
rigorous analytic result. In practice, the value of $f$ is now empirically
determined by requiring that the derived RM masses are consistent with those
predicted from the BH mass-bulge stellar velocity dispersion ($M_{\rm
BH}-\sigma_*$) relation of local inactive galaxies
\citep[e.g.,][]{Onken_etal_2004}: $f\approx 1.4$ (5.5) for $W={\rm FWHM}\
(\sigma_{\rm line})$. This $f$ value is of course then the averaged value for
the subset of RM AGNs with bulge stellar velocity dispersion measurements.
The uncertainty in $f$ and the simplification of it as a single constant
remain one of the major uncertainties in RM mass determinations, as further
discussed in \S\ref{s:vir_f}.

The uncertainty of the RM masses is typically a factor of a few, or $\sim
0.4-0.5$ dex \citep[e.g.,][]{Peterson_2010}, based on comparisons between RM
masses and predictions from the $M_{\rm BH}-\sigma_*$ relation, and
accounting for other potential systematics.

\noindent{\em The $R-L$ relation}\quad Perhaps the most remarkable finding of
RM observations is a tight correlation between the measured BLR size and the
adjacent optical continuum luminosity $L_{\rm opt}$ (usually measured at
restframe 5100\,\AA), $R\propto L^\alpha$, over $\sim 4$ orders of magnitude
in luminosity. This is known as the BLR size-luminosity relation, or the
$R-L$ relation
\citep[e.g.,][]{Kaspi_etal_2000,Kaspi_etal_2005,Bentz_etal_2009a}. To first
order all broad line quasars have similar spectral energy distributions
(SEDs) from X-ray to optical\footnote{It is also important to recognize that
AGN SEDs can vary significantly from object to object, and some of these
variances in SED must introduce certain scatter in the observed $R-L$
relation based on measurable continuum instead of the ionizing continuum, and
may cause systematic changes of BLR structure with SED properties (which
might affect the virial coefficient $f$).}, so $L_{\rm opt}$ is proportional
to the ionizing continuum $L_{\rm ion}$. The ionization parameter in a
photoionized medium is $U=Q(H)/(4\pi r^2cn_e)$, where $Q(H)\propto L$ is the
number of ionizing photons from the central source per second, $c$ is the
speed of light, and $n_e$ is electron density. Thus a slope of $\alpha=0.5$
in the $R-L$ relation is expected, if $U$ and the electron density are more
or less constant in BLRs. Alternatively, a slope of $\alpha=0.5$ is also
predicted if the BLR size is set by dust sublimation
\citep[e.g.,][]{Netzer_Laor_1993}.

Early RM work reported a slope of $\alpha\sim 0.7$
\citep[e.g.,][]{Kaspi_etal_2000}. Later work which carefully accounted for
host starlight contamination to $L_{\rm opt}$ reported $\alpha\approx 0.5$
\citep[e.g.,][]{Bentz_etal_2009a}, closer to naive expectations from
photoionization. The intrinsic scatter of the $R-L$ relation is estimated to
be $\sim 0.15$ dex \citep[$\sim 0.11$ dex with the best quality RM
data,][]{Peterson_2010}. The latest version of the $R-L$ relation based on
\hbeta\ RM measurements is \citep{Bentz_etal_2009a}:
\begin{equation}
\log\frac{R}{{\rm light\ days}} = -21.3 + 0.519\log\frac{\lambda L_\lambda(5100 \textit{\AA})}{{\rm erg\,s^{-1}}}.
\end{equation}
The tightness of the $R-L$ relation has led to suggestions to use this
relation as an absolute luminosity indicator to use quasars as a cosmology
probe \citep[e.g.,][]{Watson_etal_2011,Czerny_eteal_2012}, although RM
measurements of BLR sizes and quantification of the $R-L$ relation beyond
$z\sim 0.3$ are yet to come \citep[e.g.,][]{Kaspi_etal_2007}.

\noindent{\em Single-epoch (SE) virial BH mass estimators}\quad The observed
$R-L$ relation provides a much less expensive way to estimate the size of the
BLR based on the luminosity of the quasar. Subsequently this relation has
been used to develop the so-called ``single-epoch virial black hole mass
estimators''\footnote{Such methods are also known as BH mass scaling methods
\citep[e.g.,][]{Vestergaard_etal_2011}. Occasionally, this method is referred
to as the ``photoionization method''
\citep[e.g.,][]{Peterson_2011,Salviander_Shields_2012}. This may be a little
ambiguous, since in practice this empirical method is based on RM results
rather than photonionization calculations, even though the observed $R-L$
relation is consistent with naive photoionization predictions. The
``photoionization method'' better refers to those that estimate the BLR size
using photoionization arguments (see \S\ref{s:other_methods}). } (SE virial
masses or SE masses in short hereafter): one estimates the BLR size from the
measured quasar luminosity using the $R-L$ relation, and the width of the
broad emission line, which are then combined to give an estimate of the BH
mass using calibration coefficients determined from the sample of AGNs with
RM mass estimates. Specifically, these estimators take the form:
\begin{equation}\label{eqn:virial_mass}
\log\left(\frac{M_{\rm SE}}{M_\odot}\right)= a + b\log\left(\frac{L}{10^{44}\, {\rm erg\,s^{-1}}}\right) + c\log\left(\frac{W}{\rm
km\,s^{-1}}\right)\ ,
\end{equation}
where $L$ and $W$ are the quasar continuum (or line) luminosity and width for
the specific line, and coefficients $a$, $b$ and $c$ are calibrated against
RM AGNs. In Eqn.\ (\ref{eqn:virial_mass}), the coefficient on line width is
usually taken to be $c=2$, as expected from viral motion. Other values of $c$
are suggested, however, depending on the definition of line width, and will
be discussed further in \S\ref{s:bias}. Based on the general similarity of
quasar SEDs (both continuum and line strength), different luminosities have
been used, including continuum luminosities in X-ray, restframe UV and
optical, as well as line luminosities, in various versions of these
single-epoch virial estimators
\citep[e.g.,][]{Vestergaard_2002,McLure_Jarvis_2002,McLure_Dunlop_2004,Wu_etal_2004,Greene_Ho_2005,
Vestergaard_Peterson_2006,Kollmeier_etal_2006,Onken_Kollmeier_2008,Wang_etal_2009,Vestergaard_Osmer_2009,Greene_etal_2010b,Rafiee_Hall_2011b,Shen_etal_2011,
Shen_Liu_2012,Trakhtenbrot_Netzer_2012}. In general continuum luminosities
are preferred over line luminosities given their tighter correlations with
BLR size, but in some cases line luminosities are preferred where the
continuum may be significantly contaminated by host starlight
\citep[e.g.,][]{Greene_Ho_2005}, or by the nonthermal emission from a jet in
radio-loud objects \citep[e.g.,][]{Wu_etal_2004}. As for the choice of line
width, both FWHM and line dispersion ($\sigma_{\rm line}$) are utilized in
these calibrations.

The uncertainty of these various single-epoch virial estimators can be
inferred from the residuals in the calibrations against the RM masses, and is
estimated to be on the order of $\sim 0.5$ dex
\citep[e.g.,][]{McLure_Jarvis_2002,Vestergaard_Peterson_2006}. This is
similar to the uncertainty of RM masses, and can hardly be smaller since this
method is rooted in the RM technique. Reasons and consequences of such
substantial mass uncertainties will be elaborated in \S\ref{s:bias}. From now
on I will refer to RM and SE masses collectively as virial BH masses.

The virial estimators (RM and SE) currently are the best method in estimating
quasar BH masses. Therefore after a brief discussion on alternative methods
(\S\ref{s:other_methods}), I will focus on these virial estimators in the
rest of the review.

\subsection{Other Methods to Estimate Quasar BH Masses}\label{s:other_methods}

There are several other methods to estimate the mass of quasars. They are
much less popular than the RM method and its extension, the SE virial method.
Nevertheless there is certain merit in further developing some of these
methods, for instance, to provide complementary mass estimates and
consistency checks. Therefore I give a brief discussion on these alternative
methods.

\noindent{\em Photoionization method}\quad Historically Dibai was the first
to systematically measure BH masses for quasar samples since the 1970s
\citep[e.g.,][]{Dibai_1977,Dibai_1980,Dibai_1984}. Adopting Woltjer's
postulation that the BLR gas is in virial equilibrium in the gravitational
potential of the central BH \citep{Woltjer_1959}, Dibai used Eqn.\
(\ref{eq:rm_mass}) to estimate the BH mass, where the width of the line is
used to indicate the virial velocity. To estimate the BLR size $R$, Dibai
used the photoionization argument:
\begin{equation}\label{eqn:dibai}
L(\textrm{\hbeta})=\frac{4\pi}{3}R^3j(n_eT_e)\epsilon_V\ ,
\end{equation}
where $L(\textrm{\hbeta})$ is the luminosity of the \hbeta\ line, $j(n_eT_e)$
is the volume emissivity in the \hbeta\ line from photoionized gas, and
$\epsilon_V$ is the volume filling factor of BLR clouds. Adopting constant
values of $n_e\approx 10^9\,{\rm cm^{-3}}$, $T_e\approx 10^4\,$K and
$\epsilon_V\approx 10^{-3}$, Dibai estimated BH masses for more than $\sim
70$ nearby Seyfert 1 galaxies and quasars, and made the first plot of the
distribution of AGNs in the mass-luminosity plane. Technically speaking,
Dibai's method is also a single-epoch method, and it has an effective $R-L$
relation of $R\propto L(\textrm{\hbeta})^{1/3}$, shallower than the observed
$R-L$ relation. Despite the simplifications in Dibai's approach (sometimes
unphysical), many of his BH mass estimates of local AGNs are consistent with
today's RM masses to within 0.3 dex \citep[e.g.,][]{Bochkarev_Gaskell_2009}.
Dibai's method also motivated some later quasar BH mass estimations based on
the same argument. For example, \citet[][]{Wandel_Yahil_1985} used the same
method with modifications to the volume filling factor, and derived a
radius-luminosity relation $R\propto L(\textrm{\hbeta})^{1/2}$.

Along a completely independent path, photoionization arguments based on the
ionization parameter $U$ were used to estimate the BLR size
\citep[e.g.,][]{Netzer_1990,Wandel_etal_1999}, building on earlier
development of photoionization equilibrium theory in the 1970s
\citep[e.g.,][]{Davidson_1972,Davidson_Netzer_1979}. This approach emphasizes
more on the role of the ionization continuum, and provides an intuitive
understanding of the observed $R-L$ relation. But just as Dibai's method, all
these photoionization-based methods require assumptions or indirect
constraints on the physical conditions of the BLR gas (such as density,
covering factor, etc) to infer the BLR size, therefore today they are not as
popular as the more empirical, but more accurate RM-based methods discussed
in \S\ref{s:virial_mass}.

\noindent{\em Accretion disk model fitting (SED fitting)}\quad Another method
to infer the mass of the BH is by fitting the SED of quasars. The development
of accretion disk theory over the last four decades \citep[for a recent
review, see, e.g.,][]{Abramowicz_2013} has enabled predictions of the
emitting continuum spectrum of accreting BHs. By fitting the observed quasar
continuum SED, one can constrain the model parameters (such as BH mass,
accretion rate, BH spin, inclination) with adequate accretion disk models.
Many studies have used this SED fitting method to infer BH masses in AGNs
\citep[e.g.,][]{Malkan_1983,Sun_Malkan_1989,Wandel_Petrosian_1988,Laor_1990,
Rokaki_etal_1992,Tripp_etal_1994,Ghisellini_etal_2010,Calderone_etal_2012},
usually assuming a standard thin accretion disk model
\citep[][]{Shakura_Sunyaev_1973}. One main concern here is that standard
accretion disk models, while can successfully produce broad-band features
\citep[such as the ``big blue bump'', e.g.,][]{Shields_1978}, do not yet have
the capability to fully explain the AGN SED
\citep[e.g.,][]{Koratkar_Blaes_1999,Lawrence_2012}, and the resulting BH mass
constraints may be sensitive to these deviations from standard accretion disk
models. Given that there are parameter degeneracies and model
assumptions/simplifications in the SED fitting procedure and the requirement
for good multiwavelength coverage in UV-optical (where most of the disk
emission comes from), this method generally cannot provide an accuracy of
better than a factor of $\sim$ five in BH mass estimates at the moment
\citep[e.g.,][]{Laor_1990,Calderone_etal_2012}. But it would be interesting
to compare this method with the virial methods (\S\ref{s:virial_mass}) for
larger samples.

\noindent{\em Microlensing in gravitationally lensed quasars -- alternative
routes to BLR sizes}\quad Since resolving the BLRs requires $\mu$as to tens
of $\mu$as angular resolution, RM will remain the primary method to measure
the size of BLR in the next decade or two. Another indirect method to measure
BLR size is via microlensing in gravitationally lensed quasars. The Einstein
radius in the source (quasar) plane of a point-mass lens with mass $M$ is
$r_E=\sqrt{4GMD_sD_{LS}/(D_Lc^2)}$, where $D_s$, $D_L$ and $D_{LS}$ are the
angular diameter distances to the source, to the lens, and from the lens to
the sourse. For typical values $z_{L}=0.5$ and $z_{S}=2$ we have $r_E\approx
0.01\sqrt{M/M_\odot}\,{\rm pc}$, corresponding to an angular scale
$\theta_E=r_E/D_S\approx \sqrt{M/M_\odot}\,\mu$as. This scale is comparable
to the size of the BLR and accretion disk. Due to the relative transverse
motion between the lens and source, large magnification can happen on short
time scales ($t_{\rm cross}=r_{\rm source}/v_{\perp}$) during
caustic-crossing in the source plane \citep[for a primer on gravitational
lensing, see, e.g.,][]{Schneider_etal_1992}. Since the continuum emission
(from the inner accretion disk) and BLR emission have different spatial
scales\footnote{The BLR and the accretion disk are probably not disjointed,
i.e., some portion of the BLR gas could originate from the outer part of the
accretion disk or from a wind launched from the disk, and high-ionization
lines could have different origins from low-ionization lines, as in some
models
\citep[e.g.,][]{Collin-Souffrin_etal_1988,Elvis_2000,Risaliti_Elvis_2010}.},
microlensing will cause differential variability for the continuum and
different broad lines, which can be used to constrain the size (and geometry)
of the emitting regions, such as the accretion disk and the BLR
\citep[e.g.,][]{Irwin_etal_1989,Lewis_etal_1998,Popovic_etal_2001,Richards_etal_2004,
Morgan_etal_2010,Dai_etal_2010,Mosquera_Kochanek_2011,Sluse_etal_2011,Sluse_etal_2012,Guerras_etal_2012}.
In particular, the latest study by \citet{Guerras_etal_2012} found a $R-L$
relation based on microlensing BLR sizes, which is in reasonably good
agreement with that based on RM. This technique is primarily applied to
quasars that are already strongly lensed (with multiple images) for which the
microlensing probability by stars/compact objects in the foreground lens
galaxy is high, and the time delays between different images are known. The
latter is important to rule out variability due to intrinsic AGN variability,
which will often complicate the microlensing interpretation.

\noindent{\em Direct dynamical BH masses}\quad Although observationally
challenging (given the overwhelming AGN continuum that dilutes the stellar
absorption features and nongravitational forces on gas dynamics in the
nucleus), there have been several attempts to get direct dynamical
measurements of BH masses in Type 1 AGNs, using spatially resolved stellar
kinematics \citep[e.g.,][]{Davies_etal_2006,Onken_etal_2007} or gas
kinematics \citep[e.g.,][]{Hicks_Malkan_2008} down to the sphere of
influence, $R_{\rm SI}=GM_{\rm BH}/\sigma_*^2$, of the BH. The number of AGNs
with reliable dynamical BH mass measurements is still small, and it would be
important to obtain more dynamical mass measurements of AGNs to provide
critical consistency checks on virial BH masses.

\noindent{\em Other indirect methods}\quad These methods use correlations
between BH mass and other measurable quantities, calibrated with known BH
masses, to infer the BH mass in quasars. By virtue they can be applied to
non-broad-line quasars as well, if the required quantity is measurable, but
caution should be paid to possible systematics of each method. These
correlations include the well-known BH-host scaling relations found for local
inactive galaxies. For instance, the $M_{\rm BH}-\sigma_*$ relation in local
inactive galaxies is used to predict BH masses in local RM AGNs with
$\sigma_{*}$ measurements (see \S\ref{s:vir_f}). Another example is the
observed anti-correlations between BH mass (estimated with other methods) and
X-ray variability properties of AGNs, such as the break frequency in the
power spectral densities (PSDs) of X-ray light curves or variability
amplitude
\citep[e.g.,][]{Papadakis_2004,ONeil_etal_2005,McHardy_etal_2006,Zhou_etal_2010},
which have the potential to provide independent BH mass estimates to within a
factor of a few to the RM masses. The reliability and systematics of the
X-ray variability method, however, are yet to be explored with larger AGN
samples with known BH masses.

\section{Caveats, Uncertainties, and Biases of Virial BH Masses}\label{s:bias}

\subsection{Physical Concerns}\label{s:phy_concern}

\subsubsection{The virial assumption}\label{s:virial_assump}

There are evidence supporting the virial assumption in RM in at least several
AGNs
\citep[e.g.,][]{Peterson_Wandel_1999,Peterson_Wandel_2000,Onken_Peterson_2002,Kollatschny_2003}.
For these objects RM lags have been successfully measured for multiple lines
with different ionization potentials (such as \hbeta, \CIV, \HeII) and line
widths, which are supposed to arise at different distances, as in a
stratified BLR for different lines. The measured lags and line widths of
these different lines fall close to the expected virial relation $W\propto
R^{-1/2}$, although such a velocity-radius scaling does not necessarily rule
out other BLR models where the dynamics is not dominated by the gravity of
the central BH \citep[e.g., see discussions in][]{Krolik_2001}. A more
convincing argument is based on velocity-resolved RM, where certain dynamical
models (such as outflows) can be ruled out based on the difference (or lack
thereof) in the lags from the blue and red parts of the line
\citep[e.g.,][]{Gaskell_1988}. On the other hand, non-virial motions (such as
infall and/or outflows) may indeed be present in some BLRs, as inferred from
recent velocity-resolved RM in a handful of AGNs
\citep[e.g.,][]{Denney_etal_2009a,Bentz_etal_2010,Grier_etal_2013}.
Fortunately, even if the BLR is in a non-virial state, one might still expect
that the velocity of the BLR clouds (as measured through the line width) does
not deviate much from the virial velocity. Thus using Eqn. (\ref{eq:rm_mass})
does not introduce a large bias, and in principle this detail is accounted
for by the virial coefficient $f$ in individual sources.

\begin{figure}
\includegraphics[width=0.45\textwidth]{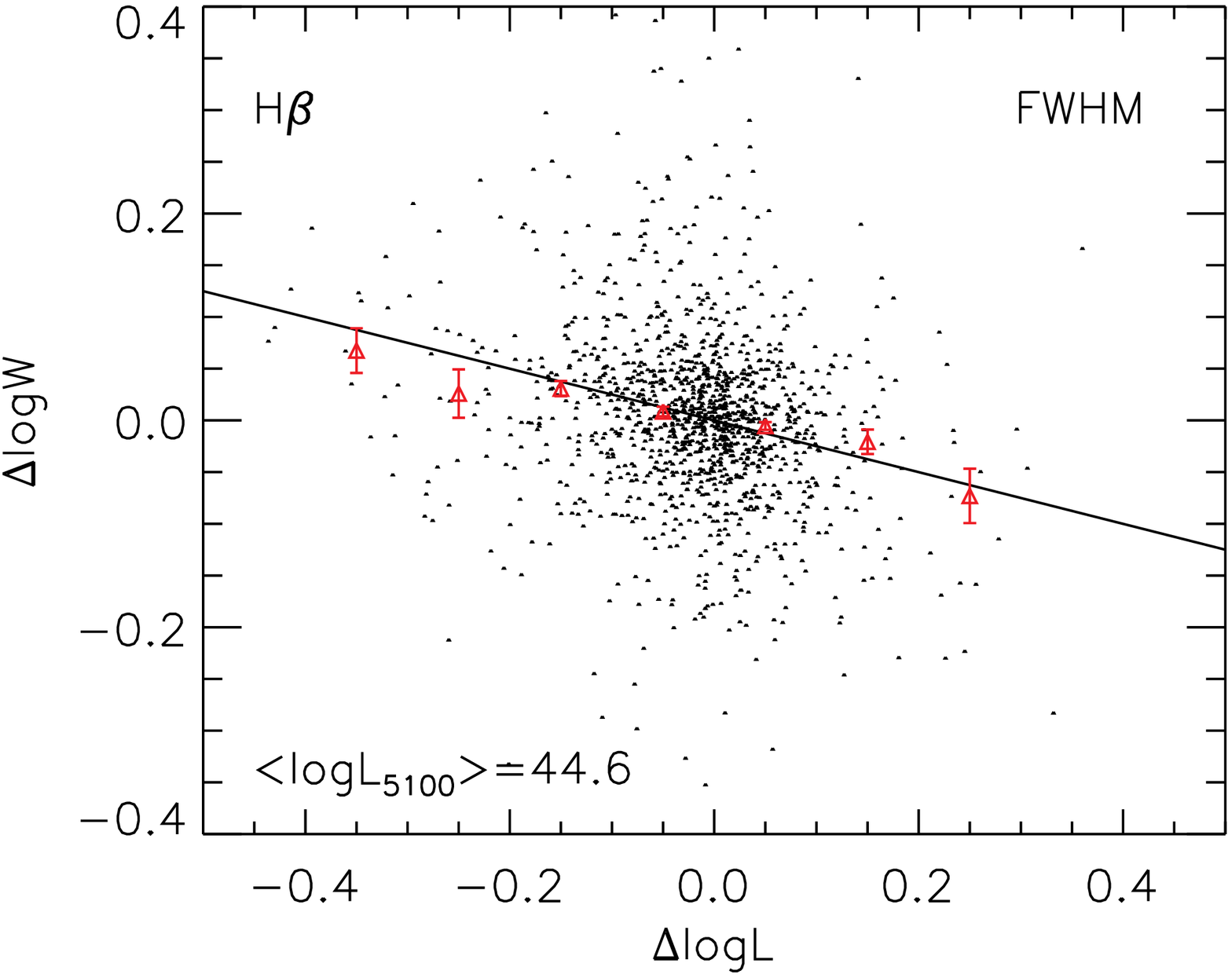}
\includegraphics[width=0.45\textwidth]{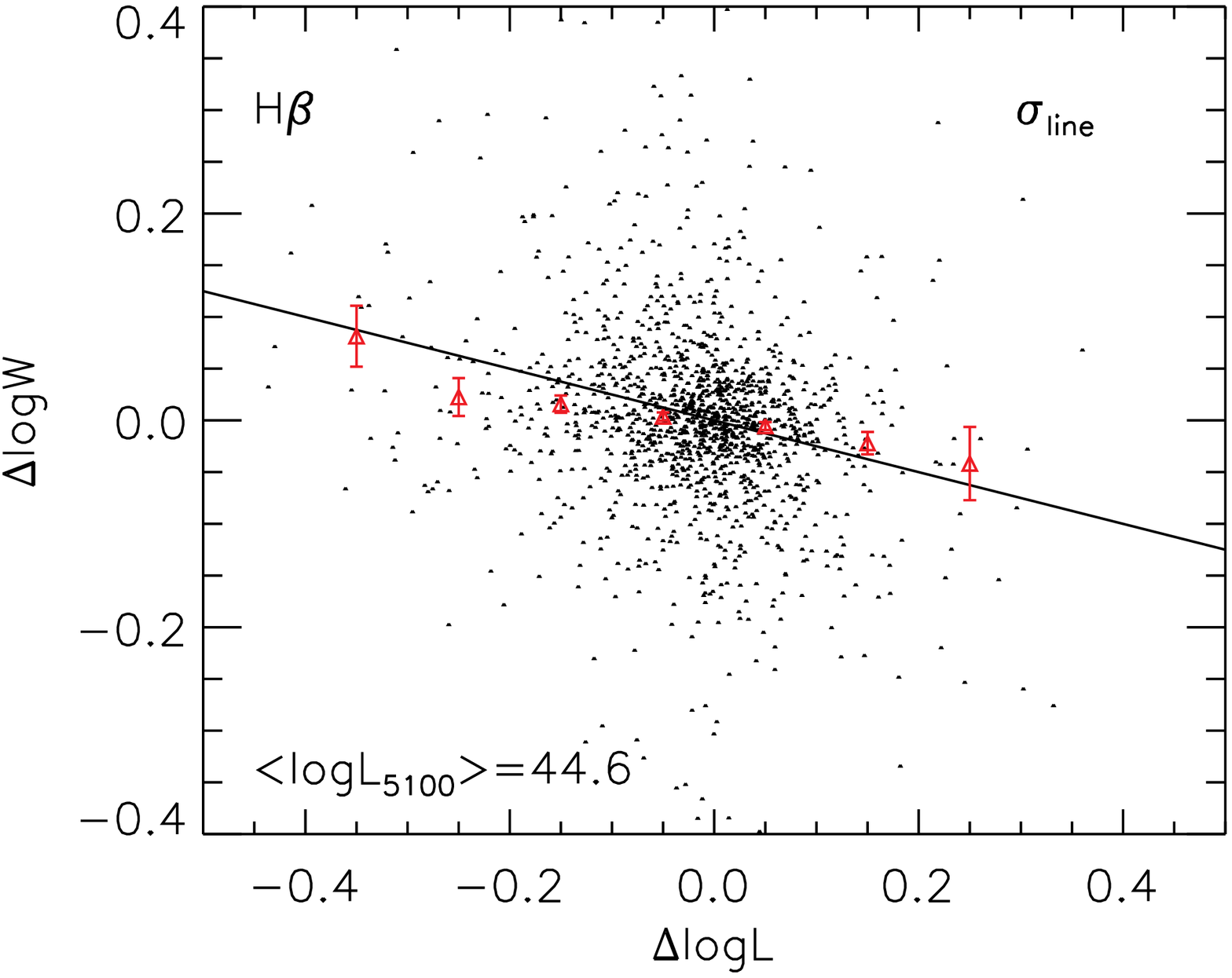}\\
\includegraphics[width=0.45\textwidth]{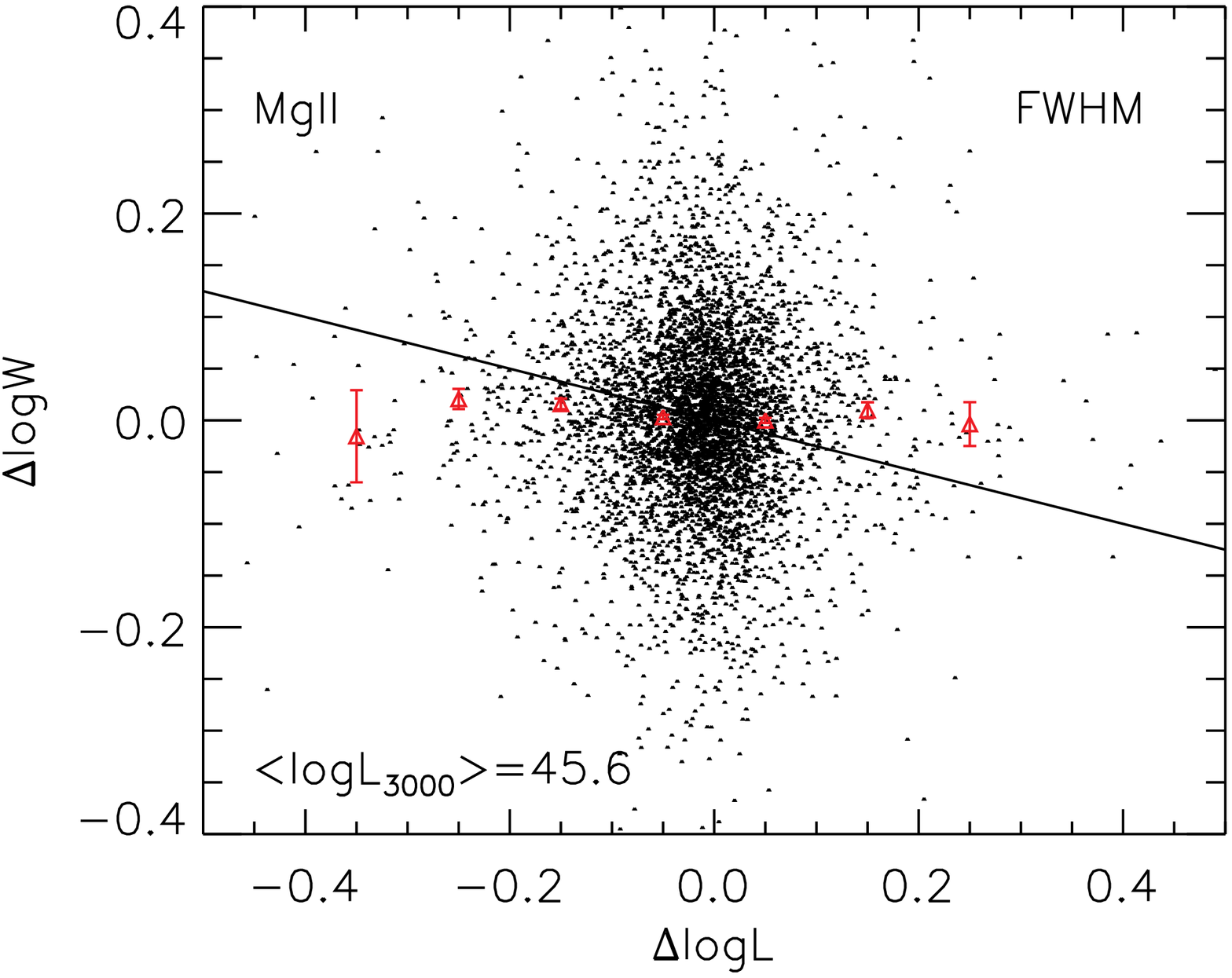}
\includegraphics[width=0.45\textwidth]{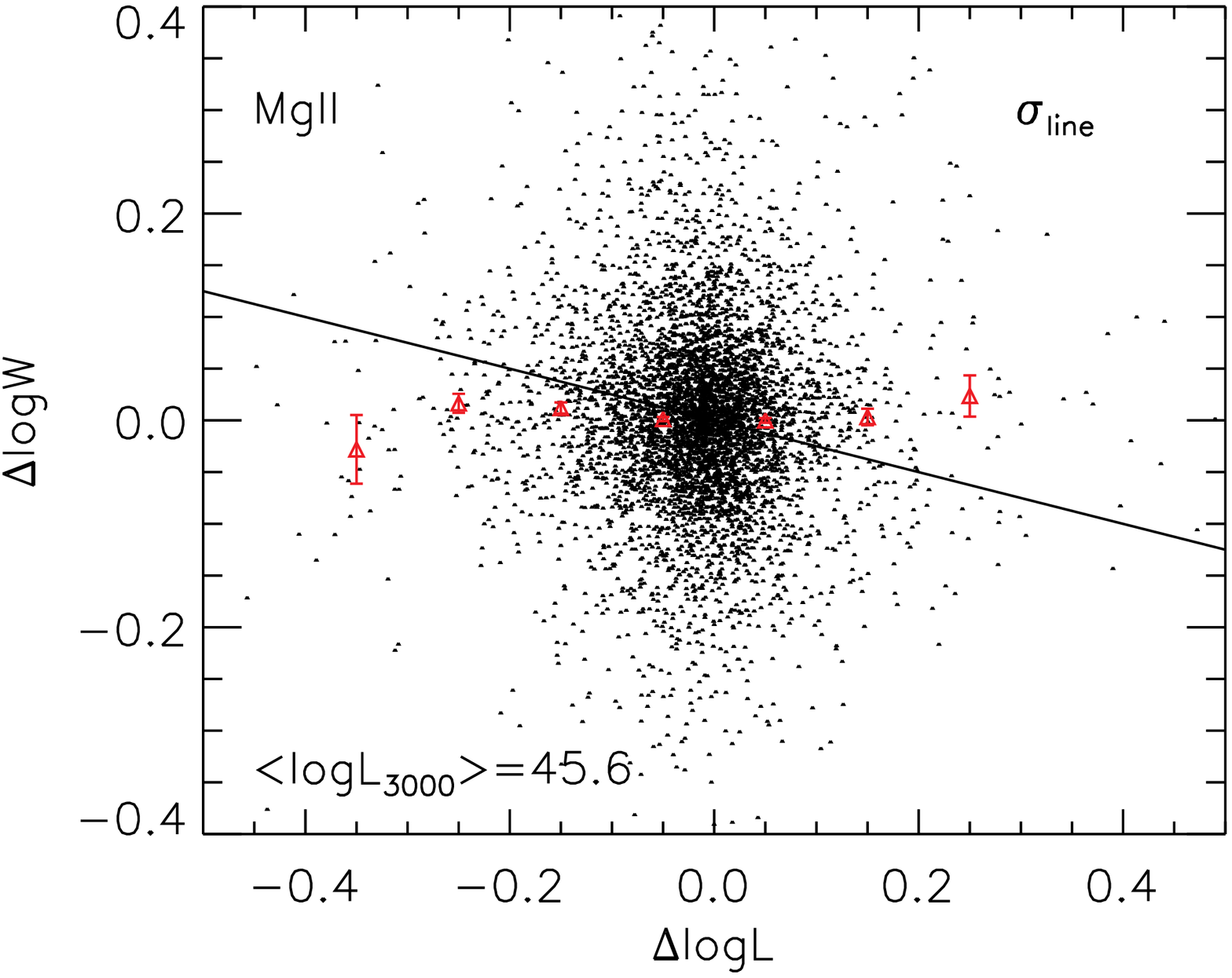}\\
\includegraphics[width=0.45\textwidth]{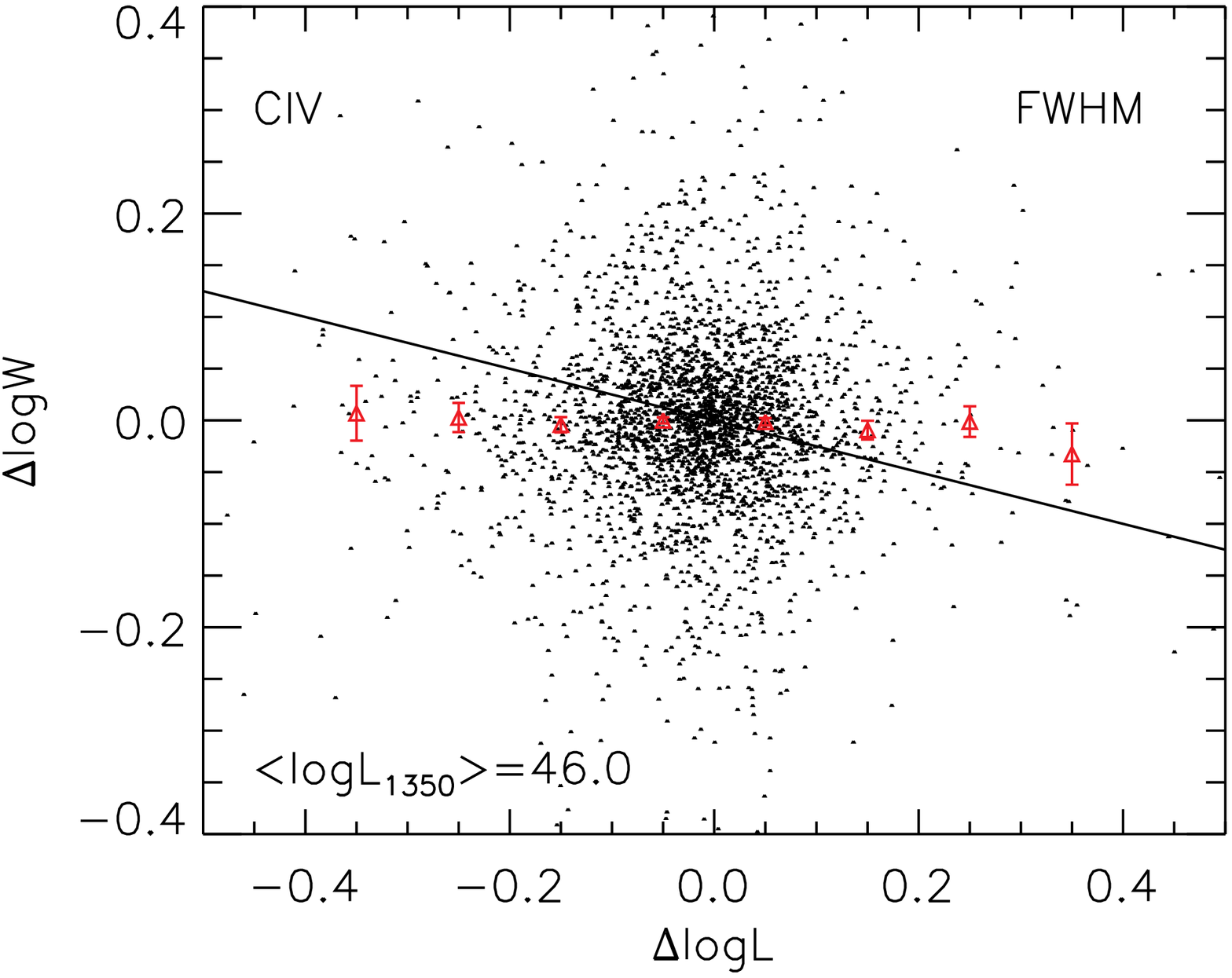}
\includegraphics[width=0.45\textwidth]{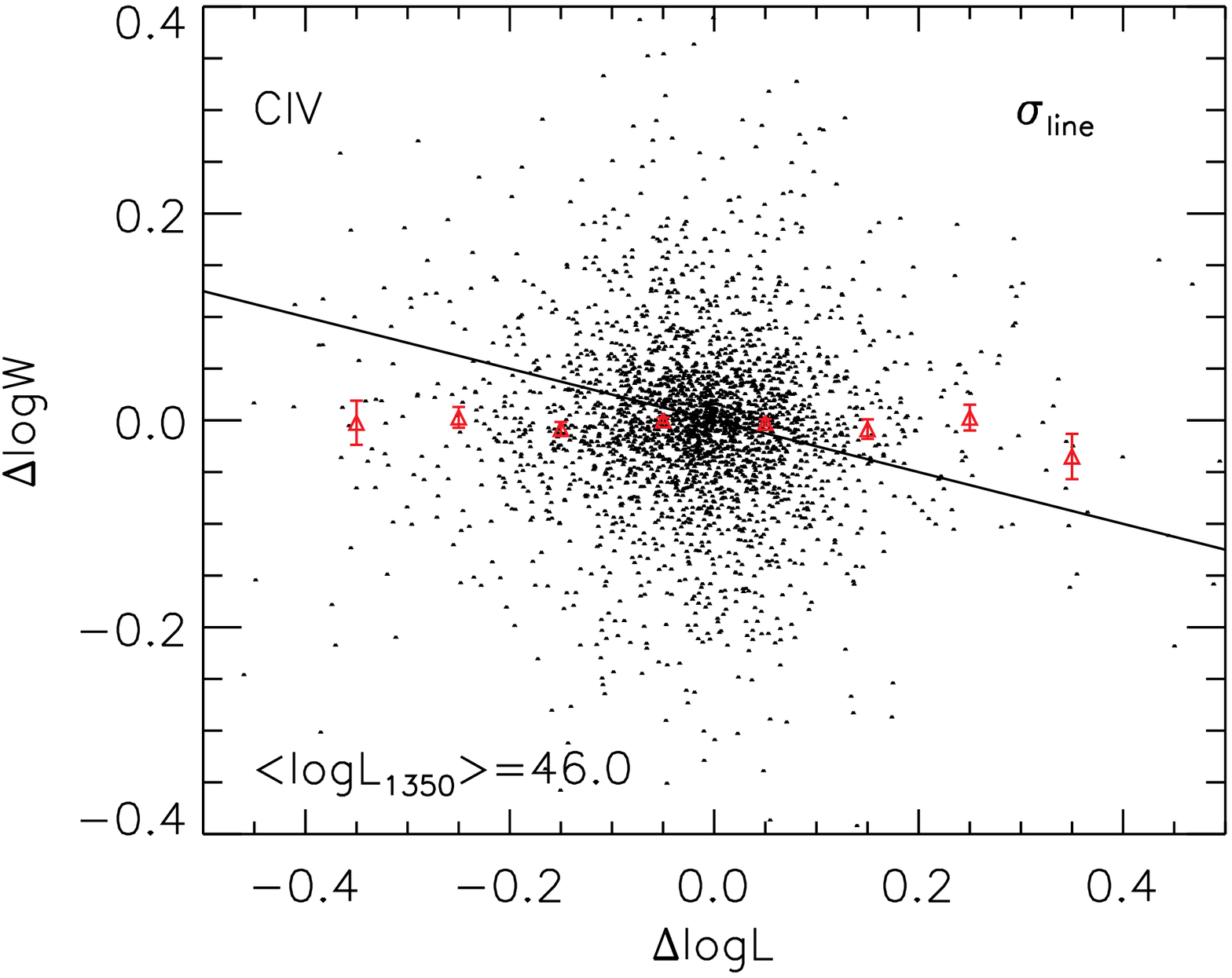}
\caption{A test of the virial assumption using two-epoch spectroscopy from
SDSS for \hbeta\ (upper), \MgII\ (middle) and \CIV\ (bottom). Plotted here are the
changes in line width as a function of changes in continuum luminosity ($L_{5100}$, $L_{3000}$, and $L_{1350}$ for \hbeta, \MgII, and
\CIV, respectively) between the two
epochs. The left column is for FWHM and the right column is for $\sigma_{\rm line}$. The dots
are for all objects with measurement S/N$>3$ at both epochs for both $L$ and
$W$ (not for $\Delta\log L$ and $\Delta\log W$). These objects tend to cluster around zero values because the typical continuum and line
variabilities
of SDSS quasars are small. The red triangles are the median values in each $\Delta\log L$ bin, where the
error bars indicate the uncertainty in the mean. A perfect virial relation would imply
$\Delta\log W=-0.25\Delta\log L$, as indicated by
the solid line in each panel. Note that I have neglected the chromatic nature of quasar variability, which
would predict an even steeper relation between $\Delta\log W$ and $\Delta\log L$ (see \S\ref{s:virial_assump} for details). The low-redshift
$z\lesssim 0.7$ SDSS quasars with median
luminosity $\langle\log (L_{5100}/{\rm erg\,s^{-1}})\rangle=10^{44.6}$ show the expected
virial relation between $\Delta\log L$ and $\Delta\log W$, which is not the case for the high-luminosity
SDSS quasars at $z>0.7$ based on \MgII\ or \CIV. }\label{fig:twoepoch_hbeta}
\end{figure}

A further test of the virial assumption on the single-epoch virial estimators
is to see if the line width varies in accordance to the changes in luminosity
for the same object. The picture here is that when luminosity increases
(decreases) the BLR expands (shrinks), and the line width should decrease
(increase), given enough response time. This test is important, because if
the line width does not change accordingly to luminosity changes, the SE mass
will change for the same object, introducing a luminosity-dependent bias in
the mass estimates (see \S\ref{s:bias_l}). This test is challenging in
practice, given the limited dynamic range in continuum variations and the
presence of measurement errors. Nevertheless, in several AGNs with
high-quality RM data, such anti-correlated variations of line width and BLR
size (or continuum luminosity) have been seen
\citep[e.g.,][]{Peterson_etal_2004,Park_etal_2012}, once the lag between
continuum and line variations is taken into account. While this lends some
further support for RM and SE virial estimators, it should be noted that: 1)
not all RM AGNs show this expected behavior, given insufficient data quality;
2) it makes a difference which line width measurements (i.e., FWHM vs
$\sigma$, rms vs mean spectra) and which BLR size estimates (i.e., $\tau$ vs
continuum luminosity) are used.

It is also not clear if the above results based on a few RM AGNs apply to the
general quasar population. Fig.\ \ref{fig:twoepoch_hbeta} shows a test of the
co-variation of line width and continuum luminosity using thousands of SDSS
quasars with spectra at two epochs (She, Shen, et al., in prep). While for
the majority of these quasars the two epochs do not span a large dynamic
range in luminosity, the large number of objects provide good statistical
constraints on the average trend. In Fig.\ \ref{fig:twoepoch_hbeta} the black
dots are measurements for individual objects, and they cluster near the
center because most quasars do not vary much between the two epochs. The
measurement uncertainties on $\Delta\log L$ and $\Delta\log W$ are large, so
I bin the results in $\Delta\log L$ bins and plot the medians and
uncertainties in the median in each bin in red triangles. The measurement
uncertainties in $\Delta\log L$ and $\Delta\log W$ are comparable for all
three lines, but only for the low-luminosity and low-$z$ ($z\lesssim 0.7$)
\hbeta\ sample is the median relation consistent with the virial relation
(the solid lines in Fig.\ \ref{fig:twoepoch_hbeta}). For the other samples at
$z>0.7$ based on \MgII\ and \CIV, the line width does not seem to respond to
luminosity changes as expected from the virial relation. This difference
could be a luminosity effect, but more detailed analyses are needed (She,
Shen, et~al., in prep).

Another important point to make is that there is a well known fact that
quasar spectra get harder (bluer) as they get brighter \citep[e.g.,][and
references therein]{VandenBerk_etal_2004}. This means that the variability
amplitude in the ionizing continuum should be larger than that at longer
wavelengths (i.e., the observed continuum). Thus we should see a somewhat
steeper slope in the line width change versus the (observed) continuum
luminosity change plot for a single object
\citep[e.g.,][]{Peterson_etal_2002}. This, however, would be in an even
larger disagreement with the trends we see in Fig.\ \ref{fig:twoepoch_hbeta}.

%

\subsubsection{The virial coefficient $f$}\label{s:vir_f}

To relate the observed broad line width to the underlying virial velocity
(e.g., Eqn.\ \ref{eq:rm_mass}) requires the knowledge of the (emissivity
weighted) geometry and kinematics of the BLR. In principle RM can provide
such information, and determine the value of $f$ from first principles.
Unfortunately the current RM data are still not good enough for such purposes
in general, although in a few cases alternative approaches have been invented
lately to account for the effect of $f$ in directly modeling the RM data
using dynamical BLR models
\citep[e.g.,][]{Brewer_etal_2011,Pancoast_etal_2012}. Early studies made
assumptions about the geometry and structure of the BLR in deriving RM masses
\citep[e.g.,][]{Netzer_1990,Wandel_etal_1999,Kaspi_etal_2000} or SE virial
masses \citep[e.g.,][]{McLure_Dunlop_2004}. Now the average value of $f$ is
mostly determined empirically by requiring that the RM masses are consistent
with those predicted from the $M_{\rm BH}-\sigma_*$ relation of local
inactive galaxies. Such an exercise was first done by
\citet{Onken_etal_2004}, who used 16 local AGNs with both RM measurements and
stellar velocity dispersion measurements to derive $\langle
f\rangle\approx1.4$ if FWHM is used, or $\langle f\rangle\approx 5.5$ if
$\sigma_{\rm line}$ is used. Later this was repeated with new RM data
\citep[e.g.,][]{Woo_etal_2010}, who derived a similar value of $\langle
f\rangle\approx 5.2$ (using $\sigma_{\rm line}$).

However, in recent years it has become evident that the scaling relations
between BH mass and bulge properties are not as simple as we thought: it
appears that different types of galaxies follow somewhat different scaling
relations, and the scatter seems to increase towards less massive systems
\citep[e.g.,][and references
therein]{Hu_2008,Greene_etal_2008,Graham_2008,Graham_Li_2009,Hu_2009,Gultekin_etal_2009,Greene_etal_2010b,McConnell_Ma_2012}.
Therefore, depending on the choice of the specific form of the $M_{\rm
BH}-\sigma_*$ relation used and the types of galaxies hosting RM AGNs in the
calibration, the derived average $f$ value could vary significantly. For
instance, \citet{Graham_etal_2011} derived a $\langle f\rangle$ value that is
is only half of the values derived by \citet{Onken_etal_2004} and
\citet{Woo_etal_2010}. \citet{Park_etal_2012} performed a detailed
investigation on the effects of different regression methods and sample
selection in determining the $M_{\rm BH}-\sigma_*$ relation and in turn the
$\langle f\rangle$ value, and concluded that the latter is the primary cause
for the discrepancy in the reported $\langle f\rangle$ values. Given the
small sample sizes of RM AGNs with host property measurements and the
uncertainties in the BH-host scaling relations in inactive galaxies, the
uncertainty of $\langle f\rangle$ is still $\sim $ a factor of 2 or more, and
will remain one of the main obstacles to estimate accurate RM (or SE) BH
masses in terms of the overall normalization. One may also expect that the
actual $f$ value is different in individual sources, either from the
diversity in BLR structure or from orientation effects (since the line width
only reflects the line-of-sight velocity, see \S\ref{s:ori_rad}). Thus using
a constant $f$ value in these RM masses and SE virial estimators introduces
additional scatter in these mass estimates.

Perhaps a more serious concern is the assumption that the BH-host scaling
relations are the same in active and inactive galaxies. While there is a
clear correlation between bulge properties and the RM masses in RM AGNs
\citep[e.g.,][]{Bentz_etal_2009c}, it could be offset from that for inactive
galaxies if the actual $\langle f\rangle$ value is different. Such a scenario
is plausible if the BH growth and host bulge formation are not always
synchronized. The only way to tackle this problem is to infer $f$ from
directly constrained BLR geometry/kinematics with exquisite velocity-resolved
RM data that map the line response (transfer function) in detail, and this
must be done for a large number of AGNs to explore its diversity.

\subsubsection{FWHM versus line dispersion}\label{s:line_width}

Both FWHM and $\sigma_{\rm line}$ are commonly used in SE virial mass
estimates as the proxy for the virial velocity (when combined with the virial
coefficient $f$). Both definitions have advantages and disadvantages. FWHM is
a quantity that is easier to measure, less susceptible to noise in the wings
and treatments of line blending than $\sigma_{\rm line}$, while $\sigma_{\rm
line}$ is less sensitive to the treatment of narrow line removal and peculiar
line profiles. Overall FWHM is preferred over $\sigma_{\rm line}$ in terms of
easiness of the measurement and repeatability. As $\sigma_{\rm line}$
measurements depend sensitively on data quality and different methods used
\citep[e.g.,][]{Denney_etal_2009b,Rafiee_Hall_2011a,Rafiee_Hall_2011b,Assef_etal_2011},
the SE virial masses (e.g., Eqn.\ \ref{eqn:virial_mass}) based on
$\sigma_{\rm line}$ could differ significantly for the same objects.

Physically one may argue $\sigma_{\rm line}$ is more trustworthy to use than
FWHM, although the evidence to date is only suggestive.
\citet[][]{Collin_etal_2006} compared the virial products based on both
$\sigma_{\rm line}$ and FWHM with those expected from the $M_{\rm
BH}-\sigma_*$ relation, for 14 RM AGNs. All their line width measurements
were based on the rms or mean spectra of the RM AGNs. They found that the
average scale factor (i.e., the virial coefficient $f$) between virial
products to the $M_{\rm BH}-\sigma_*$ masses depends on the shape of the line
if FWHM is used, while it is more or less constant if $\sigma_{\rm line}$ is
used. Based on this, they argue that $\sigma_{\rm line}$ is a better
surrogate to use in estimating RM masses. Additionally, $\sigma_{\rm line}$
measured in rms spectra seems to follow the expected virial relation better
than FWHM in some RM AGNs \citep[e.g.,][]{Peterson_etal_2004}, although such
evidence is circumstantial.

It is important to note that for a given line, the ratio of FWHM to
$\sigma_{\rm line}$ is not necessarily a constant \citep[e.g.,][but cf.,
Decarli et al.\ 2008a]{Collin_etal_2006,Peterson_2011}, while a Gaussian line
profile leads to ${\rm FWHM}/\sigma_{\rm line}\approx 2.35$. For \hbeta,
${\rm FWHM}/\sigma_{\rm line}$ seems to increase when the line width
increases. This might be related to the Populations A and B sequences
developed by Sulentic and collaborators \citep[][]{Sulentic_etal_2000a},
which is an extension of earlier work on the correlation space of AGNs
\citep[the so-called ``eigenvector 1'',
e.g.,][]{Boroson_Green_1992,Wang_etal_1996}. A direct consequence is that
there will be systematic differences in $M_{\rm SE}$ whether FWHM or
$\sigma_{\rm line}$ is used for the same set of quasars, especially for
objects with extreme line widths. In general a ``tilt'' between the FWHM and
$\sigma_{\rm line}$-based virial masses is expected
\citep[e.g.,][]{Rafiee_Hall_2011a,Rafiee_Hall_2011b}. Currently directly
measuring $\sigma_{\rm line}$ from single-epoch spectra is much more
ambiguous and methodology-dependent than measuring FWHM. If one accepts that
$\sigma_{\rm line}$ is a more robust virial velocity indicator, it is
possible to convert the measured FWHM to $\sigma_{\rm line}$ using the
relation found for high S/N data \citep[e.g.,][]{Collin_etal_2006}, or
empirically determine the dependence of SE mass on FWHM (i.e., coefficient
$c$ in Eqn.\ \ref{eqn:virial_mass}) using RM masses as calibrators
\citep[e.g.,][]{Wang_etal_2009}, which generally leads to values of $c<2$.

The choice of line width indicators is still an open issue. It will be
important to revisit the arguments in, e.g., \citet[][]{Collin_etal_2006},
using not only more but also better-quality RM data, as well as to
investigate the behaviors of FWHM and $\sigma_{\rm line}$ (and perhaps
alternative line width measures) for large quasar samples.

\subsubsection{Broad line profiles}\label{s:line_prof}

As briefly mentioned in \S\ref{s:virial_mass}, part of the reason that we are
struggling with $f$ and line width definitions is because of the
simplifications of a single BLR size and using only one line profile
characteristic to infer the underlying BLR velocity structure. If we have a
decent understanding of the BLR dynamics and structure (geometry, kinematics,
emissivity, ionization, etc.), then in principle we can solve the inverse
problem of inferring the virial velocity from the broad line profile.
Unfortunately, the detailed BLR properties are yet to be probed with
velocity-resolved reverberation maps, and the solution of this inverse
problem may not be unique (e.g., different BLR dynamics and structure may
produce similar line profiles).

Nevertheless, there have been efforts to model the observed broad line
profiles with simple BLR models. The best known example is the disk-emitter
model
\citep[e.g.,][]{Chen_etal_1989,Eracleous_Halpern_1994,Eracleous_etal_1995},
where a Keplerian disk with a turbulent broadening component is used to model
the double-peaked broad line profile seen in $\sim 10-15\%$ radio-loud
quasars (and several percent of radio-quite quasars). The line profile then
can place constraints on certain geometrical parameters, such as the
inclination of the disk, thus has relevance in the $f$ value for individual
objects \citep[e.g.,][]{LaMura_etal_2009}. Another example is using simple
kinematic BLR models to explain the trend of the line shape parameter
FWHM$/\sigma_{\rm line}$ as a function of line width
\citep[e.g.,][]{Kollatschny_Zetzl_2011,Kollatschny_Zetzl_2013}, as mentioned
earlier in \S\ref{s:line_width}. These authors found that a turbulent
component broadened by a rotation component can explain the observed trend of
line shape parameter, and their model provides conversions between the
observed line width and the underlying virial (rotational) velocity. More
complicated BLR models can be built \citep[e.g.,][]{Goad_etal_2012}, which
has the potential to underpin a physical connection between the BLR structure
and the observed broad line characteristics. While all these exercises are
worth further investigations, it is important to build self-consistent models
that are also verified with velocity-resolved RM.


\subsubsection{Effects of host starlight and dust reddening}\label{s:host_contam}

The luminosity that enters the $R-L$ relation and the SE mass estimators
(Eqn.\ \ref{eqn:virial_mass}) refers to the AGN luminosity. At low AGN
luminosities, the contamination from host starlight to the 5100\AA\
luminosity can be significant. This motivated the alternative uses of Balmer
line luminosities in Eqn.\ (\ref{eqn:virial_mass})
\citep[e.g.,][]{Greene_Ho_2005}. Using line luminosity is also preferred for
radio-loud objects where the continuum may be severely contaminated by the
nonthermal emission from the jet \citep[e.g.,][]{Wu_etal_2004}.
\citet{Bentz_etal_2006} and \citet{Bentz_etal_2009a} showed that properly
accounting for the host starlight contamination at optical luminosities in RM
AGNs leads to a slope in the $R-L$ relation that is closer to the naive
expectation from photoionization. Similarly, using host-corrected $L_{\rm
5100}$ can lead to reduced scatter in the \hbeta-$L_{\rm 5100}$ SE
calibration against RM AGNs \citep[e.g.,][]{Shen_Kelly_2011}.

The average contribution of host starlight to $L_{5100}$ has been quantified
by \citet{Shen_etal_2011}, using low-redshift SDSS quasars. They found that
significant host contamination ($\gtrsim 20\%$) is present for $\log L_{\rm
5100,total}<10^{44.5}\,{\rm erg\,s^{-1}}$, and provided an empirical
correction for this average contamination. Variations in host contribution
could be substantial for individual objects though.

For UV luminosities ($L_{3000}$, $L_{1350}$ or $L_{1450}$), the host
contamination is usually negligible, although may be significant for rare
objects with excessive ongoing star formation. A more serious concern,
however, is that some quasars may be heavily reddened by dust internal or
external to the host. The so-called ``dust-reddened'' quasars
\citep[e.g.,][]{Glikman_etal_2007} have UV luminosities significantly dust
attenuated, and corrections are required to measure their intrinsic AGN
luminosities. It is possible that optical quasar surveys (such as SDSS) are
missing a significant population of dust-reddened quasars.

\subsubsection{Effects of orientation and radiation
pressure}\label{s:ori_rad}

If the BLR velocity distribution is not isotropic, orientation effects may
affect the RM and SE mass estimates. Specific BLR geometry and kinematics,
such as a flattened BLR where the orbits are confined to low latitudes, will
lead to orientation-dependent line width. Some studies report a correlation
between the broad line FWHM and the source orientation inferred from radio
properties\footnote{Some recent studies
\citep[e.g.,][]{Fine_etal_2011,Runnoe_etal_2012} argue that the dependence of
line FWHM on source orientation is different for low-ionization and
high-ionization lines, such that the \CIV-emitting gas velocity field may be
more isotropic than \hbeta\ and \MgII.}
\citep[e.g.,][]{Wills_Browne_1986,Jarvis_McLure_2006}, in favor of a
flattened BLR geometry. Similar conclusions were achieved in
\citet{Decarli_etal_2008a} based on somewhat different arguments. Since we
use the average virial coefficient $\langle f\rangle$ in our RM and SE mass
estimates, the true BH masses in individual sources may be over- or
underestimated depending on the actual inclination of the BLR
\citep[e.g.,][]{Krolik_2001,Decarli_etal_2008a,Fine_etal_2011,Runnoe_etal_2012}\footnote{Of
some relevance here is the interpretation of the apparently small BH masses
in a sub-class of Type 1 AGNs called narrow-line Seyfert 1s (NLS1s), where
the \hbeta\ FWHM is narrower than 2000$\,{\rm km\,s^{-1}}$ along with other
unusual properties (such as strong iron emission and weak \OIII\ emission).
Some argue \citep[e.g.,][]{Decarli_etal_2008b} that NLS1s are preferentially
seen close to face-on, hence their virial BH masses based on FWHM are
underestimations of true masses. However, NLS1s also differ from normal Type
1 objects in ways that are difficult to explain with orientation effects
(such as weak \OIII\ and strong X-ray variability). Orientation may play some
role in the interpretation of NLS1s (especially for a minority of radio-loud
NLS1s), but is unlikely to be a major factor. }. The distributions of broad
line widths in bright quasars are typically log-normal, with dispersions of
$\sim 0.1-0.2$ dex over $\sim 5$ magnitudes in luminosity
\citep[e.g.,][]{Shen_etal_2008b,Fine_etal_2008,Fine_etal_2010}. A thin
disk-like BLR geometry with a large range of inclination angles cannot
account for such narrow distributions of line width, indicating either the
inclination angle is limited to a narrow range for Type 1 objects, and/or
there is a significant random velocity component (such as turbulent motion)
of the BLR. This limits the scatter in BH mass estimates caused by
orientation effects to be $< 0.2-0.4$ dex.

So far we have assumed that the dynamics of the BLR is dominated by the
gravity of the central BH. The possible effects of radiation pressure, which
also has a $\propto R^{-2}$ dilution as gravity, on the BLR dynamics have
been emphasized by, e.g., \citet[][]{Krolik_2001}. On average the possible
radiation effects are eliminated in the empirical calibration of the $\langle
f\rangle$ value (see \S\ref{s:vir_f}), but neglecting such effects may
introduce scatter in individual sources and luminosity-dependent trends. Most
recently \citet{Marconi_etal_2008} modified the virial mass estimation by
adding a luminosity term:
\begin{equation}
M_{\rm BH,M08}=f\frac{W^2R}{G}+g\left(\frac{L_{5100}}{10^{44}\,{\rm erg\,s^{-1}}}\right)M_\odot\ ,
\end{equation}
where the last term describes the effect of radiation pressure on the BLR
dynamics with a free parameter $g$. By allowing this extra term,
\citet{Marconi_etal_2008} re-calibrated the RM masses using the $M_{\rm
BH}-\sigma_*$ relation, and the SE mass estimator using the new RM masses.
This approach improves the rms scatter between single-epoch masses and RM
masses, from $\sim 0.4$ dex to $\sim 0.2$ dex, and removes the slight
systematic trend of the SE mass scatter with RM masses seen in
\citet{Vestergaard_Peterson_2006}. However, it is also possible that the
reduction of scatter between the SE and RM masses is caused by the addition
of fitting freedoms. Since the intrinsic errors on the RM masses are unlikely
to be $<0.3$ dex, optimizing the SE masses relative to RM masses to smaller
scatter may lead to blown-up errors when apply the optimized scaling relation
to other objects. It would be interesting to split the RM sample in
\citet{Marconi_etal_2008} in half and use one half for calibration and the
other half for prediction, and see if similar scatter can be achieved in both
subsets. The relevance of radiation pressure is also questioned by
\citet{Netzer_2009}, who used large samples of Type 1 and Type 2 AGNs from
the SDSS to show that the radiation-pressure corrected viral masses lead to
inconsistent Eddington ratio distributions in Type 1s and Type 2s, even
though the \OIII\ luminosity distribution is consistent in the two samples.
However, \citet{Marconi_etal_2009} argues that the difference in the
``observed'' Eddington ratio distributions does not mean that radiation
pressure is not important, rather it could result from a broad range of
column densities which are not properly described by single values of
parameters in the radiation-pressure-corrected mass formula. These studies
then revealed that using the simple corrected formula as provided in
\citet{Marconi_etal_2008} does not provide a satisfactory recipe to account
for radiation pressure in RM or SE mass estimates, and the relevance of
radiation pressure and a practical method to correct for its effect are
therefore still under active investigations
\citep[e.g.,][]{Netzer_Marziani_2010}.

\subsubsection{Comparison among different line estimators}\label{s:comp_line}

There are both low-ionization and high-ionization broad lines in the
restframe UV to near-infrared of the quasar spectrum. Despite different
ionization potential and probably different BLR structure, several of them
have been adopted as SE virial mass estimators. The most frequently used
line-luminosity pairs include strong Balmer lines (\halpha\ and \hbeta) with
$L_{\rm 5100}$ or $L_{\textrm{\halpha,\hbeta}}$, \MgII\ with $L_{\rm 3000}$,
and \CIV\ with $L_{1350}$ or $L_{1450}$. Hydrogen Paschen lines in the
near-IR can also be used if such near-IR spectroscopy exists.

There have been SE calibrations upon specific lines against RM masses, or
against SE masses based on another line. Comparisons between different SE
line estimators using various quasar samples are often made in the
literature: some claim consistency, while others report discrepancy. As
emphasized in \citet{Shen_etal_2008b}, it is important to use a consistent
method in measuring luminosity and line width with that used for the
calibrations if one wants to make a fair comparison using external samples.
Failure to do so may lead to unreliable conclusions
\citep[e.g.,][]{Dietrich_Hamann_2004}.

\begin{figure}
\includegraphics[width=0.45\textwidth]{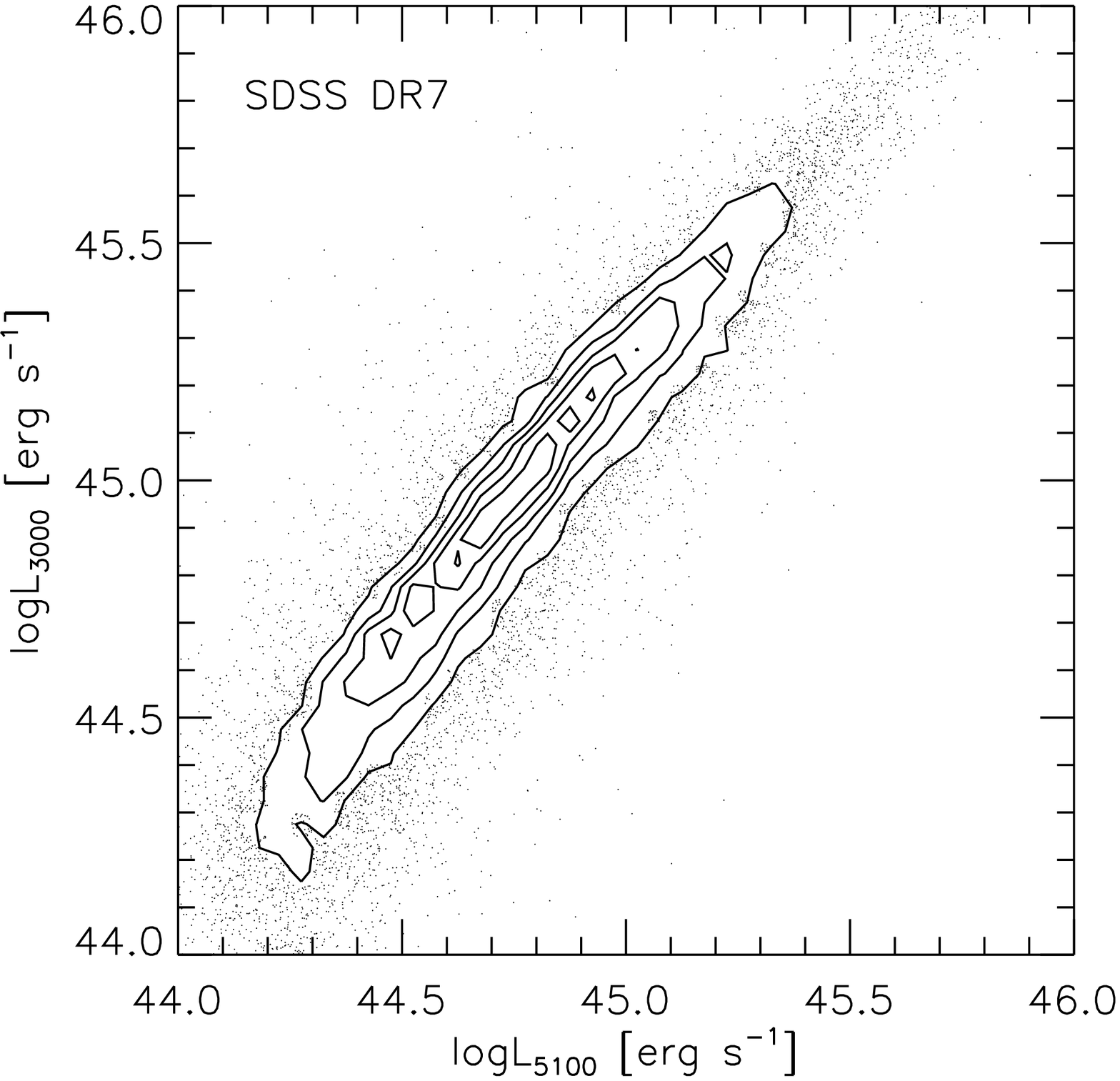}
\includegraphics[width=0.45\textwidth]{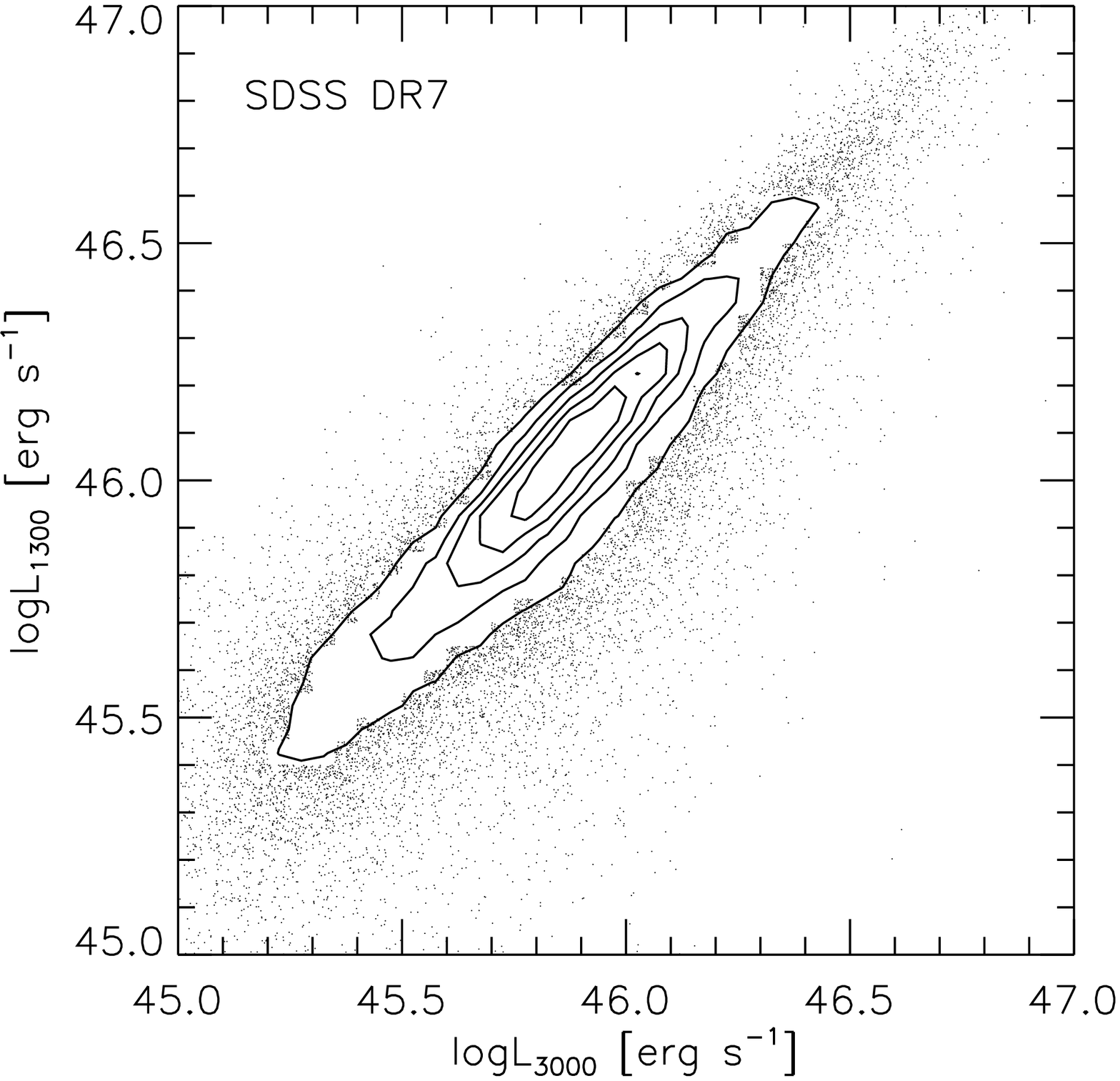}\\
\includegraphics[width=0.45\textwidth]{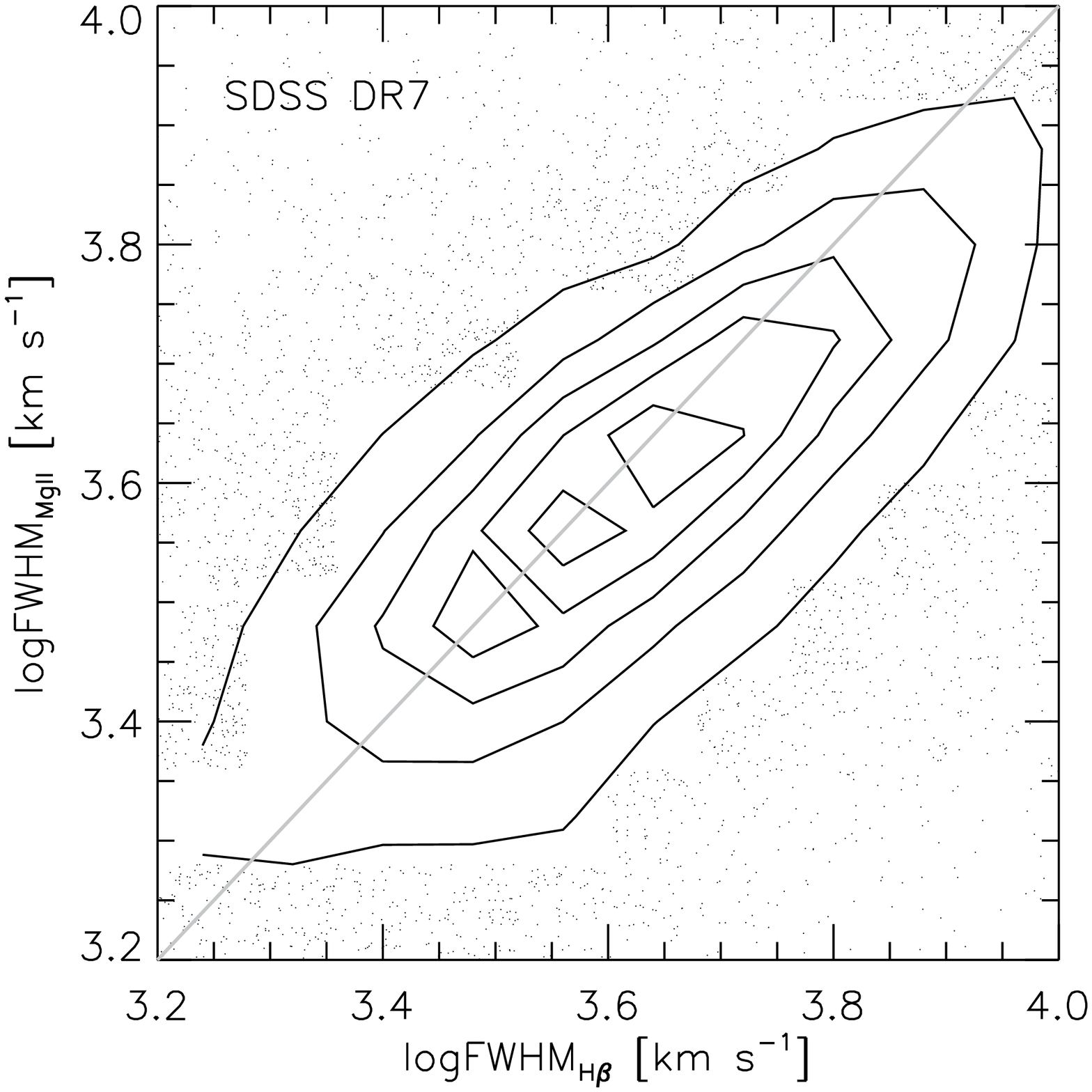}
\includegraphics[width=0.45\textwidth]{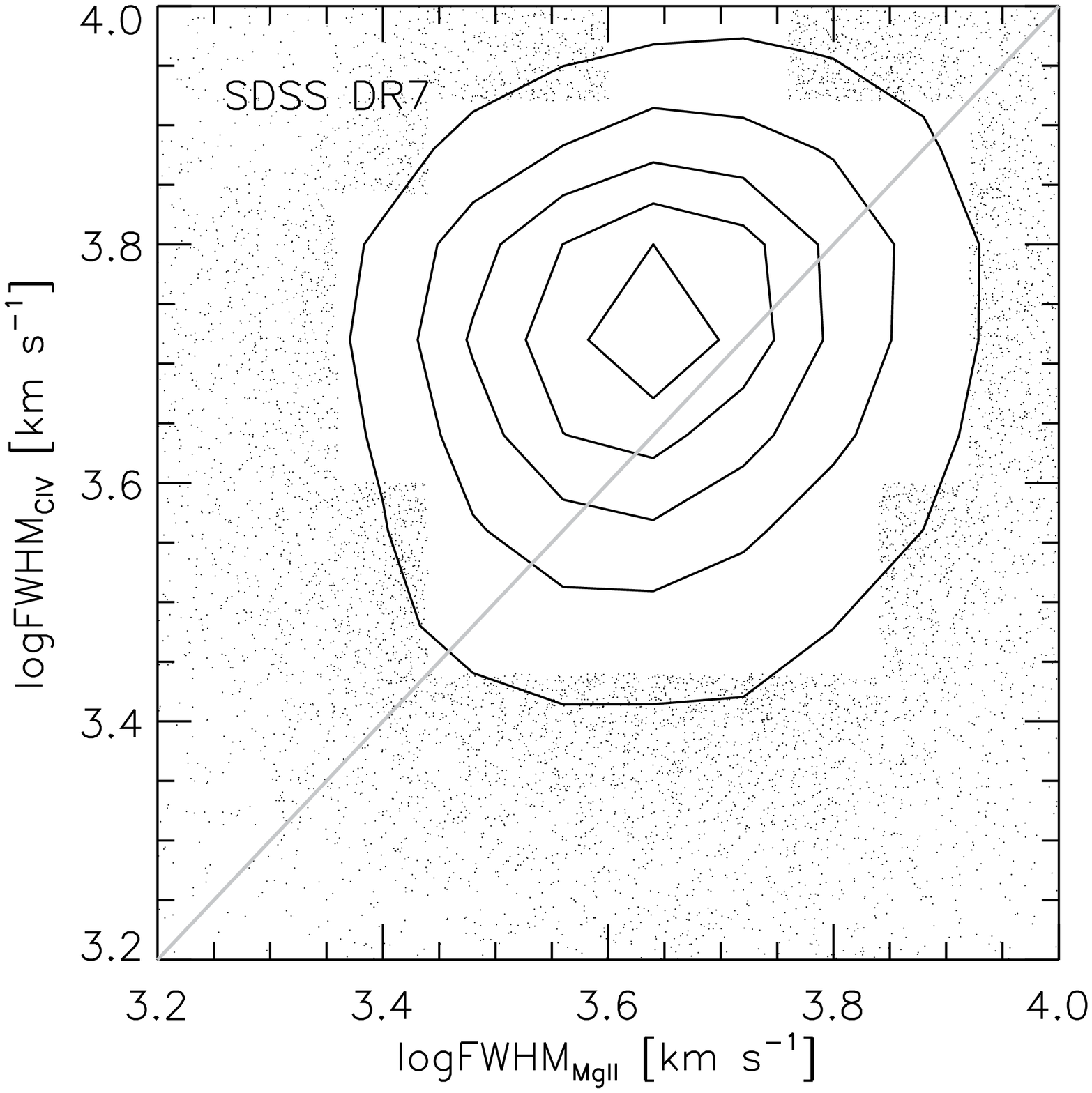}
\caption{Comparisons between different continuum luminosities and line FWHMs, using
SDSS quasar spectra that cover two lines. Shown here are the local point density contours.
Measurements are from \citet{Shen_etal_2011}. The upper panels show the correlations
between continuum luminosities, and the bottom panels show the correlations between line
FWHMs. While the \MgII\ FWHM correlates with \hbeta\ FWHM reasonably well, the correlation between
the \CIV\ FWHM and \MgII\ FWHM is poor \citep[also see, e.g.,][]{Shen_etal_2008b,Fine_etal_2008,Fine_etal_2010}.}\label{fig:sdss_comp}
\end{figure}

\begin{figure}
\includegraphics[width=0.45\textwidth]{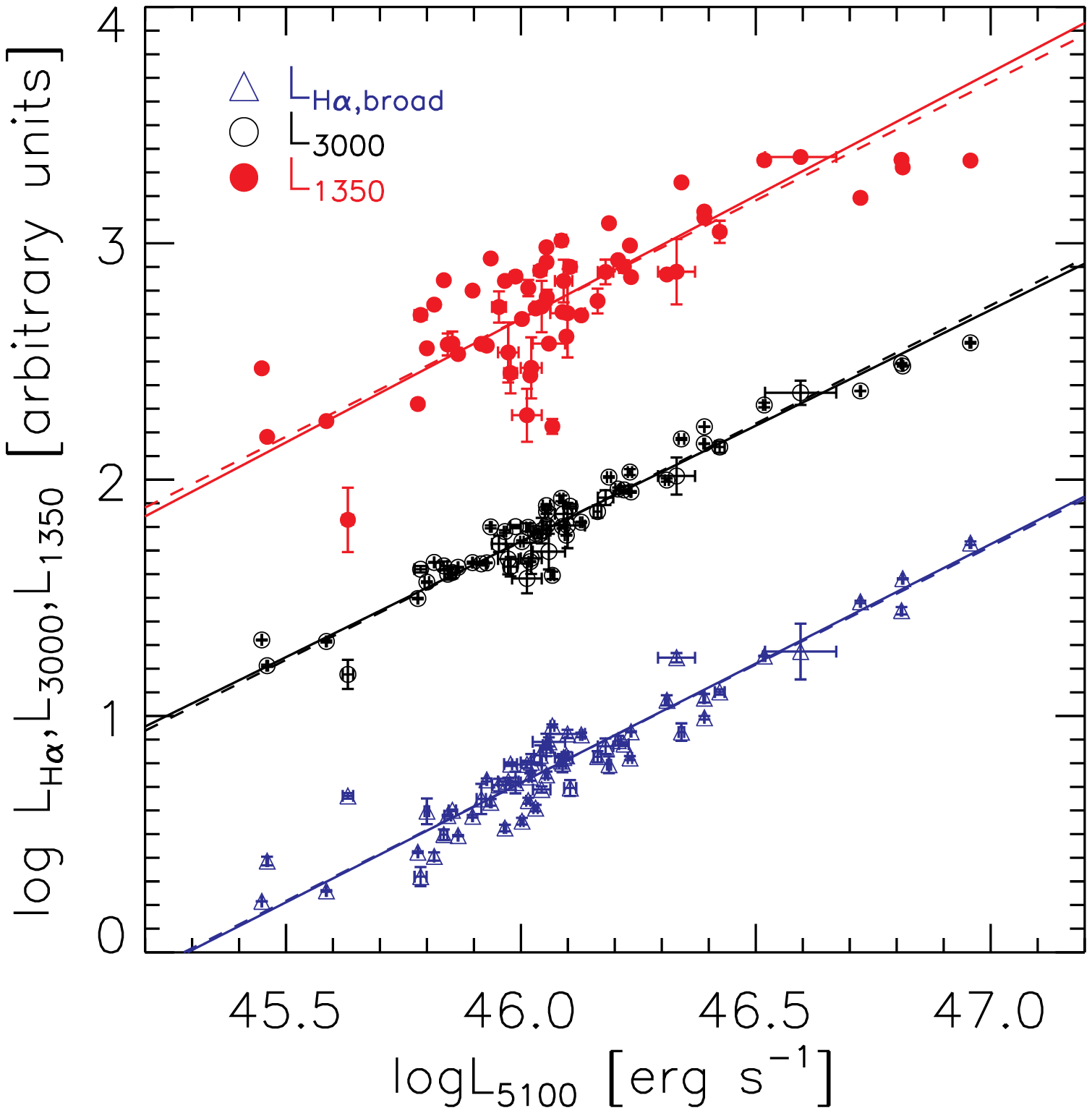}
\includegraphics[width=0.445\textwidth]{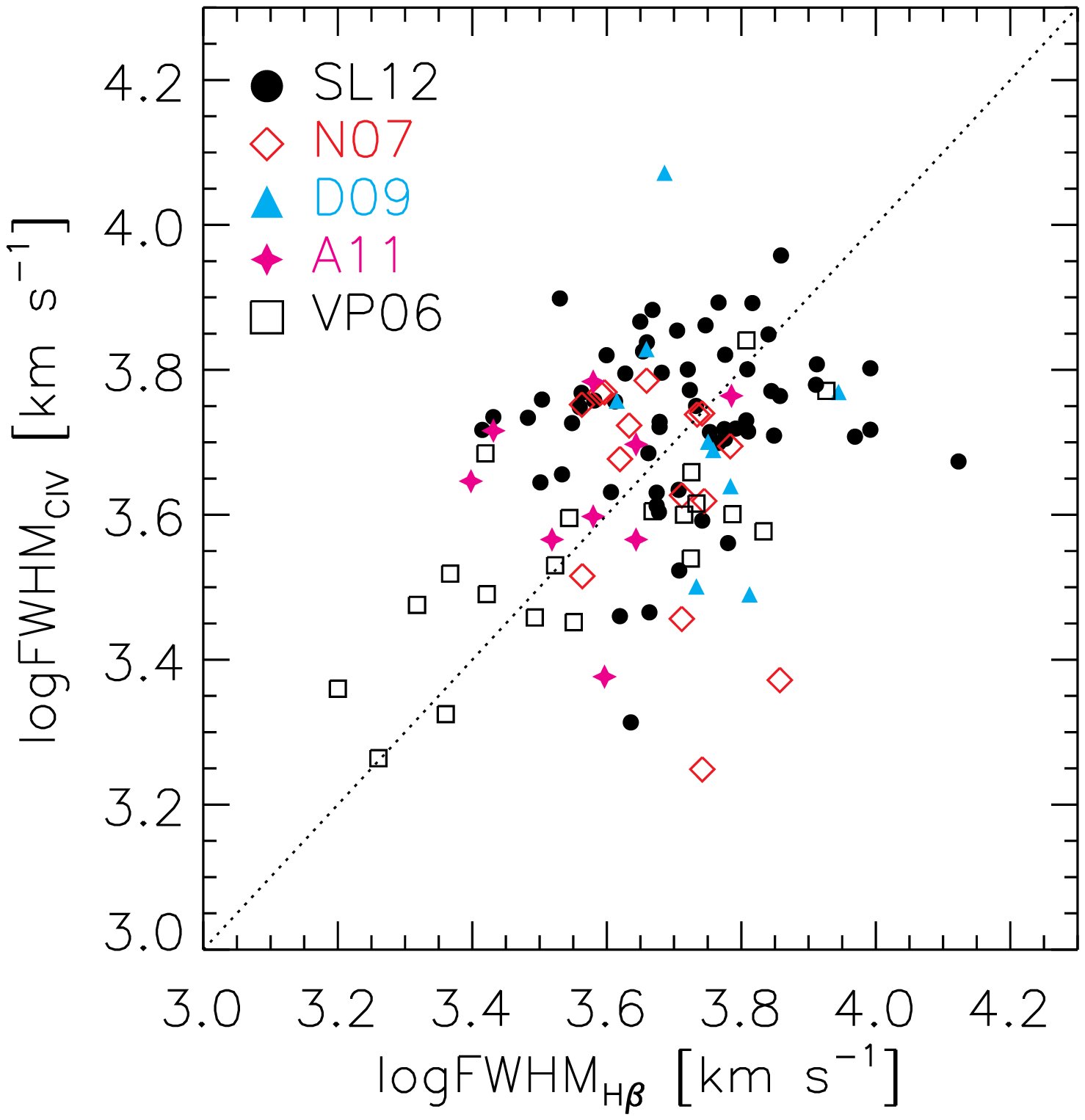}
\caption{{\em Left:} correlations between different luminosities using
the quasar sample in \citet{Shen_Liu_2012}, which covers all four lines (\CIV, \MgII, \hbeta, \halpha) in the same object, for the high-luminosity
regime
$L_{5100}>10^{45.4}\,{\rm erg\,s^{-1}}$. The solid lines are the bisector linear regression results using
the BCES estimator \citep[e.g.,][]{Akritas_Bershady_1996}, and the dashed lines indicate a linear correlation
of unity slope. {\em Right: }comparison between \CIV\ FWHM and \hbeta\ FWHM using different samples
from the literature [\citet[][60 objects; SL12]{Shen_Liu_2012}, \citet[][9
objects; A11]{Assef_etal_2011}, \citet[][21 objects;
VP06]{Vestergaard_Peterson_2006}, \citet[][15 objects;
N07]{Netzer_etal_2007}, and \citet[][9 objects; D09]{Dietrich_etal_2009}]. Only for the low-redshift and
low-luminosity VP06 sample is there a significant correlation between the two FWHMs.
}\label{fig:SL12_comp}
\end{figure}

The continuum luminosities at different wavelengths and several line
luminosities are all correlated with each other, with different levels of
scatter. Fig.\ \ref{fig:sdss_comp} shows some correlations between different
continuum luminosities using the spectral measurements of SDSS quasars from
\citet{Shen_etal_2011}. To compare $L_{1350}$ and $L_{5100}$ directly, one
needs either UV+optical or optical+near-IR to cover both restframe
wavelengths. Fig.\ \ref{fig:SL12_comp} (left) shows such a comparison from a
recent sample of quasars with optical spectra from SDSS and near-IR spectra
from \citet{Shen_Liu_2012}, which probes a higher luminosity range
$L_{5100}>10^{45.4}\,{\rm erg\,s^{-1}}$ than the SDSS sample. Correlations
between these luminosities are still seen at the high-luminosity end. For the
SDSS quasar population, different luminosities correlate with each other
well, but this may be somewhat affected by the optical target selection of
SDSS quasars that may preferentially miss dust-reddened quasars (see
\S\ref{s:host_contam}). In other words, the intrinsic dispersion in the
UV-optical SED may be larger for the general quasar population. For instance,
\citet{Assef_etal_2011} found a much larger dispersion in the
$L_{1350}/L_{5100}$ ratio for a gravitationally lensed quasar sample, which
is selected differently from the SDSS. This large dispersion in the
$L_{1350}/L_{5100}$ ratio will lead to more scatter between the \hbeta\ and
\CIV\ based SE masses.

It is also important to compare the widths of different lines. Since \hbeta\
is the most studied line in reverberation mapping and the $R-L$ relation was
measured using BLR radius for \hbeta\
\citep[e.g.,][]{Kaspi_etal_2000,Kaspi_etal_2005,Bentz_etal_2009a}, it is
reasonable to argue that the SE mass estimators based on the Balmer lines are
the most reliable ones. The width of the broad \halpha\ is well correlated
with that of the broad \hbeta\ and therefore it provides a good substitution
in the absence of \hbeta\ \citep[e.g.,][]{Greene_Ho_2005}. The widths of
\MgII\ are found to correlate well with those of the Balmer lines
\citep[e.g.,][see Fig.\ \ref{fig:sdss_comp} for a comparison based on SDSS
quasars]{Salviander_etal_2007,McGill_etal_2008,Shen_etal_2008b,Shen_etal_2011,Wang_etal_2009,Vestergaard_etal_2011,Shen_Liu_2012}.
But such a correlation may not be linear: despite different methods to
measure line widths, most recent studies favor a slope shallower than unity
in the correlation between the two FWHMs (e.g., see Fig.\
\ref{fig:sdss_comp}). Given this correlation it is practical to use the
\MgII\ width as a surrogate for \hbeta\ width in a \MgII-based SE mass
estimators, and some recent \MgII\ calibrations can be found in, e.g.,
\citet[][]{Vestergaard_Osmer_2009,Shen_Liu_2012,Trakhtenbrot_Netzer_2012}.
However, one intriguing feature regarding the \MgII\ line is that the
distribution of its line widths seem to have small dispersions in large
quasar samples \citep[e.g.,][]{Shen_etal_2008b,Fine_etal_2008}. It appears as
if the \MgII\ varies at a less extent compared with \hbeta\ \citep[cf.,][and
references therein]{Woo_2008}. It is also recently argued that for a small
fraction of quasars ($\sim 10\%$) in the NLS1 regime (e.g., small \hbeta\
FWHM and strong \FeII\ emission), \MgII\ may have a blueshifted, non-virial
component, and an overall larger FWHM than \hbeta, that will bias the virial
mass estimate \citep[e.g.,][]{Marziani_etal_2013}. This is consistent with
the general trend found between \MgII\ and \hbeta\ FWHMs using SDSS quasars
\citep[e.g.,][]{Wang_etal_2009,Shen_etal_2011,Vestergaard_etal_2011}, and may
be connected to the disk wind scenario for \CIV\ discussed below.

The correlation between \hbeta\ (or \MgII) and \CIV\ widths is more
controversial. While some claim that these two do not correlate well
\citep[e.g.,][]{Bachev_etal_2004,Baskin_Laor_2005,Netzer_etal_2007,Shen_etal_2008b,Fine_etal_2010,Shen_Liu_2012,Trakhtenbrot_Netzer_2012},
others claim there is a significant correlation
\citep[e.g.,][]{Vestergaard_Peterson_2006,Assef_etal_2011}. Fig.\
\ref{fig:SL12_comp} (right) shows a compilation of \CIV\ and \hbeta\ FWHMs
from the literature, which are derived for quasars in different luminosities
and redshift ranges. Only the low-luminosity (and low-$z$) RM sample in
\citet{Vestergaard_Peterson_2006} shows a significant correlation. It is
often argued that sufficient data quality is needed to secure the \CIV\ FWHM
measurements, although measurement errors are unlikely to account for all the
scatter seen in the comparison between \CIV\ and \hbeta\ FWHMs -- the
correlation between the two is still considerably poorer than that between
\MgII\ and \hbeta\ FWHMs for the samples in Fig.\ \ref{fig:SL12_comp} when
restricted to high-quality data. \citet{Shen_Liu_2012} suggested that the
reported strong correlation between \CIV\ and \hbeta\ FWHMs is probably
caused by the small sample statistics, or only valid for low-luminosity
objects.

The high-ionization \CIV\ line also differs from low-ionization lines such as
\MgII\ and the Balmer lines in many ways \citep[for a review, see
][]{Sulentic_etal_2000b}. Most notably it shows a prominent blueshift
(typically hundreds, up to thousands of ${\rm km\,s^{-1}}$) with respect to
the low-ionization lines
\citep[e.g.,][]{Gaskell_1982,Tytler_Fan_1992,Richards_etal_2002}, which
becomes more prominent when luminosity increases. There is also a systemic
trend (albeit with large scatter) of increasing \CIV\ FWHM and line asymmetry
when the \CIV\ blueshift increases, a trend not present for low-ionization
lines \citep[e.g.,][]{Shen_etal_2008b,Shen_etal_2011}. The \CIV\ blueshift is
predominantly believed to be an indication of outflows in some form, and
integrated in the disk-wind framework discussed below \citep[but see ][for a
different interpretation]{Gaskell_2009}. These properties of \CIV\ motivated
the idea that \CIV\ is likely more affected by a non-virial component than
low-ionization lines \citep[e.g.,][]{Shen_etal_2008b}, probably from a
radiatively-driven (and/or MHD-driven) accretion disk wind
\citep[e.g.,][]{Konigl_Kartje_1994,Murray_etal_1995,Proga_etal_2000,Everett_2005},
especially for high-luminosity objects. A generic two-component model for the
\CIV\ emission is then implied
\citep[e.g.,][]{Collin-Souffrin_etal_1988,Richards_etal_2011,Wang_etal_2011}.
A similar argument is proposed by \citet{Denney_2012} based on the \CIV\ RM
data of local AGNs, where she finds that there is a component of the \CIV\
line profile that does not reverberate, which is likely associated with the
disk wind (although alternative interpretations exist). This may also explain
the poorer correlation between \CIV\ width and \hbeta\ (or \MgII) width for
more luminous quasars, where the wind component is stronger (see further
discussion in \S\ref{s:RM_limit}). Therefore \CIV\ is likely a biased virial
mass estimator \citep[e.g.,][and references
therein]{Baskin_Laor_2005,Sulentic_etal_2007,Netzer_etal_2007,Shen_etal_2008b,Marziani_Sulentic_2012}.

Although in principle certain properties of \CIV\ (such as line shape
parameters) can be used to infer the \CIV\ blueshift and then correct for the
\CIV-based SE mass, such corrections are difficult in practice given the
large scatter in these trends and typical spectral quality. Proponents on the
usage of \CIV\ line often emphasize the need for good-quality spectra and
proper measurements of the line width. But the fact is \CIV\ is indeed more
problematic than the other lines, and there is no immediate way to improve
the \CIV\ estimator for high-redshift quasars, although some recent works are
showing some promising trends that may be used to improve the \CIV\ estimator
\citep[e.g.,][]{Denney_2012}.

There have also been proposals for using the \CIII, \AlIII, or \SiIII\ lines
in replacement of \CIV\
\citep[e.g.,][]{Greene_etal_2010,Marziani_Sulentic_2012}.
\citet{Shen_Liu_2012} found that the FWHMs of \CIV\ and \CIII\ are correlated
with each other, and hence \CIII\ may not be a good line either \citep[also
see][]{Ho_etal_2012}. On the other hand, \AlIII\ and \SiIII\ are more
difficult to measure given their relative weakness compared to \CIV\ and
\CIII\ as well as their blend nature, hence are not practical for large
samples of quasars. Another possible line to use is \lya. Although \lya\ is
more severely affected by absorption, intrinsically it may behave similarly
as the Balmer lines. Such an investigation is ongoing.

To summarize, currently the most reliable lines to use are the Balmer lines,
although this conclusion is largely based on the fact that these are the most
studied and best understood lines, and does not mean there is no problem with
them. \MgII\ can be used in the absence of the Balmer lines, although the
lack of RM data for \MgII\ poses some uneasiness in its usage as a SE
estimator. \CIV\ has local RM data (though not enough to derive a $R-L$
relation on its own), but the application of \CIV\ to high-redshift and/or
high-luminosity quasars should proceed with caution. In light of the
potential problems with \CIV, efforts have been underway to acquire near-IR
spectroscopy to study the high-$z$ quasar BH masses using \MgII\ and Balmer
lines
\citep[e.g.,][]{Shemmer_etal_2004,Netzer_etal_2007,Marziani_etal_2009,Dietrich_etal_2009,Greene_etal_2010,
Trakhtenbrot_etal_2011,Assef_etal_2011,Shen_Liu_2012,Ho_etal_2012,Matsuoka_etal_2013}.

\subsubsection{Effects of AGN variability on SE masses}

Quasars and AGNs vary on a wide range of timescales. It is variability that
made reverberation mapping possible in the first place. One might be
concerned that the SE masses may subject to changes due to quasar
variability. Several studies have shown, using multi-epoch spectra of
quasars, that the scatter due to luminosity changes (and possibly
corresponding changes in line width) does not introduce significant ($\gtrsim
0.1$ dex) scatter to the SE masses
\citep[e.g.,][]{Wilhite_etal_2007,Denney_etal_2009b,Park_etal_2012a}. This is
expected, since the average luminosity variability amplitude of quasars is
only $\sim 0.1-0.2$ magnitude over month-to-year timescales
\citep[e.g.,][]{Sesar_etal_2007,MacLeod_etal_2010,MacLeod_etal_2012}, thus
the difference in SE masses from multi-epochs will be dominated by
measurement errors (in particular those on line widths).

However it is legitimate to consider the consequence of uncorrelated
stochastic variations between line width and luminosity on SE masses, whether
or not such uncorrelated variations are due to actual physical effects, or
due to improper measurements of the continuum luminosity and line widths.
Examples are already given in \S\ref{s:virial_assump}, and more detailed
discussion will be provided in \S\ref{s:conse}.

\subsubsection{Limitations of the RM AGN sample}\label{s:RM_limit}

\begin{figure}
\centering
\includegraphics[width=0.9\textwidth]{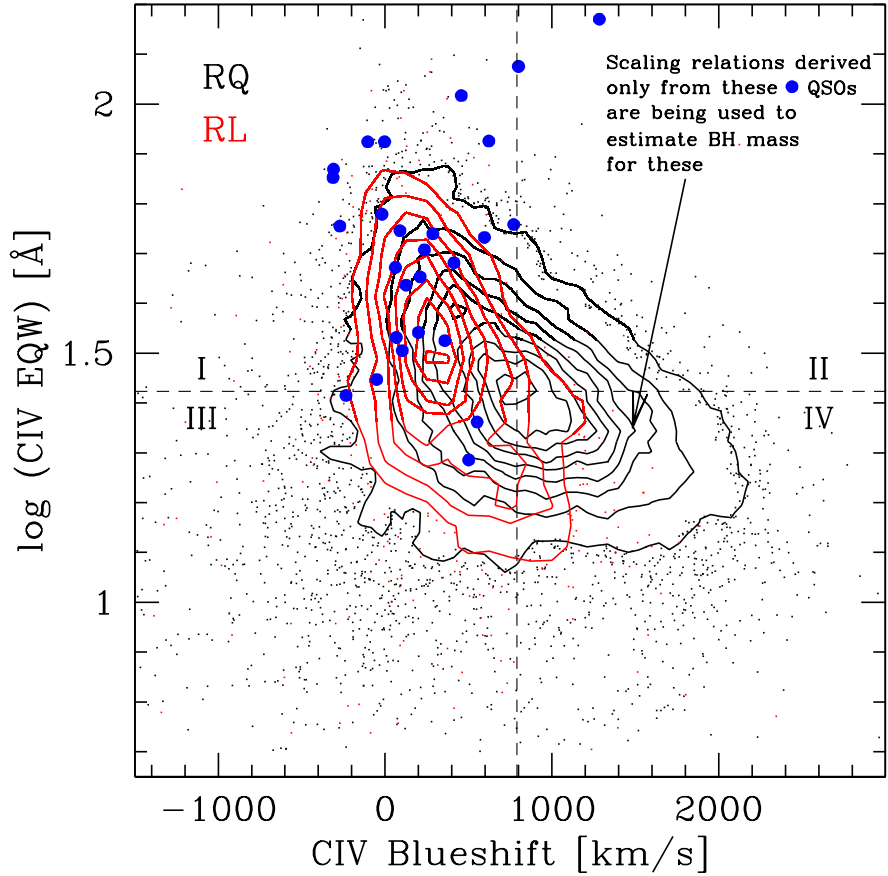}
\caption{An updated version of Fig.\ 18 in \citet{Richards_etal_2011}, showing the biased
distribution of the local RM AGNs in the parameter space for \CIV\ (blueshift relative to
\MgII\ versus the rest equivalent width). The contours and dots are
$1.5\lesssim z\lesssim 2.2$ SDSS quasars, and the blue filled circles are RM AGNs. The black and
red contours show the results for radio-quiet and radio-loud populations respectively.
Most of the RM AGNs occupy quadrant I, where the \CIV\ line is dominated by the virial component in
the two-component picture in \citet{Richards_etal_2011}. In this picture, quadrant IV is for \CIV\ lines
dominated by the non-virial wind component. The average quasar luminosity increases from quadrant I
to quadrant IV. Figure courtesy of G. Richards.
}\label{fig:civ_para}
\end{figure}

Last but not least, the current RM sample is by no means representative of
the general quasar/AGN population. It is a highly heterogeneous sample that
poorly samples the high-luminosity regime of quasars, and most objects are at
$z<0.3$. This alone calls into question the reliability of extrapolations of
locally-calibrated SE relations against these RM AGNs to high-$z$ and/or
high-luminosity quasars.

The distribution of the RM AGNs in the spectral parameter space of quasars is
also highly biased relative to the general population.
\citet[][]{Richards_etal_2011} developed \citep[building on earlier ideas by,
e.g.,][]{Collin-Souffrin_etal_1988,Murray_etal_1995,Proga_etal_2000,Elvis_2000,Leighly_Moore_2004,Leighly_2004}
a generic picture of two-component BLR structure for \CIV, composed of a
virial component, and a non-virial wind component which is filtering the
ionizing continuum from the inner accretion disk. This generic picture is
able to explain, phenomenologically, many characteristics of the continuum
and \CIV\ line properties, such as the \CIV\ blueshift and the Baldwin effect
\citep[i.e., the anti-correlation between \CIV\ equivalent width and adjacent
continuum luminosity,][]{Baldwin_1977}. Fig.\ \ref{fig:civ_para} shows the
distribution of RM objects in the parameter space of \CIV\ spectral
properties, where most RM AGNs occupy the regime dominated by the virial
component. Part of this is driven by luminosity, since more luminous quasars
have on average larger \CIV\ blueshift (\S\ref{s:comp_line}). It will be
important to explore this under-represented regime with \CIV\ RM at
high-redshift, which has just begun \citep[e.g.,][]{Kaspi_etal_2007}.
Although this is an immediate concern for \CIV, \citet{Richards_etal_2011}
made a fair argument that the BLR properties for \hbeta\ and \MgII\ may also
be biased in the RM sample relative to all quasars, if the non-virial wind
component is also affecting the BLR of \hbeta\ and \MgII\ by filtering the
ionizing continuum.

To date most of the RM lag measurements are for \hbeta, and lag measurements
are either lacking for \MgII\ \citep[but see][and references therein, for
\MgII\ RM attempts and tentative results]{Metzroth_etal_2006,Woo_2008} or
insufficient for \CIV\ to derive a direct $R-L$ relation based on these two
UV lines. The total number of RM AGNs is also small, $\sim 50$, not enough to
probe the diversity in BLR structure and other general quasar properties. The
current sample size and inhomogeneity of RM AGNs pose another major obstacle
to develop precise BH mass estimators based on RM and its extension, SE
virial methods.

\subsection{Practical Concerns}\label{prac_concern}

\subsubsection{How to measure the continuum luminosity and line
widths}\label{s:howto_mea}

Usually the continuum and line properties are measured either directly from
the spectrum, or derived from $\chi^2$ fits to the spectrum with some
functional forms for the continuum and for the lines. Arguably functional
fits are better suited for spectroscopic samples with moderate to low
spectral quality. As briefly mentioned earlier (\S\ref{s:comp_line}), it is
essential to measure the continuum and line width properly when using the
existing SE calibrations. Different methods sometimes do yield systematically
different results, in particular for the line width measurements. Some
studies fit the broad lines with a single component
\citep[e.g.,][]{McLure_Dunlop_2004}, while others use multiple components to
fit the broad line. But if one wants to use the calibration in, e.g.,
\citet{McLure_Dunlop_2004}, then it is better to be consistent with their
fitting method. Some comparisons between the broad line widths from different
fitting recipes can be made using the catalog provided in
\citet{Shen_etal_2011}. Take \hbeta\ for example, since this broad line is
not always a single Gaussian or Lorentzian, the line widths from the
single-component and multiple-component fit could differ significantly in
some cases.

The detailed description of spectral fitting procedure can be found in many
papers
\citep[e.g.,][]{McLure_Dunlop_2004,Greene_Ho_2005,Shen_etal_2008b,Shen_etal_2011,Shen_Liu_2012}.
In short, the spectrum is first fit with a power-law plus an iron emission
template\footnote{Empirical iron emission templates in the rest-frame UV to
optical can be found in, e.g.,
\citet[][]{Boroson_Green_1992,Vestergaard_Wilkes_2001,Tsuzuki_etal_2006}.
Using different iron templates may lead to small systematic offsets
($\lesssim 0.05$ dex) in the measured continuum luminosity and line width
\citep[e.g.,][]{Nobuta_etal_2012}. Occasionally a Balmer continuum component
is added in the pseudo-continuum fit to improve the fit around the ``small
blue bump'' region near $3000$\AA\
\citep[e.g.,][]{Grandi_1982,Dietrich_etal_2002}, but such a component is
generally difficult to constrain from spectra with limited wavelength
coverage \citep[see discussions in, e.g.,][]{Wang_etal_2009,Shen_Liu_2012}.}
in several spectral windows free of major broad lines. The best-fit
``pseudo-continuum'' is then subtracted from the original spectrum, leaving
the emission line spectrum. The broad line region is then fit with a mixture
of functions (such as multiple Gaussians or Gauss-Hermite polynomials). The
continuum luminosity and line width are then extracted from the best-fit
model. The measurement errors from the multiple component fits are often
estimated using some Monte Carlo methods
\citep[e.g.,][]{Shen_etal_2011,Shen_Liu_2012}: mock spectra are generated
either by adding noise to the original spectrum, or by adding ``scrambled''
residuals from the data minus best-fit model back to the model. The mock
spectra are then fit with the same fitting procedure, and the formal errors
are estimated from the distributions of the measured quantity from the mocks.
This mock-based error estimation approach takes into account both the noise
of the spectrum and ambiguities in decomposing different components in the
fits.

Below are some additional notes regarding continuum and line measurements.

\begin{itemize}

\item[$\bullet$] {\em Narrow line subtraction} Since the narrow line
    region (NLR) dynamics is not dominated by the central BH gravity, we
    want to subtract strong narrow line component before we measure the
    broad line width from the spectrum. This is particularly important
    for FWHM measurements, while for $\sigma_{\rm line}$ the effects of
    narrow lines are less important. For \hbeta\ and \halpha, reliable
    constraints on the velocity and width of the narrow components can be
    obtained from the adjacent narrow lines such as \OIIIab\ and \SIIab.
    For \MgII\ and \CIV, this is not so simple mainly for two reasons: 1)
    there are usually no adjacent strong narrow lines such as \OIII\ to
    provide constraints on the narrow line component; and even if \OIII\
    can be covered in other wavelengths there is no guarantee the NLR
    properties are the same for \OIII\ and for \MgII/\CIV. 2) Although
    some quasars do show evidence of narrow component \MgII\ and \CIV, it
    is unclear if this applies to the general quasar population.
    \citet{Shen_Liu_2012} found that for the 60 high-luminosity
    ($L_{5100}>10^{45.4}\,{\rm erg\,s^{-1}}$)
 quasars in their sample with optical and near-IR spectroscopy covering
    \CIV\ to \OIII, the contribution of the narrow line component to
    \CIV\ is too small to affect the estimated broad \CIV\ FWHM
    significantly. However, for less luminous objects, the relative
    importance of the narrow line component to \CIV\ might be larger
    \citep[e.g.,][]{Bachev_etal_2004,Sulentic_etal_2007}.

\item[$\bullet$] {\em Remedy for absorption} Sometimes there are
    absorption features superposed on the spectrum, which is most
    relevant for \CIV, and then \MgII. Not accounting for these
    absorption features will bias the continuum and line measurements.
    While for narrow or moderately-broad absorption troughs, manual or
    automatic treatments can greatly minimize their effects
    \citep[e.g.,][]{Shen_etal_2011}, there is no easy way to fit objects
    that are heavily absorbed (such as broad absorption line quasars).

\item[$\bullet$] {\em Effects of low signal-to-noise ratio (S/N)} The
    quality of the continuum luminosity and line width measurements
    decreases as the quality of the spectrum degrades. In addition to
    increased measurement errors, low S/N data may also lead to biases in
    the spectral measurements. \citet{Denney_etal_2009b} performed a
    detailed investigation on the effects of S/N on the measured \hbeta\
    line width using many single-epoch spectra of two RM AGNs (NGC 5548
    and PG1229+204). They tested both direct measurements and
    Gauss-Hermite polynomial fits to the spectrum, and found that the
    best-fit line width is systematically underestimated at low S/N for
    both direct measurements and functional fits. The only exception is
    that their Gauss-Hermite fits to degraded NGC 5548 spectra tend to
    overestimate the FWHM at lower S/N. However, this is mainly caused by
    the fact that the Gauss-Hermite model is often unable to accurately
    fit the complex \hbeta\ line profile of NGC 5548. Using
    multiple-Gaussian model fits, and for a much larger sample of SDSS
    quasars, \citet{Shen_etal_2011} also investigated the effects of S/N
    on the model fits by artificially degrading high S/N spectra (see
    their Figs.\ 5-8). They found that the exact magnitude of the bias
    depends on the line profile as well as the strength of the line. The
    continuum is usually unbiased as S/N decreases. The FWHMs and
    equivalent widths (EWs) are biased by less than $\pm20\%$ for high-EW
    objects as S/N is reduced to as low as $\sim3/$pixel. For low-EW
    objects, the FWHMs and EWs are biased low/high by $>20\%$ for
    S/N$\lesssim 5/$pixel. But the direction of the bias in FWHM is not
    always underestimation.

\end{itemize}

\subsection{Consequences of the Uncertainties in SE Mass
Estimates}\label{s:conse}

Given the many physical and practical concerns discussed in
\S\S\ref{s:phy_concern} and \ref{prac_concern}, one immediately realizes that
these mass estimates, especially those SE mass estimates, should be
interpreted with great caution. Almost everyone acknowledges the large
uncertainties associated with these mass estimates, but only very few are
taking these uncertainties seriously. Since at present there is no way to
know whether or not the extrapolation of these SE methods to high-$z$ and/or
high-luminosity quasars introduces significant biases, let us assume naively
that these SE estimators provide unbiased mass estimates in the average
sense, and focus on the statistical uncertainties (scatter) of these
estimators.

In mathematical terms, we have:

\begin{equation}\label{eqn:m_e}
m_e|m=m+G(0,\sigma_{\rm SE})\ ,
\end{equation}
where $m_e\equiv \log M_{\rm BH,SE}$ is the SE mass estimate, $m\equiv M_{\rm
BH}$ is the true BH mass, and $G(\mu,\sigma)$ is a Gaussian random deviate
with mean $\mu$ and dispersion $\sigma$. I use $x|y$ to denote a random value
of $x$ at fixed $y$ drawn from the conditional probability distribution
$p(x|y)$. Eqn.\ (\ref{eqn:m_e}) thus means that the distribution of SE masses
given true BH mass, $p_0(m_e|m)$, is a lognormal with mean equal to $m$ and
dispersion $\sigma_{\rm SE}$. It is then clear that this equation stipulates
our assumption that the SE mass is on average an unbiased estimate of the
true mass, but with a statistical scatter of $\sigma_{\rm SE}\sim 0.5$ (dex),
i.e., the formal uncertainty of SE masses.

\subsubsection{The Malmquist-type bias (Eddington bias)}\label{s:bias_m}

Now let us assume that we have a mass-selected sample of objects with known
true BH masses, and ``observed'' masses based on the SE estimators. By
``mass-selected'' I mean there is no selection bias caused by a flux (or
luminosity) threshold -- all BHs are observed regardless of their luminosity.
If we further assume that the distribution of true BH masses in this sample
is bottom-heavy, then a statistical bias in the SE masses naturally arises
from the errors of SE masses
\citep[e.g.,][]{Shen_etal_2008b,Kelly_etal_2009a,Shen_Kelly_2010,Kelly_etal_2010},
because there are more intrinsically lower-mass objects scattering into a SE
mass bin due to errors than do intrinsically higher-mass objects. This
statistical bias can be shown analytically assuming simple analytical forms
of the distribution of true BH masses. Suppose the underlying true mass
distribution is a power-law, $dN/dM_{\rm BH}\propto M_{\rm BH}^{\gamma_M}$,
then Bayes's theorem tells us the distribution of true BH masses at given SE
mass is (recall $p_0(m_e|m)$ is the conditional probability distribution of
$m_e$ given $m$):
\begin{eqnarray}\label{eqn:bias_1}
p_1(m|m_e)=p_010^{\gamma_M m}\left[\int p_010^{\gamma_Mm}dm\right]^{-1}=(2\pi\sigma_{\rm SE}^2)^{-1/2}\exp
\left\{-\frac{[m-(m_e+\ln(10)\gamma_M\sigma_{\rm SE}^2)]^2}{2\sigma_{\rm SE}^2}\right\}\ .
\end{eqnarray}
Thus the expectation value of true mass at given SE mass is:
\begin{equation}
\langle m \rangle_{m_e} = m_e+\ln(10)\gamma_M\sigma_{\rm SE}^2\ .
\end{equation}

Therefore for bottom-heavy ($\gamma_M<0$) true mass distributions, the
average true mass at given SE mass is smaller by $-\ln(10)\gamma_M\sigma_{\rm
SE}^2$ dex than the SE mass. This has an important consequence that the
quasar black hole mass function (BHMF) constructed using SE virial masses
will be severely overestimated at the high-mass end
\citep[e.g.,][]{Kelly_etal_2009a,Kelly_etal_2010,Shen_Kelly_2011}.

This statistical bias due to the uncertainty in the mass estimates and a
non-flat true mass distribution is formally known as the Eddington bias
\citep{Eddington}. Historically this has also been referred to as the
Malmquist bias in studies involving distance estimates
\citep[e.g.,][]{Lynden-Bell_1988}, which bear some resemblance to the
familiar Malmquist bias in magnitude-limited samples
\citep[e.g.,][]{Malmquist}. For this reason, this bias was called the
``Malmquist'' or ``Malmquist-type'' bias in \citet[][]{Shen_etal_2008b} and
\citet{Shen_Kelly_2010}, and I adopted this name here as well. Perhaps a
better name for this class of biases is the ``Bayes correction'', which then
also applies to the generalization of statistical biases caused by threshold
data and correlation scatter (or measurement errors). The
luminosity-dependent bias discussed next, and the Lauer et~al.\ bias
\citep[][]{Lauer_etal_2007b} discussed in \S\ref{s:app:evo}, can also be
described by this name.

\subsubsection{Luminosity-dependent bias in SE virial BH
masses}\label{s:bias_l}

Now let us take one step further, and consider the conditional probability
distribution of $m_e$ at fixed true mass $m$ and fixed luminosity $l\equiv
\log L$, $p(m_e|m,l)$. If the SE mass distribution at given true mass is
independent on luminosity, then we have $p(m_e|m,l)=p(m_e|m)$. This means
that the SE mass is always unbiased in the mean regardless of luminosity.
However, one may consider such a situation where $p(m_e|m,l)\neq p(m_e|m)$,
which means the distribution of SE masses will be modified once one limits on
luminosity. This is an important issue, since essentially all statistical
quasar samples are flux-limited samples (except for heterogeneous samples,
such as the local RM AGN sample), and frequently the SE mass distribution in
finite luminosity bins is measured and interpreted.

Below I will explore this possibility and its consequences in detail. To help
the reader understand these issues, here is an outline of the discussion that
follows: 1) I will first formulate the basic equations to understand the
(mathematical) origin of the uncertainty in SE mass, $\sigma_{\rm SE}$; 2) I
will then provide physical considerations to justify this formulation; 3) The
conditional probability distribution of SE mass at fixed true mass and
luminosity $p(m_e|m,l)$ is then derived, and I demonstrate the two most
important consequences: the luminosity-dependent bias, and the narrower
distribution of SE masses at fixed true mass and luminosity than the SE mass
uncertainty $\sigma_{\rm SE}$; 4) I then discuss current observational
constraints on the luminosity-dependent bias and demonstrate its effect using
a simulated flux-limited quasar sample.

\noindent{\em 1) Understanding the origin of the uncertainty $\sigma_{\rm
SE}$ in SE masses}

I will use Gaussians (lognormal) to describe most distributions and neglect
higher-order moments, mainly because the current precision and our
understanding of SE masses are not sufficient for more sophisticated
modeling. Assuming the distributions of luminosity and line width at given
true mass $m$ both follow lognormal distributions, we can write such
distributions as
\begin{equation}\label{eqn:lw_dist}
l|m=\langle l\rangle_m+ G_1(0,\sigma_l),\ w|m=\langle w\rangle_m + G_2(0,\sigma_w)\ ,
\end{equation}
where notations are the same as in Eqn.\ (\ref{eqn:m_e}), $w\equiv \log W$,
and $\langle\rangle_m$ indicates the expectation value at $m$. The
dispersions in luminosity and line width at this fixed true mass should be
understood {\em as due to both variations in single objects (i.e.,
variability) and object-by-object variance.} The SE mass estimated using $l$
and $w$ are then (e.g., Eqn.\ \ref{eqn:virial_mass}):
\begin{equation}\label{eqn:virial_mass2}
m_e|m = bl + cw + {\rm constant}\ ,
\end{equation}
where the last term ``constant'' absorbs coefficient $a$ and other constants
from SE mass calibrations. Now let us consider the following two scenarios:

\begin{itemize}

\item[A.] In the ideal case where the SE method gives a perfect mass
    estimate, we have
\begin{equation}
G_1(0,\sigma_l)=-\frac{c}{b}G_2(0,\sigma_w)\ ,
\end{equation}
i.e., the deviations in luminosity and line width from their mean values
at given true mass are perfectly correlated. The resulting $m_e$
distribution thus peaks at $m$ with zero width, i.e., $m_e|m=m=b\langle
l\rangle_m+c\langle w\rangle_m + {\rm constant}$.

\item[B.] In the realistic case, certain amount of the deviations in $l$
    and $w$ from their mean values are uncorrelated with each other
    (either intrinsic or due to measurement errors). Without loss of
    generality, we can rewrite Eqn.\ (\ref{eqn:lw_dist}) as
    \begin{equation}\label{eqn:sigma_me}
      l|m=\langle l\rangle_m+ G_1(0,\sigma^{\prime}_l)+G_0(0,\sigma_{\rm corr}),
      \ w|m=\langle w\rangle_m + G_2(0,\sigma^{\prime}_w)-\frac{b}{c}G_0(0,\sigma_{\rm corr})\ ,
    \end{equation}
 where the {\em total} dispersions in the distributions of $l$ and $w$
 are
 \begin{equation}\label{eqn:total_dist}
 \sigma_l=\sqrt{\sigma^{\prime 2}_l+\sigma^2_{\rm corr}}\ ,\quad
 \sigma_w=\sqrt{\sigma^{\prime 2}_w+(\sigma_{\rm corr}b/c)^2}\ .
 \end{equation}
 Eqn.\ (\ref{eqn:sigma_me}) stipulates that some portions of the
 dispersions in $l$ and $w$, described by $\sigma_{\rm corr}$, are
 correlated and do not contribute to the dispersion (scatter) of $m_e$ at
 $m$. On the other hand, the remaining dispersions in $l$
 ($\sigma^\prime_l$) and in $w$ ($\sigma^\prime_w$) are stochastic terms,
 and they combine to cause the dispersion of $m_e$ at $m$:
 \begin{equation}
    m_e|m = m + bG_1(0,\sigma^\prime_l) + cG_2(0,\sigma^\prime_w)=m+G(0,\sigma_{\rm SE})\ ,
 \end{equation}
where
\begin{equation}\label{eqn:sigma_se}
\sigma_{\rm
SE}=\sqrt{(b\sigma^\prime_l)^2+(c\sigma^\prime_w)^2}
\end{equation}
is the formal uncertainty in SE mass, i.e., the scatter in $m_e$ at given
true mass $m$.

\end{itemize}

Eqn.\ (\ref{eqn:sigma_me}) through (\ref{eqn:sigma_se}) provide a general
description of SE mass error budget from luminosity and line width, and form
the basis of the following discussion. From now on I will only consider the
realistic case B.

\noindent{\em 2) Physical considerations on the variances
$\sigma_{l}^\prime$, $\sigma_{w}^\prime$ and $\sigma_{\rm corr}$}

Most of the studies to date have implicitly assumed $\sigma^\prime_l=0$ in
Eqn.\ (\ref{eqn:sigma_me}), with the few exceptions in e.g.,
\citet{Shen_etal_2008b}, \citet{Shen_Kelly_2010} and \citet{Shen_Kelly_2011}.
$\sigma^\prime_l=0$ imposes a strong requirement that {\em all} the
variations in luminosity are compensated by line width such that the
uncertainty in SE masses now completely comes from the $\sigma^\prime_w$ part
in line width dispersion. While this is what we hope for the SE method, there
are physical and practical reasons to expect a non-zero $\sigma^\prime_l$, as
discussed in, e.g., \citet{Shen_Kelly_2011}. Specifically we have the
following considerations:

\textbf{(a)} the stochastic continuum luminosity variation and response of
the BLR (hence the response in line width) are not synchronized, as resulting
from the time lag in the reverberation of the BLR. The rms continuum
variability on timescales of the BLR light-cross time is $\sim 0.05$ dex
using the ensemble structure function in, e.g.,
\citet[][]{MacLeod_etal_2010};

\textbf{(b)} even with the same true mass, individual quasars have different
BLR properties, and presumably the measured optical-UV continuum luminosity
is not as tightly connected to the BLR as the ionizing luminosity. Both will
lead to stochastic deviations of luminosity and line width from the perfect
correlation (source-by-source variation in the virial coefficient $f$,
scatter in the $R-L$ relation, etc.). The level of this luminosity
stochasticity is unknown but is at least $0.2-0.3$ dex given the scatter in
the $R-L$ relation alone, and thus it is a major contributor to
$\sigma^\prime_l$;

\textbf{(c)} although not explicitly specified in Eqn.\
(\ref{eqn:sigma_me}), there are uncorrelated measurement errors in
luminosity and line width; typical measurement error in luminosity for
SDSS spectra (with S/N$\sim 5-10/$pixel, e.g., see fig.\ 4 of Shen et~al.
2011) is $\sim 0.02$ dex (statistical only), but increases rapidly at low
S/N;

\textbf{(d)} and finally, what we measure as line width does not perfectly
trace the virial velocity. This is a concern for essentially all three lines,
and for both of the two common definitions of line width (FWHM and
$\sigma_{\rm line}$). Two particular concerns arise. First, single-epoch
spectra do not provide a line width that describes the reverberating part of
the line only, thus some portion of the line width may not respond to
luminosity variations. Second, if a line is affected by a non-virial
component (say, \CIV\, for instance), and if this component strengthens and
widens when luminosity increases, the total line width would not response to
the luminosity variation as expected. As in (b), this contribution to the
uncompensated (by line width) luminosity variance $\sigma^\prime_l$ is
unknown, but could be as significant as in (b).

One extreme of (d) would be that line width has nothing to do with the virial
velocity except for providing a mean value in the calibrations of Eqn.\
(\ref{eqn:virial_mass}), as suggested by \citet[][]{Croom_2011}, i.e.,
$\sigma_{\rm corr}=0$. In this case while the average SE masses are unbiased
by calibration, the luminosity-dependent SE mass bias at given true mass is
maximum (see below). Note that this $\sigma_{\rm corr}=0$ case was already
considered in \citet{Shen_etal_2008b} and \citet{Shen_Kelly_2010} when
demonstrating the luminosity-dependent bias, and is only a special case of
the above generalized formalism.

On the other hand, $\sigma_{\rm corr}> 0$ would mean that line width does
respond to luminosity variations to some extent, justifying the inclusion of
line width in Eqn.\ (\ref{eqn:virial_mass}). This was indeed seen at least
for some local, low-luminosity objects, although not so much for the
high-luminosity SDSS sample, based on the tests described in
\S\ref{s:virial_assump}; additional evidence is provided in, e.g.,
\citet[][]{Kelly_Bechtold_2007} and \citet{Assef_etal_2012}, again for the
low-luminosity RM AGN sample. Therefore, the most realistic scenario is that
at fixed true mass, some portions of the dispersions in luminosity (or
equivalently, Eddington ratio) and in line width are correlated with each
other, and they cancel out in the calculation of SE masses; the remaining
portions of the dispersions in $l$ and $w$ are stochastic in nature and they
combine to contribute to the SE mass uncertainty (as in Eqn.\
\ref{eqn:sigma_se}). In other words, we expect $\sigma^\prime_l>0$,
$\sigma^\prime_w>0$, and $\sigma_{\rm corr}>0$. For simplicity I take
constant values for these scatters in the following discussion, but it is
possible that they depend on true BH mass.

\noindent{\em 3) The distribution of SE mass at fixed true mass and
luminosity $p(m_e|m,l)$}

Now that we have formulated the distributions of $l$, $w$ and $m_e$ at fixed
$m$ (e.g., Eqns.\ \ref{eqn:virial_mass2} and \ref{eqn:sigma_me}), we can
derive the conditional probability distribution of $m_e$ at fixed $m$ and
$l$, $p(m_e|m,l)$. It is straightforward to show\footnote{Here I give one
possible derivation. For brevity I will drop $m$ in all probability
distributions, but it should be understood that all these distributions are
at fixed $m$. Consider the $\sigma^\prime_w=0$ case first, where we want to
derive $p(m_e|l)$. Using Bayes's theorem, $p(m_e|l)\propto p(m_e)p(l|m_e)$.
We have $p(m_e)\propto e^{-(m_e-m)^2/[2(b\sigma^\prime_l)^2]}$ (i.e., all
variance in $m_e$ comes from $\sigma^\prime_l$ since $\sigma^\prime_w=0$),
and $\displaystyle p(l|m_e)\propto e^{-[l - (\frac{m_e-m}{b}+\langle
l\rangle_m)]^2/(2\sigma_{\rm corr}^2)}$ (i.e., $l$ can only vary due to
$\sigma_{\rm corr}$ at fixed $m_e$). Therefore $p(m_e|l)\propto
e^{-(m_e-\langle m_e\rangle_{m,l})^2/(2\sigma^{\prime2}_{ml})}$, where
$\langle m_e\rangle_{m,l}$ is the same as in Eqn.\ (\ref{eqn:dist_ml}), and
$\sigma^{\prime2}_{ml}=(b\sigma^\prime_l)^2\sigma_{\rm
corr}^2/(\sigma^{\prime2}_l+\sigma^2_{\rm corr})$. Now add back in the
$\sigma^{\prime}_w$ term, which will convolve $p(m_e|l)$ with a Gaussian
distribution. Then the general distribution $p(m_e|l)$ for arbitrary values
of $\sigma^{\prime}_w$ is also a Gaussian, with the same mean, but a
dispersion that is broadened by $c\sigma^{\prime}_w$ (i.e., the same as in
Eqn.\ \ref{eqn:dist_ml}). } (again using Bayes's theorem) that this
distribution is also a Gaussian distribution, with mean and dispersion:
\begin{equation}\label{eqn:dist_ml}
\langle m_e\rangle_{m,l}=m+\frac{b\sigma^{\prime2}_l}{\sigma^{\prime2}_l + \sigma^2_{\rm corr}}(l-\langle l\rangle_m)\ ,\quad
\sigma_{ml}^2=\frac{(b\sigma^\prime_l)^2\sigma_{\rm corr}^2}{\sigma^{\prime2}_l+\sigma^2_{\rm corr}} + (c\sigma^\prime_w)^2\ .
\end{equation}
Therefore we can generate the distribution of $m_e$ at fixed $m$ and $l$ as:
\begin{equation}\label{eqn:beta}
m_e|m, l = m + \beta(l-\langle l\rangle_m) + G(0,\sigma_{ml})\ ,
\end{equation}
where (using Eqns.\ \ref{eqn:total_dist} and \ref{eqn:sigma_se})
\begin{equation}\label{eqn:sigma_ml}
\beta=\displaystyle\frac{b\sigma^{\prime2}_l}{\sigma^{\prime2}_l +
\sigma^2_{\rm corr}}=\frac{b\sigma^{\prime2}_l}{\sigma^2_l}\ ,\quad
\sigma_{ml}^2 = \sigma_{\rm SE}^2 - \beta^2\sigma_l^2=(b\sigma^\prime_l)^2+(c\sigma^\prime_w)^2-\beta^2\sigma_l^2\ .
\end{equation}
Physically $\sigma_{ml}$ is the dispersion of SE mass at fixed true mass and
fixed luminosity. The parameter $\beta$ denotes the magnitude (slope) of the
luminosity-dependent bias, and we have\footnote{For the sake of completeness,
I note that $\beta>b$ could happen, if the line width were actually
positively correlated to luminosity in the $\sigma_{\rm corr}$ terms in Eqn.\
(\ref{eqn:sigma_me}). Of course such a scenario is counter-intuitive
(regarding the virial assumption) and thus unlikely, and it means one should
not use line width at all in estimating SE masses.}
$0<\beta<b$, where the lower and upper boundaries correspond to the two
extreme cases $\sigma^\prime_{l}=0$ and $\sigma_{\rm corr}=0$. A larger
$\beta$ means a stronger luminosity-dependent bias. Given the values of
$\sigma_l^\prime$, $\sigma_w^\prime$ and $\sigma_{\rm corr}$, and a SE
calibration ($b$ and $c$), all other quantities can be derived using Eqns.\
(\ref{eqn:total_dist})-(\ref{eqn:sigma_ml}).

\begin{figure}[!ht]
\includegraphics[width=1.\textwidth]{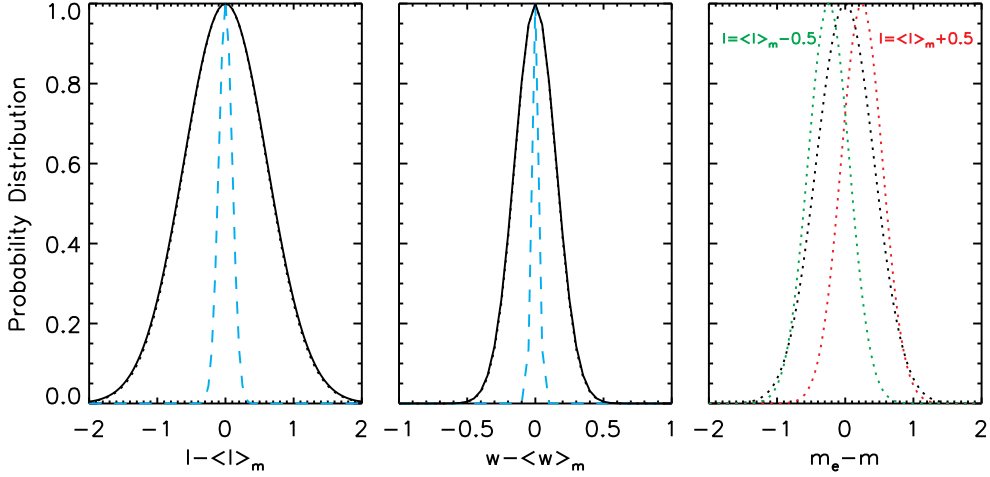}
\caption{Simulated distributions of $l$ (left), $w$ (middle), and $m_e$ (right) at fixed true mass $m$, following the
description in \S\ref{s:bias_l} (e.g., see Eqns.\ \ref{eqn:sigma_me}-\ref{eqn:sigma_ml}). The example shown here assumes $b=0.5$, $c=2$ in Eqn.\
(\ref{eqn:sigma_me}), e.g., the
typical values for SE mass estimators. {\em Left}: The distribution of luminosity $l\equiv \log L$. The black dotted line
indicates the dispersion in $\sigma^\prime_l=0.6$, the cyan dashed line indicates the dispersion in $\sigma_{\rm corr}=0.1$,
and the black solid line indicates the total dispersion in $\sigma_l=\sqrt{\sigma^{\prime2}_l+\sigma^2_{\rm corr}}$ (which essentially overlaps
with the dotted line given that $\sigma_l$ is dominated by $\sigma^\prime_l$). {\em Middle}: The distribution of line width $w\equiv \log W$. The
black dotted
line indicates the dispersion
in $\sigma_{w}^\prime=0.15$, the cyan dashed line indicates the dispersion in $-\sigma_{\rm corr}b/c$ (i.e., the part that correlates with
luminosity),
and the black solid line indicates the total dispersion in $\sigma_w=\sqrt{\sigma_{w}^{\prime2} + (\sigma_{\rm corr}b/c)^2}$. Again the
solid line and the dotted line are almost on top of each other. {\em Right}: The distribution of SE mass $m_e\equiv \log M_{\rm SE}$. The black
dotted
line
indicates the total dispersion of $m_e$ at fixed m, $\sigma_{\rm SE}=\sqrt{(b\sigma^\prime_l)^2+(c\sigma^\prime_w)^2}=0.42$. The
green (red) dotted line indicates the distribution of $m_e$ at fixed $m$ and fixed $l=\langle l\rangle_m-0.5$ ($\langle l\rangle_m+0.5$),
which is a Gaussian described by Eqns.\ (\ref{eqn:beta})-(\ref{eqn:sigma_ml}). In this example most of the dispersions in luminosity and
line width are uncorrelated with each other, leading to a large uncertainty in SE masses $\sigma_{\rm SE}=0.42$ dex. The inferred
luminosity-dependent
bias has a slope of $\beta=0.49$. The dispersion of $m_e$ at fixed luminosity is only $\sigma_{ml}=0.3$ dex (Eqn.\ \ref{eqn:sigma_ml}), much smaller
than the uncertainty in SE masses, $\sigma_{\rm SE}$.
}\label{fig:dist}
\end{figure}

Fig.\ \ref{fig:dist} shows a demonstration with $\sigma^{\prime}_l=0.6$,
$\sigma_{\rm corr}=0.1$, $\sigma^{\prime}_w=0.15$, $b=0.5$ and $c=2$. In this
example we have $\beta=0.49$, $\sigma_{\rm SE}=0.42$, and $\sigma_{ml}=0.3$.
The left two panels show the distributions of luminosity and line width at
fixed true mass, from the stochastic term
($\sigma_l^\prime,\sigma_w^\prime$), the correlated term ($\sigma_{\rm
corr}$) and the total dispersion ($\sigma_l,\sigma_w$). The right panel shows
the distributions of SE masses at this fixed true mass for all luminosities
(black dotted line) and for fixed luminosities (green and red dotted lines).
The distributions of SE masses at fixed luminosity are both narrower and
biased compared with the distribution without luminosity constraint.

There are two important conclusions that can be drawn from Eqns.\
(\ref{eqn:dist_ml})-(\ref{eqn:sigma_ml}):

\begin{itemize}

\item[$\bullet$] At fixed true mass $m$, the SE mass $m_e$ is
    over-/underestimated when luminosity $l$ is higher/lower than the
    average value at fixed $m$. This is the ``luminosity-dependent bias''
    first introduced in \citet{Shen_etal_2008b} and subsequently
    developed in \citet{Shen_Kelly_2010} and \citet{Shen_Kelly_2011}. The
    magnitude of this bias, determined by $\beta$, depends on how much of
    the dispersion in luminosity at fixed true mass is stochastic with
    respect to line width (i.e., not compensated by responses in $w$).

\item[$\bullet$] Secondly, the variance in SE masses at fixed true mass
    and fixed luminosity, $\sigma_{ml}^2$, is generally smaller than the
    uncertainty of SE masses, $\sigma_{\rm SE}^2$, by an amount of
    $(\beta\sigma_l)^2$. A simple way to understand this is that the
    uncertainty (variance) of SE mass at fixed true mass comes from the
    stochastic variance terms in both luminosity and line width.
    Therefore when one reduces the variance in either luminosity or line
    width by fixing either variable, the variance in SE mass is also
    reduced. This has important consequences in interpreting the observed
    distribution of SE masses for quasars in flux-limited samples or in
    narrow luminosity bins. One {\em cannot} simply argue \citep[as did
    in, e.g.,][]{Kollmeier_etal_2006,Steinhardt_Elvis_2010b} that the
    uncertainty in SE masses is small because the distribution of SE
    masses for samples with restricted luminosity ranges is narrow. The
    example shown in Fig.\ \ref{fig:dist} clearly demonstrates that one
    can easily get a much narrower SE mass distribution at fixed true
    mass and luminosity, than the nominal SE mass uncertainty
    $\sigma_{\rm SE}$. On the other hand, if enforcing
    $\sigma_l^\prime=0$, then $\sigma_{\rm
    SE}=c\sigma^\prime_w<c\sigma_w$, and there will be tension between
    the observed narrow distribution in line width
    \citep[e.g.,][]{Shen_etal_2008b,Fine_etal_2008}, which indicates
    $\sigma_w\lesssim 0.15$ dex for \MgII, and the expectation that
    $\sigma_{\rm SE}>0.3$ dex.

\end{itemize}

\begin{figure}[ht]
\includegraphics[width=0.5\textwidth]{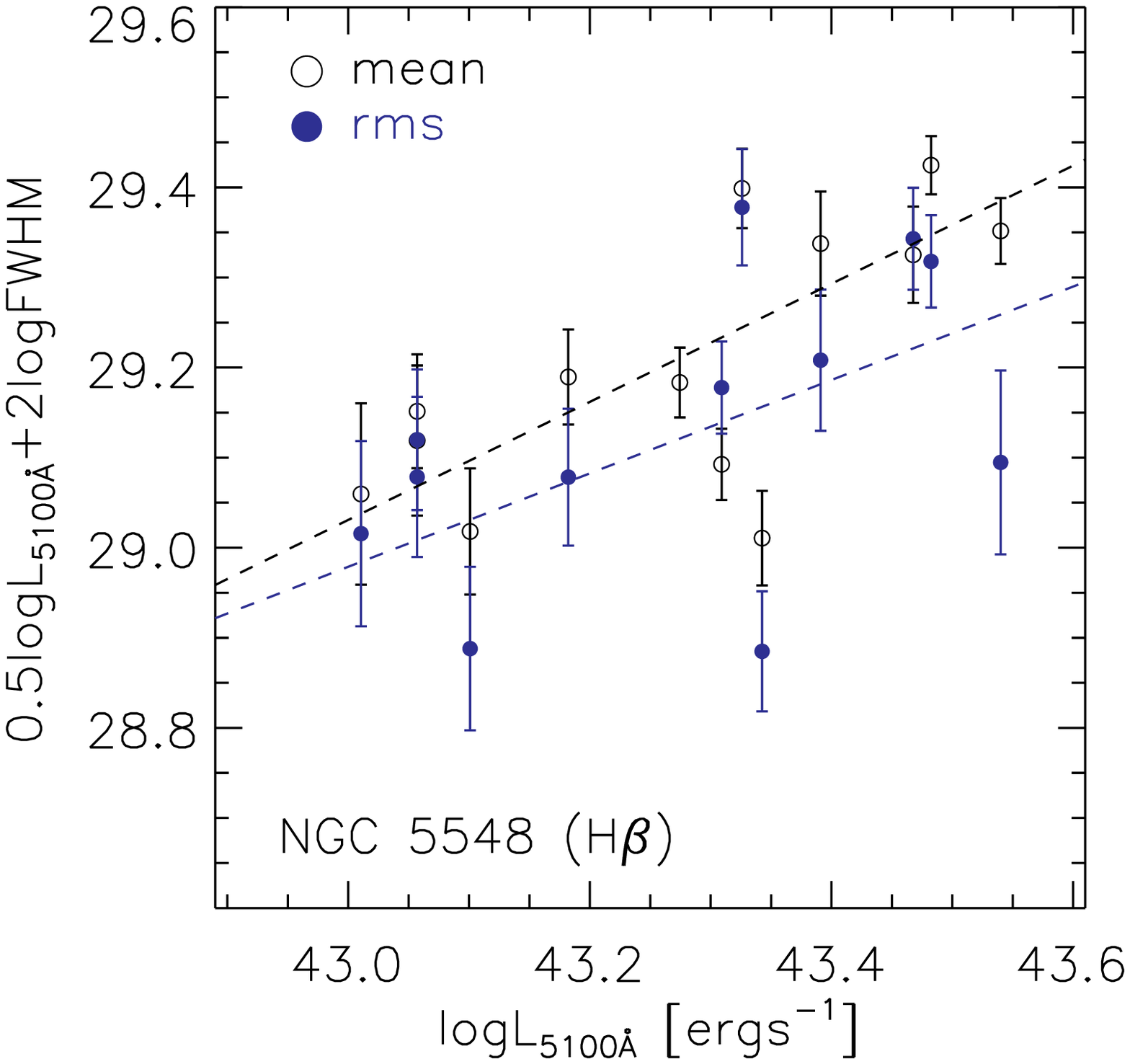}
\includegraphics[width=0.5\textwidth]{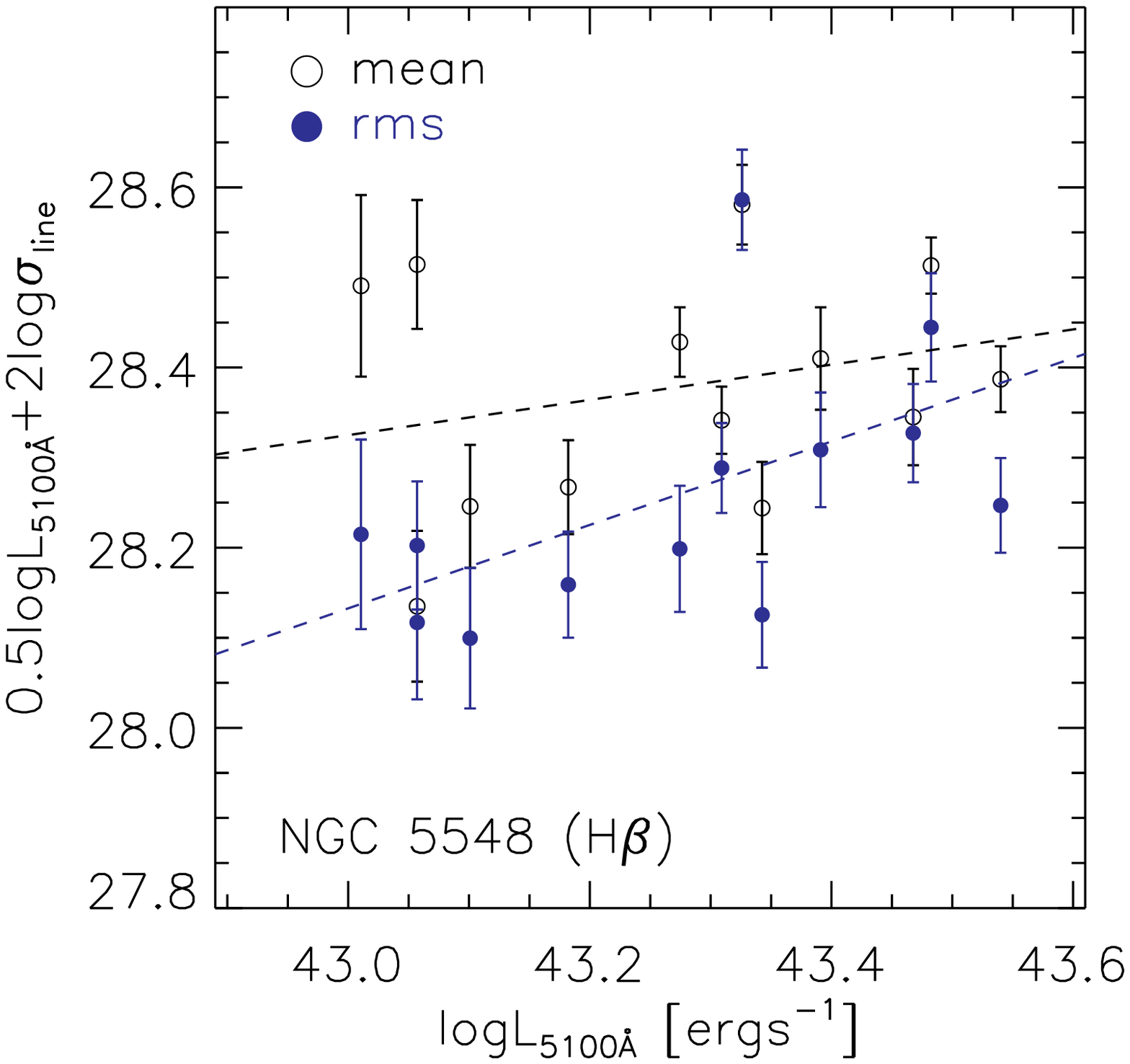}
\caption{A test on a non-zero $\beta$ using RM data for NGC 5548. Plotted here is the dependence of the virial products computed from
$5100$\AA\ continuum luminosity and line width as a function
    of luminosity, for NGC 5548 and for \hbeta\ only, for FWHM (left) and $\sigma_{\rm line}$ (right), respectively.
This form of the virial product represents SE estimators with $b=0.5$ and $c=2$ in Eqn.\ (\ref{eqn:virial_mass}), and I am using luminosity
instead of time lag $\tau$ in computing the virial product since I am testing SE mass estimators.
    The measurements were taken from
    \citet{Collin_etal_2006}, which are based on both mean and rms spectra during each monitoring
    period. Error bars represent measurement errors. The error bars in luminosity have been
    omitted in the plot for clarity. The continuum luminosity has been corrected for
    host starlight using the correction provided by \citet{Bentz_etal_2009a}. The black
    and blue dashed lines are the best linear-regression fits using the Bayesian method of \citet{Kelly_2007},
    for measurements based on mean and rms spectra, respectively. The residual correlation between the virial product
    and luminosity cannot be completely removed, and a positive $\beta\sim 0.2-0.6$ is inferred in all
    cases, although the uncertainty in $\beta$ is too large to rule out a zero $\beta$ at $>3\sigma$ significance. Figure
    adapted from \citet{Shen_Kelly_2011}. }\label{fig:beta_ngc5548}
\end{figure}

\begin{figure}
\includegraphics[width=0.5\textwidth]{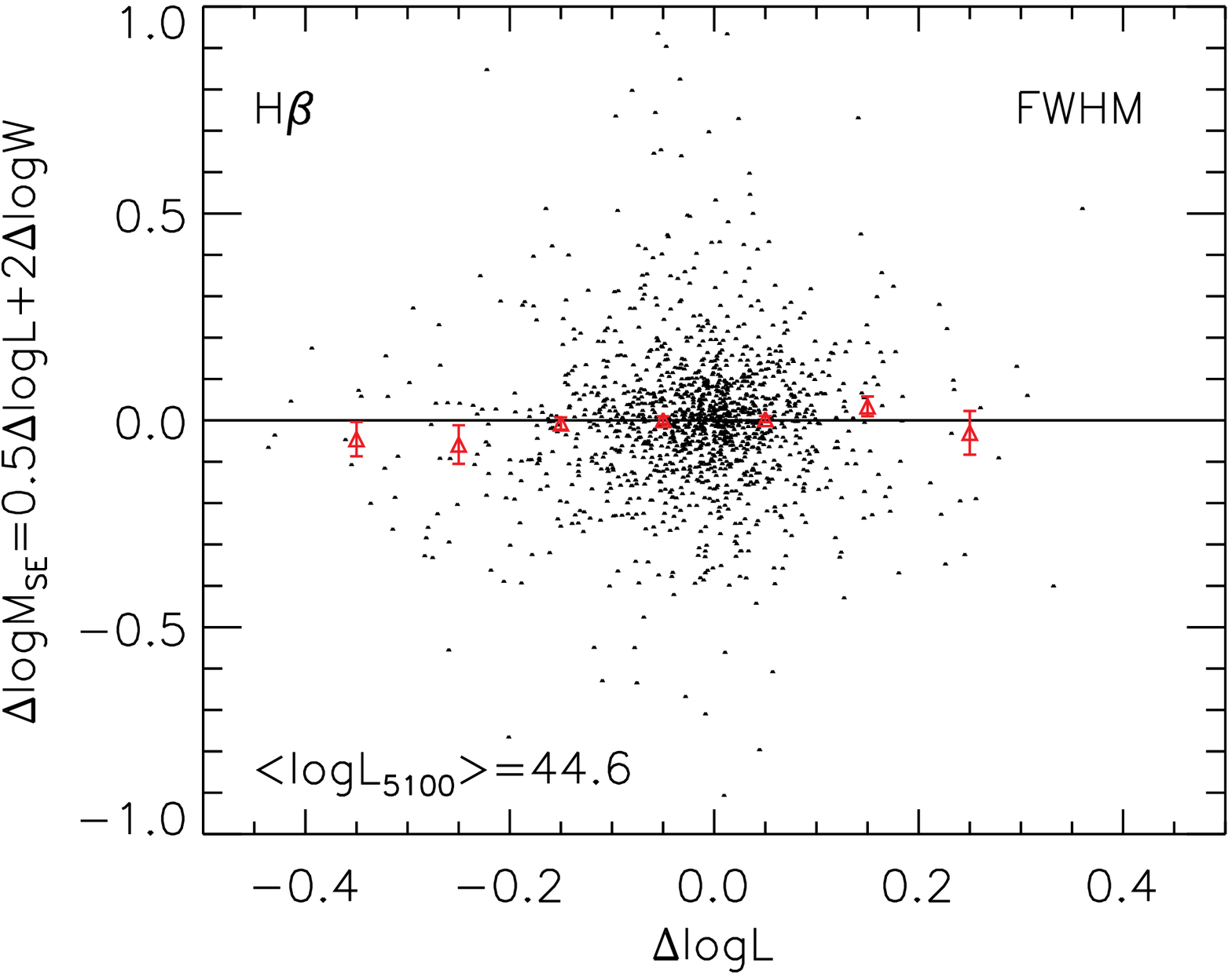}
\includegraphics[width=0.5\textwidth]{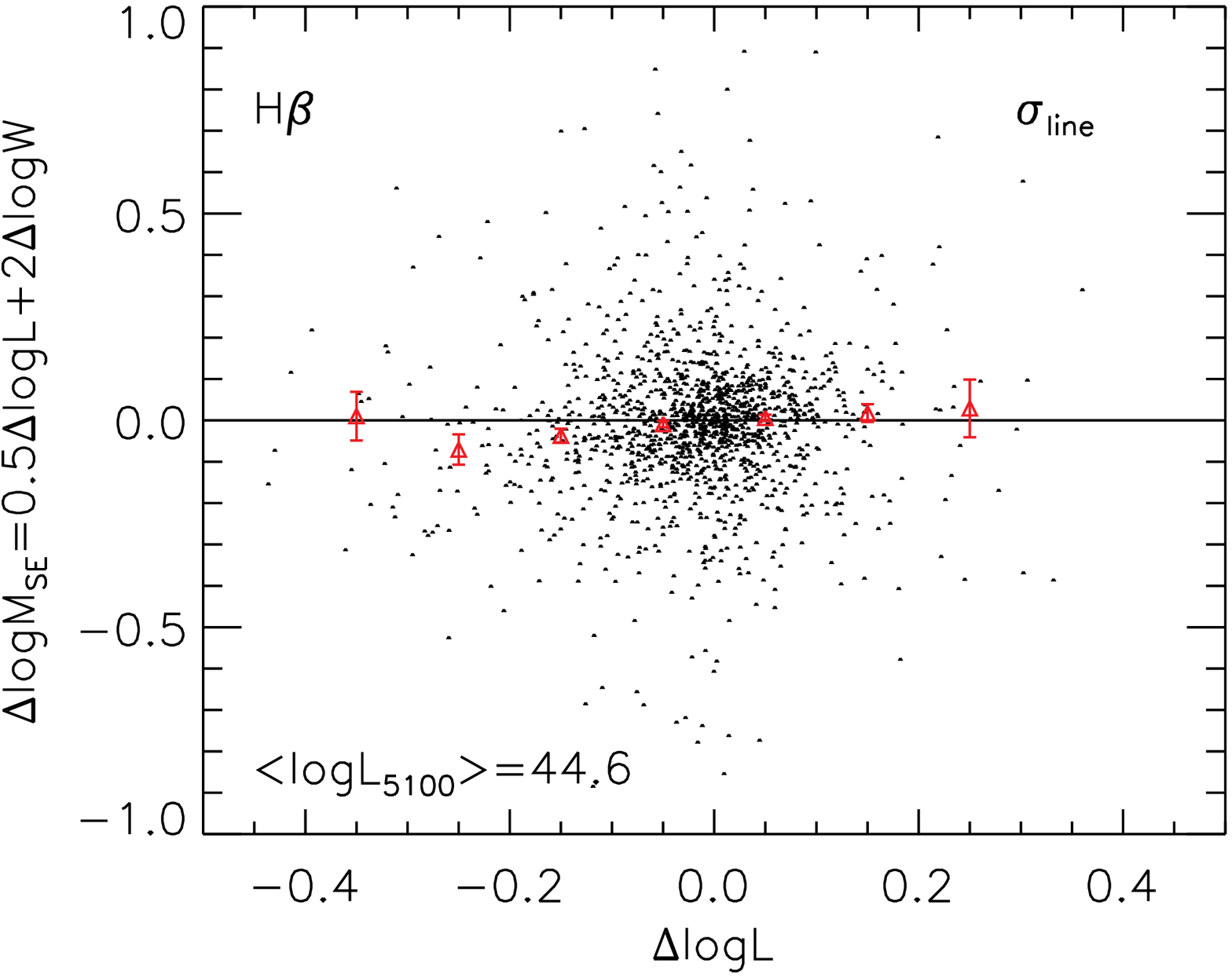}\\
\includegraphics[width=0.5\textwidth]{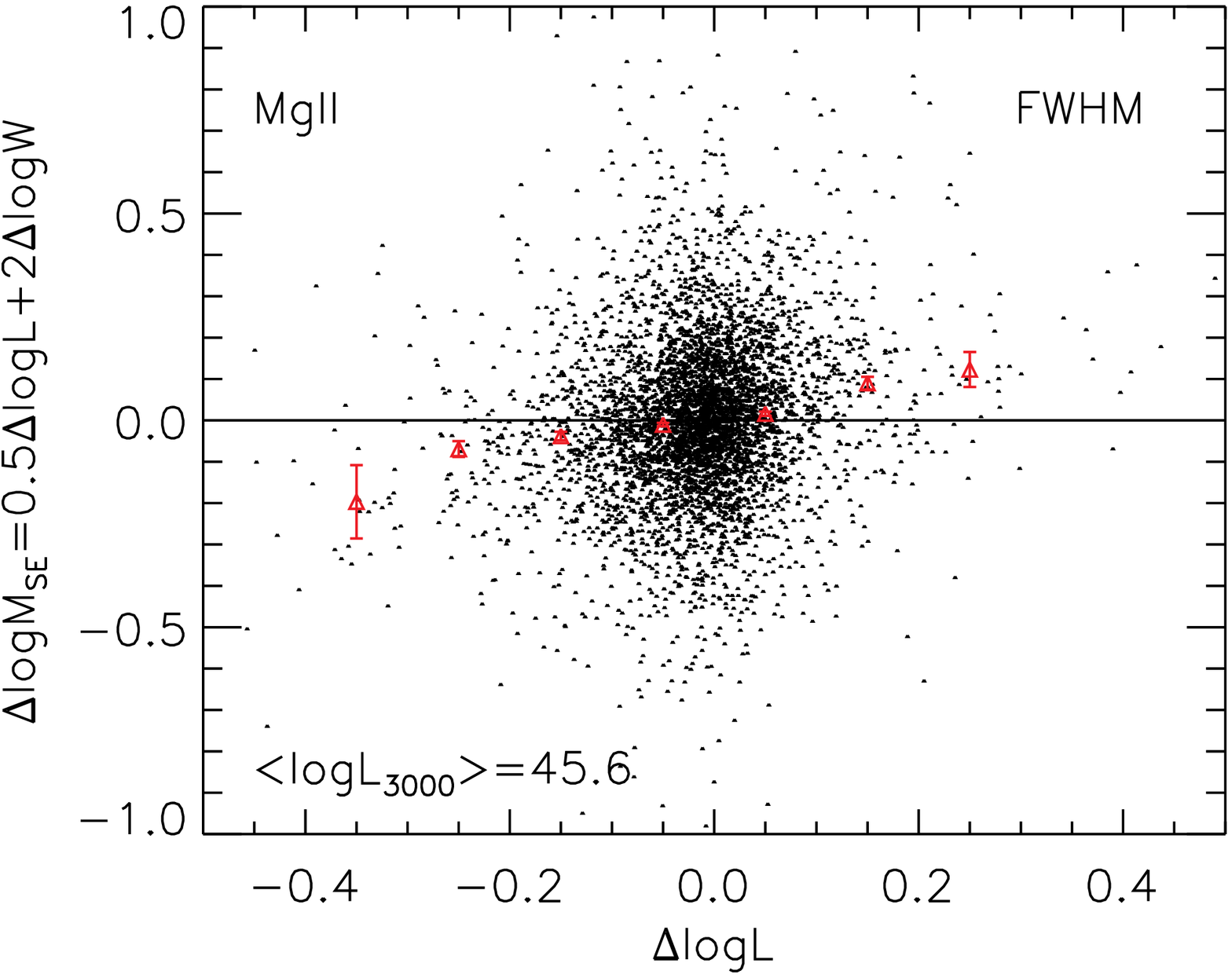}
\includegraphics[width=0.5\textwidth]{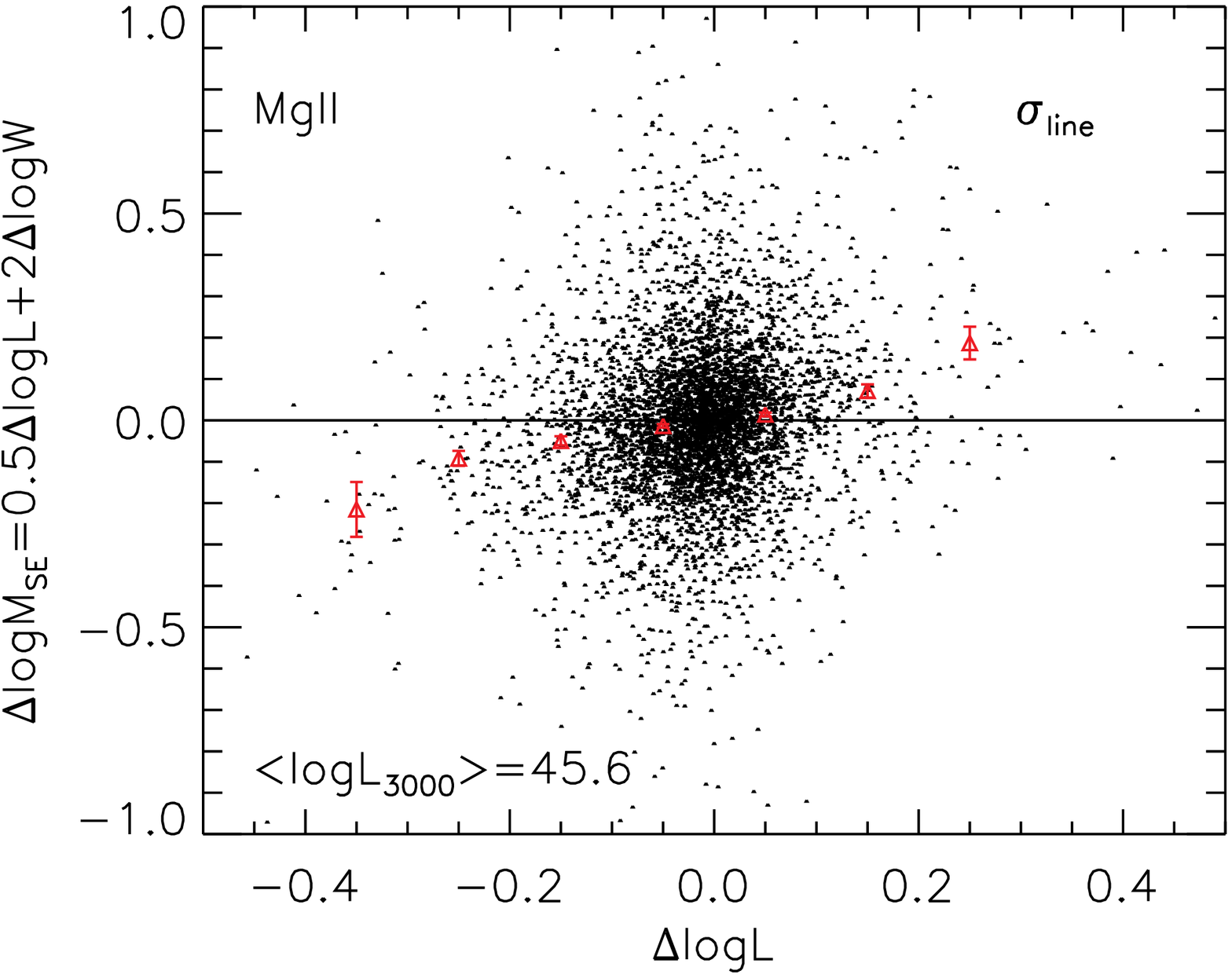}\\
\includegraphics[width=0.5\textwidth]{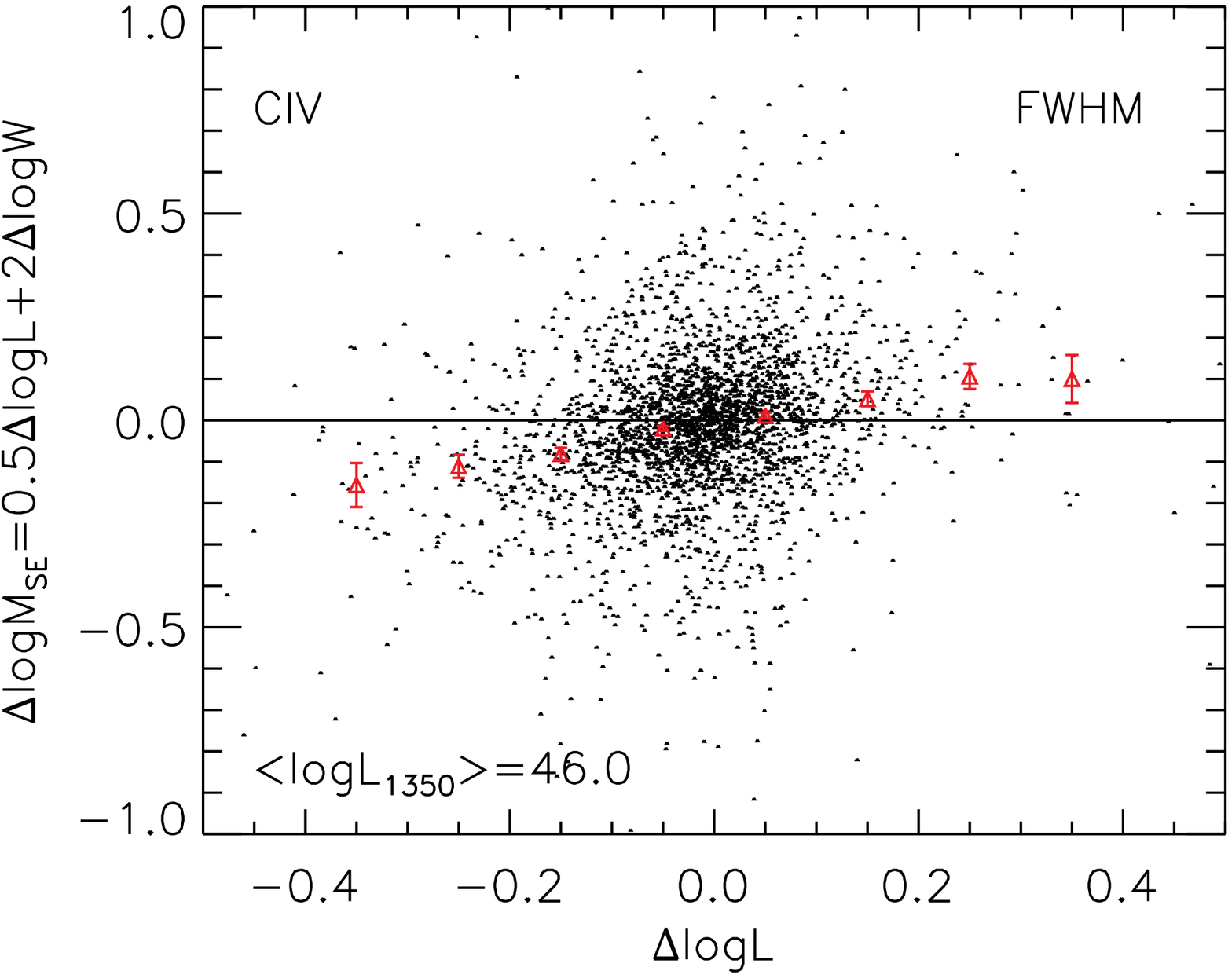}
\includegraphics[width=0.5\textwidth]{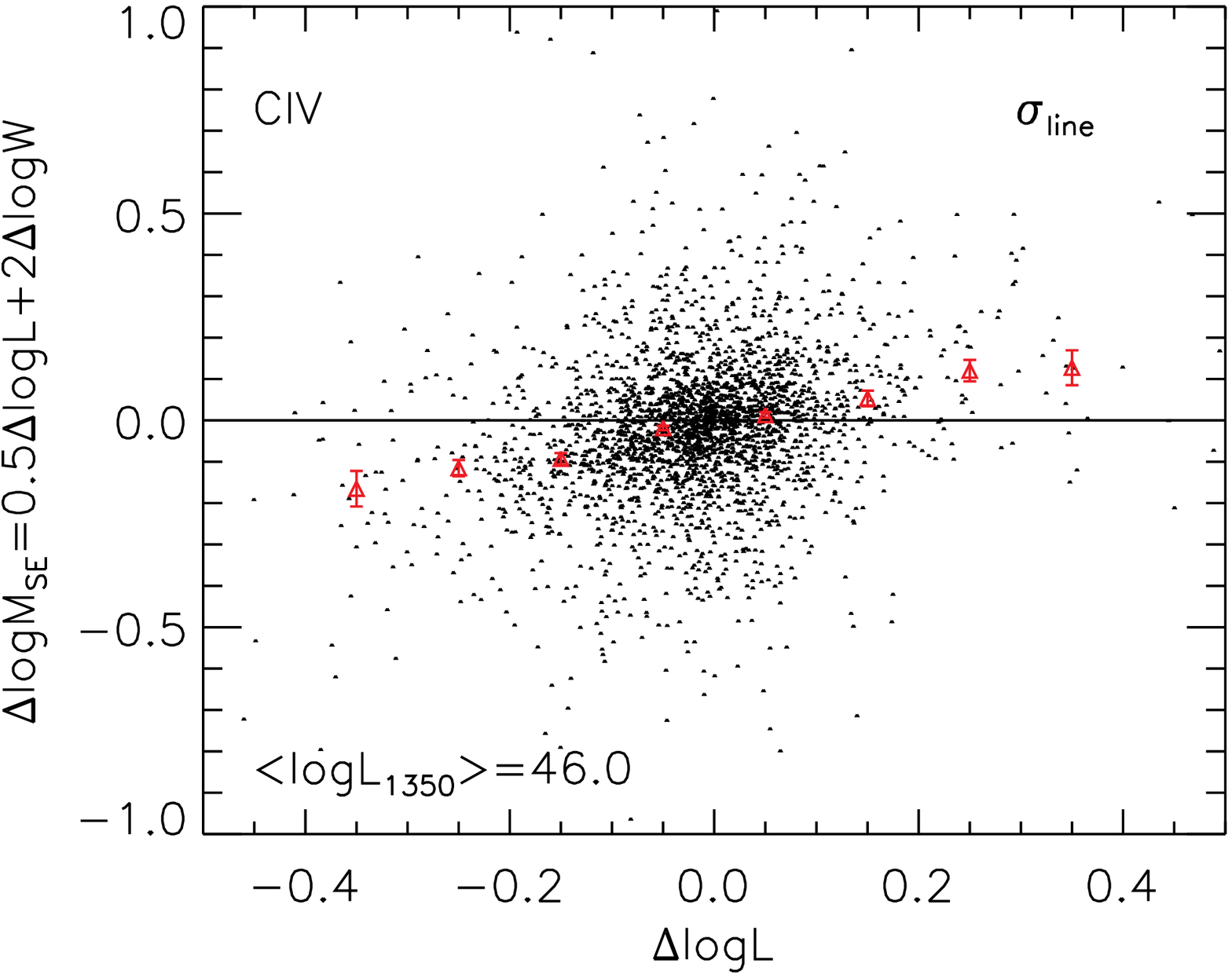}
\caption{Tests of a non-zero $\beta$ using two-epoch spectroscopy from
SDSS (She, Shen, et~al., in preparation).
The data and notations are the same as in Fig.\ \ref{fig:twoepoch_hbeta}. The SE masses remain more or less constant
as luminosity changes only for the low-luminosity and low-$z$ sample based on \hbeta. For the high-luminosity and
high-$z$ samples based on \MgII\ and \CIV, a luminosity-dependent bias in SE masses is inferred, with an error slope of
$\beta\sim 0.5$. }\label{fig:twoepoch_mass}
\end{figure}

\noindent{\em 4) Observational constraints on the luminosity-dependent bias}

The exact value of $\beta$ is difficult to determine observationally,
although some rough estimates can be made based on monitoring data of single
objects, or samples of AGNs with known ``true'' mass (using RM masses or
$M_{\rm BH}-\sigma_*$ masses). The former test constrains the stochasticity
in single objects, while the latter test explores object-by-object variance.
Using the intensively monitored \hbeta\ RM data for a single object, NGC
5548, \citet{Shen_Kelly_2011} tested the possibility of a non-zero $\beta$.
The continuum luminosity of this object varied by $\sim 0.5$ dex within a
decade, providing a test on how good line width varies in accordance to
luminosity variations for a single object and for a single line. Fig.\
\ref{fig:beta_ngc5548} shows the change of SE masses as a function of the
mean continuum luminosity in each monitoring year, computed using both FWHM
and $\sigma_{\rm line}$ from both mean and rms spectra in each year. There is
an average trend of increasing the SE masses as luminosity increases in all
four cases, although the trend is less obvious for $\sigma_{\rm line}$-based
SE masses. The inferred value of $\beta$, using the linear regression method
in \citet{Kelly_2007}, is $\sim 0.2-0.6$, although the uncertainty in $\beta$
is generally too large to rule out a zero $\beta$ at $>3\sigma$ significance.

A similar test is based on the repeated spectroscopy in SDSS (see discussion
in \S\ref{s:virial_assump}). While most objects do not have a large dynamic
range in luminosity variations in two epochs, the large number of objects
allows a reasonable determination of the average trend of SE masses with
luminosity, for the whole population of quasars. In addition, we want to
include measurement uncertainties (both statistical and systematic), which
allows us to make realistic constraints, as measurement errors will always be
present. As shown in Fig.\ \ref{fig:twoepoch_hbeta}, the line width from
single-epoch spectra does not seem to respond to luminosity variations except
for low-luminosity objects based on \hbeta, as expected from the
physical/practical reasons I described above. I plot the changes in SE masses
as a function of luminosity changes in Fig.\ \ref{fig:twoepoch_mass}. From
this figure I estimate $\beta\sim 0.5$ for the high-$L$ samples based on
\MgII\ and \CIV, and $\beta\sim 0$ for the low-$L$ sample based on \hbeta\
(She, Shen et~al., in preparation). The difference in the low-$L$ (low-$z$)
and high-$L$ (high-$z$) samples could be due to a luminosity effect, e.g.,
the correspondence between line width and luminosity variations is poorer at
higher luminosities, or due to the difference between \hbeta\ and the other
two lines (She, Shen et~al., in preparation).

\citet{Shen_Kelly_2011} also attempted to constrain $\beta$ using forward
Bayesian modeling of SDSS quasars in the mass-luminosity plane (see
\S\ref{s:app:ML}). While the results suggested a non-zero $\beta\sim
0.2-0.4$, the constraints were not very strong (see their fig.\ 11).
Combining all these tests, we can conclude the following: $\beta$ is probably
smaller than $\sim 0.5$ (i.e., line width still plays some physical role in
SE mass estimates) but unlikely zero, although the exact value is uncertain.
The value of $\beta$ likely also depends on the specific line. More
monitoring data of individual AGNs, and/or a substantially larger sample of
AGNs with RM mass (or $M_{\rm BH}-\sigma_*$ masses) spreading enough dynamic
range in luminosity at fixed mass, will be critical in better constraining
$\beta$.

\begin{figure}
\includegraphics[width=1.\textwidth]{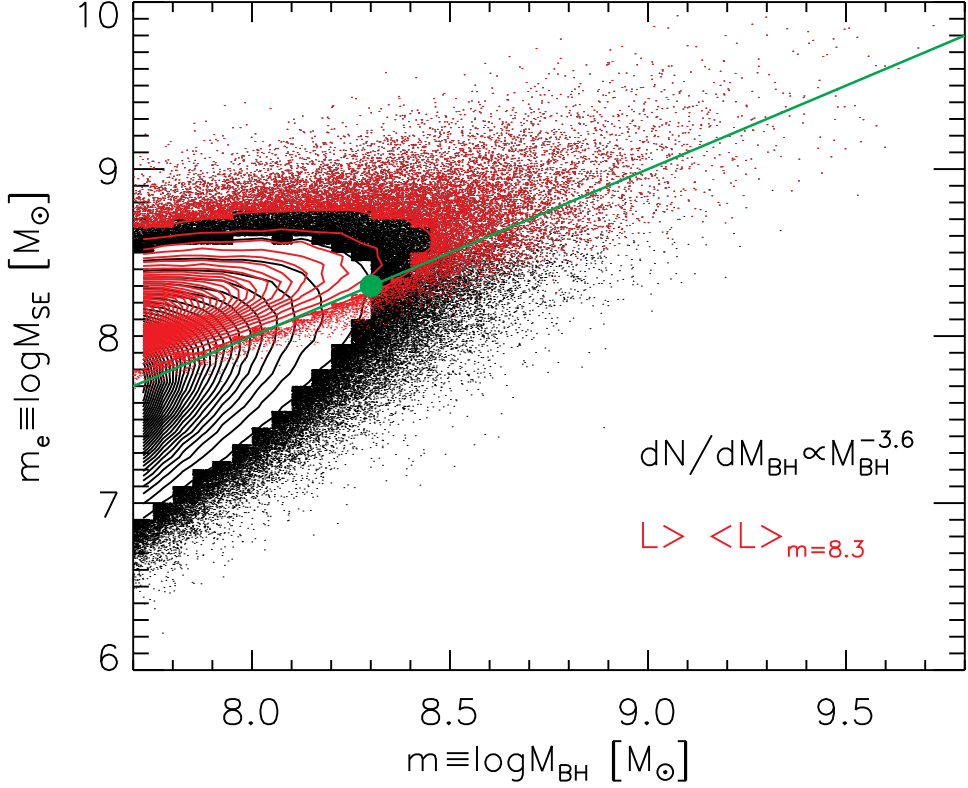}
\caption{A simulated population of BHs with true masses $M_{\rm BH}$ within $5\times 10^7-10^{10}\,M_\odot$, distributed as
a power-law $dN/dM_{\rm BH}\propto M_{\rm BH}^{-3.6}$. I have assumed a constant Eddington ratio $\lambda=0.05$
to generate the mean luminosity $\langle l\rangle_m$ at fixed true mass. I then used the dispersions specified by the
example shown in Fig.\ \ref{fig:dist} to generate instantaneous luminosity and SE masses. The resulting distributions
in luminosity, line width and SE masses are consistent with those for the SDSS quasar sample \citep[e.g.,][]{Shen_etal_2011}.
The black contours (local point density contours) and points show the distribution in the $m_e-m$ plane. The green line
shows the unity relation $m_e=m$. As expected, there is
substantial scatter in $m_e$ at fixed $m$ due to the uncertainty in SE masses ($\sigma_{\rm SE}=0.42$ dex). The red contours
and points show the distribution for a subset of quasars with $l>\langle l\rangle_{m=8.3}$, the corresponding
mean luminosity at $M_{\rm BH}=2\times 10^8\,M_\odot$ (marked by the green circle). The SE masses are biased high from their true masses for this
flux-limited sample (i.e., most points are above the unity relation) due to the substantial luminosity-dependent bias, the large dispersion in
luminosity at fixed true mass, and
the bottom-heavy true mass distribution in this example.}\label{fig:sim_bh}
\end{figure}

The effects of the luminosity-dependent bias on flux-limited samples are
discussed intensively in, e.g., \citet{Shen_etal_2008b},
\citet{Shen_Kelly_2010} and \citet{Shen_Kelly_2011}. Here I use a simple
simulation of a power-law true BH mass distribution, $dN/dM_{\rm BH}\propto
M_{\rm BH}^{-3.6}$, to demonstrate these effects in Fig.\ \ref{fig:sim_bh}.
This steep true mass distribution was chosen to reproduce the distribution at
the high-mass end for SDSS quasars \citep{Shen_etal_2008b}, which is
certainly not appropriate at the low-mass end. I use the same example as in
Fig.\ \ref{fig:dist} for all the dispersion terms in luminosity and line
width at fixed true mass. The true masses are distributed between $5\times
10^7\,M_\odot$ and $10^{10}\,M_\odot$ according to the specified power-law
distribution. The mean luminosity at fixed $m$ is determined assuming a
constant Eddington ratio $\lambda=0.05$. Then the instantaneous luminosity
and SE mass at each true mass $m$ are generated using Eqns.\
(\ref{eqn:sigma_me})-(\ref{eqn:sigma_ml}). The resulting distribution in the
$m-m_e$ plan is shown in black contours and points in Fig.\ \ref{fig:sim_bh},
where I also show the distribution of a flux-limited (luminosity-limited)
subset of BHs with $l>\langle l\rangle_{m=8.3}$, the mean luminosity
corresponding to $M_{\rm BH}=2\times 10^8\,M_\odot$. The simulated
distributions in luminosity, line width and SE virial masses are consistent
with those for SDSS quasars when a similar flux limit is imposed on the
simulated BHs. It is clear from Fig.\ \ref{fig:sim_bh} that for the
flux-limited subset, the SE masses are biased high from the true masses. This
is because in this simulation we have $\beta=0.49$, which implies a
significant luminosity-dependent bias at fixed true mass. Then the
bottom-heavy BH mass distribution and the large scatter of luminosity at
fixed true mass lead to more overestimated SE masses scattered upward than
underestimated ones scattered downward, causing a net sample bias in SE
masses \citep[][]{Shen_etal_2008b,Shen_Kelly_2010,Shen_Kelly_2011}.

\section{Applications to Statistical Quasar Samples}\label{s:app}

Despite the many caveats of SE mass estimators discussed above, they have
been extensively used in recent years to measure BH masses in quasar and AGN
samples over wide luminosity and redshift ranges, given their easiness to
use. These applications include the Eddington ratio distributions of quasars,
the demographics of quasars in terms of the black hole mass function (BHMF),
the correlations between BH mass and host properties, and BH mass dependence
of quasar properties. It is important to recognize, however, that these BH
mass estimates are not true masses, and the uncertainty in these mass
estimates has dramatic influences on the interpretation of these
measurements.

Below I discuss several major applications of the SE virial mass estimators
to statistical quasar samples. Other applications of these SE masses, such as
quasar phenomenology, while equally important, will not be covered here.

\subsection{Early Growth of SMBHs}\label{s:app:highz}

One of the strong drivers for developing the SE virial mass technique is to
estimate BH masses for high redshift quasars to better than a factor of ten
accuracy, and to study the growth of SMBHs up to very high redshift
\citep[e.g.,][]{Vestergaard_2004}. Such investigations have been greatly
improved in the era of modern, large-scale spectroscopic surveys. The SDSS
survey has been influential on this topic by providing more than tens of
thousands of optical quasar spectra and SE mass estimates up to $z\sim 5$
\citep[e.g.,][]{McLure_Dunlop_2004,Netzer_Trakhtenbrot_2007,Shen_etal_2008b,Shen_etal_2011,Labita_etal_2009a,Labita_etal_2009b}.
On the other hand, deeper and dedicated optical and near-IR spectroscopic
programs are probing the SMBH growth to even higher redshift
\citep[e.g.,][]{Jiang_etal_2007,Kurk_etal_2007,Willott_etal_2010,Mortlock_etal_2011}.

\begin{figure}
\includegraphics[width=1.\textwidth]{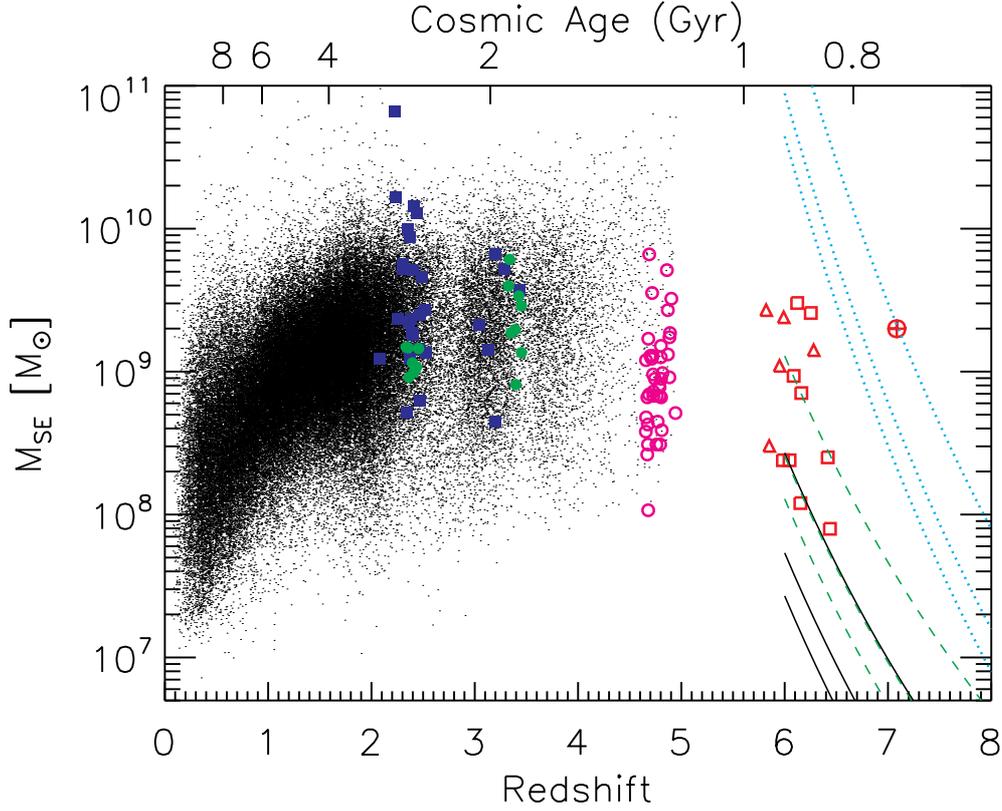}
\caption{A compilation of SE virial mass estimates from different samples of quasars.
The black dots are the SE masses for SDSS quasars from \citet{Shen_etal_2011}, based
on \hbeta\ ($z<0.7$), \MgII\ ($0.7<z<1.9$) and \CIV\ ($z>1.9$). Given the potential
caveats of \CIV-based SE masses, near-IR spectroscopy has been undertaken to estimate
SE masses for $z\gtrsim 2$ quasars based on \hbeta\ and \MgII. The different large
symbols are for SE masses based on \hbeta\ (filled symbols) and \MgII\ (open symbols) from
\citet[][filled squares]{Shemmer_etal_2004}, \citet[][filled circiles]{Netzer_etal_2007},
\citet[][open circles]{Trakhtenbrot_etal_2011}, \citet[][open triangles]{Kurk_etal_2007},
\citet[][open squares]{Willott_etal_2010}, and \citet[][open circle with cross]{Mortlock_etal_2011}.
Bear in mind the large uncertainty associated with individual SE masses and potential biases. I
also show the predicted SMBH growth at $z>6$ based on simple, continuous accretion models with constant
Eddington ratio $\lambda$ and radiative efficiency $\epsilon=0.1$, as described in Eqn.\ (\ref{eqn:edd_growth}).
The solid lines are Eddington-limited ($\lambda=1$) growth models from a seed BH at $z=20$; the dashed lines are
$\lambda=1$ growth models from a seed BH at $z=30$; the dotted lines are mildly super-Eddington ($\lambda=1.5$)
growth models from a $z=20$ seed BH. For each model I used three seed BH masses, $M_{\rm seed}=10,20,100\,M_\odot$
to accommodate reasonable ranges of seed BH mass from a Pop III star remnant at $z\sim 20-30$.
}\label{fig:mvir_evo}
\end{figure}

In Fig.\ \ref{fig:mvir_evo} I show a compilation of SE virial mass estimates
for quasars over a wide redshift range $0<z\lesssim 7$ from different
studies. The black dots show the SE masses from the SDSS DR7 quasar sample
\citep{Schneider_etal_2010,Shen_etal_2011}, which were estimated based on
\hbeta\ for $z<0.7$, \MgII\ for $0.7<z<1.9$ and \CIV\ for $1.9<z<5$. As
discussed in \S\ref{s:comp_line}, the reliability of \CIV-based SE masses for
the high-$z$ and high-luminosity quasars has been questioned, and several
studies have obtained near-IR spectra for $z\gtrsim 2$ quasars to get
\hbeta-based (filled symbols) and \MgII-based (open symbols) SE masses at
high redshift. Albeit with considerable uncertainties and possible biases in
individual SE mass estimates (all on the order of a factor of a few), these
studies show that massive, $\gtrsim 10^9\,M_\odot$ BHs are probably already
in place by $z\sim 7$, when the age of the Universe is only less than $1$
Gyr. The abundance of these massive and active BHs then evolves strongly with
redshift, showing the rise and fall of the bright quasar population with
cosmic time.

One outstanding question regarding the observed earliest quasars is how they
could have grown such massive SMBHs given the limited time they have, which
is a non-trivial problem since the discoveries of $z>4$ quasars
\citep[e.g.,][]{Turner_1991,Haiman_Loeb_2001}. One concern is if these
highest redshift quasars have their luminosities magnified by gravitational
lensing or if their luminosities are strongly beamed
\citep[e.g.,][]{Wyithe_Loeb_2002,Haiman_Cen_2002}, which will affect earlier
estimates of their BH masses using the Eddington-limit argument. The lensing
hypothesis will also lead to overestimated virial BH masses. However later
deep, high-resolution imaging of $z>4$ quasars with HST did not find any
multiple images around these objects
\citep[e.g.,][]{Richards_etal_2004b,Richards_etal_2006b}, rendering the
lensing hypothesis highly unlikely \citep[e.g.,][]{Keeton_etal_2005}. Strong
beaming can also be ruled out based on the high values of the observed
line/continuum ratio of these high-redshift quasars
\citep[e.g.,][]{Haiman_Cen_2002}.

Given the $e$-folding time introduced in \S\ref{s:intro}, $t_e=4.5\times
10^8\frac{\epsilon}{\lambda(1-\epsilon)}$ yr, and a seed BH mass $M_{\rm
seed}$ at an earlier epoch $z_i$, the final mass at $z_f\sim 6$ is
\begin{equation}\label{eqn:edd_growth}
M_{\rm BH}=M_{\rm seed}\exp\left(\frac{t_f-t_i}{t_e}\right)\ ,
\end{equation}
where $t_f$ and $t_i$ are the cosmic age at $z_f$ and $z_i$, respectively.

Assuming continuous accretion with constant radiative efficiency $\epsilon$
and luminosity Eddington ratio $\lambda$ and {\em without mergers}, I showed
in Fig.\ \ref{fig:mvir_evo} three different growth histories from a seed BH
at higher redshift. The solid lines are for a seed BH at $z=20$ and
$\lambda=1$, i.e., Eddington-limited accretion; the dashed lines are for a
seed BH at $z=30$ and $\lambda=1$; the dotted lines are for a seed BH at
$z=20$ and $\lambda=1.5$, i.e., mildly super-Eddington accretion. For each
model I used three seed BH mass, $M_{\rm seed}=10,20,100\,M_\odot$, which
encloses the reasonable ranges of predicted remnant BH mass from the first
generation of stars \citep[Pop III stars, for a review see,
e.g.,][]{Bromm_etal_2009}. Then it is clear from Fig.\ \ref{fig:mvir_evo}
that, if the accretion is Eddington-limited, it is difficult to grow $\gtrsim
10^9\,M_\odot$ BHs at $z\sim 6$ from a Pop III remnant seed BH at $z\sim
20-30$ (where such first stars were formed out of $\sim10^{6}\,M_\odot$
halos, corresponding to $\sim 3-4\sigma$ peaks in the density perturbation
field). On the other hand, if allowing mildly super-Eddington accretion, then
a $\sim 10^9\,M_\odot$ BH can be readily formed at $z\sim 7$ from a large Pop
III star remnant $M_{\rm seed}\sim 100\,M_\odot$, although more recent
simulations suggest somewhat lower masses of Pop III stars due to possible
effects of clump fragmentation and/or radiative feedback
\citep[e.g.,][]{Turk_etal_2009,Hosokawa_etal_2011,Stacy_etal_2012}, and hence
a lower typical value of the remnant mass of less than tens of solar masses.
Mildly super-Eddington accretion (up to $\lambda\sim$ a few) could happen,
for instance, if the radiation and density fields of the accretion flow are
anisotropic and most of the accretion flow is not impeded by the radiation
force. Mergers between BHs at high-$z$ can also help with the required growth
if the coalesced BH is not ejected from the halo by the gravitational recoil
from the merger. The main challenge here is whether or not such critical
accretion can maintain stable and uninterrupted for the entire time
\citep[e.g.,][]{Pelupessy_etal_2007}. But in any case, it is quite likely
that the observed $z>6$ quasars are all born in rare environments of the
early Universe, thus extreme conditions (such as large gas density, high
merger rate, etc.) may have facilitated their growth. Indeed, some
theoretical studies can successfully produce such massive BHs at $z\sim 6$
growing from a typical Pop III remnant BH seed without super-Eddington
accretion \citep[e.g.,][]{Yoo_2004,Li_etal_2007,Tanaka_Haiman_2009}. But the
detailed physics (accretion rate, mergers, BH recoils, etc.) regarding the
formation of these earliest SMBHs is still uncertain to some large extent.

While this is not seen as an immediate crisis, there are multiple pathways to
make it much easier to grow $\gtrsim 10^9\,M_\odot$ SMBHs at $z\gtrsim 6$ by
boosting either the accretion rate or the seed BH mass \citep[for a recent
review, see, e.g.,][]{Volonteri_2010,Haiman_2012}. These recipes include:

1) supercritical accretion \citep[e.g.,][]{Volonteri_Rees_2005} where the
accretion rate $\dot{M}_{\rm BH}$ greatly exceeds the Eddington limit with a
canonical radiative efficiency $\epsilon=0.1$. One possibility is that the
radiation is trapped in the accretion flow
\citep[e.g.,][]{Begelman_1979,Wyithe_Loeb_2012}, leading to a very low
$\epsilon$ and hence a much shorter $e$-folding time. Note that in such a
radiatively inefficient accretion flow (RIAF), the luminosity is still
bounded by the Eddington limit;

2) rapid formation of massive ($\sim 10^3-10^5\,M_\odot$) BH seeds from
direct collapse of primordial gas clouds
\citep[e.g.,][]{Bromm_Loeb_2003,Begelman_etal_2006,Agarwal_etal_2012} or from
a hypothetical supermassive star or ``quasi-star''
\citep[e.g.,][]{Shibata_Shapiro_2002,Begelman_etal_2008,Johnson_etal_2012} at
high redshift. Supercritical accretion may also be expected in some of these
models to grow to the final seed BH mass, which then continue to accrete in
the normal way. By increasing $M_{\rm seed}$ it requires much less $e$-folds
to grow to a $>10^9\,M_\odot$ BH. Another possible route to produce massive
BH seeds up to $\sim 10^3\,M_\odot$ is by the runaway collisional growth in a
dense star cluster formed in a high-redshift halo
\citep[e.g.,][]{Omukai_etal_2008}.

\subsection{Quasar Demographics in the Mass-Luminosity Plane}\label{s:app:ML}

BH mass estimates provide an additional dimension in the physical properties
of quasars. The distribution of quasars in the two-dimensional BH
mass-luminosity ($M-L$) plane conveys important information about the
accretion process of these active SMBHs. The first quasar mass-luminosity
plane plot was made by Dibai in the 1970s as mentioned in
\S\ref{s:other_methods}. Over the years, such a 2D plot has been repeatedly
generated based on increasingly larger quasar samples and improved BH mass
estimates
\citep[e.g.,][]{Wandel_etal_1999,Woo_Urry_2002,Kollmeier_etal_2006,Shen_etal_2008b,Shen_Kelly_2011},
and the much better statistics now allows a more detailed and deeper look
into this quasar mass-luminosity plane.

In what follows I will mainly focus on the SDSS quasar sample because this is
the largest and most homogeneous quasar samples to date. But as emphasized in
\citet{Shen_Kelly_2011}, the SDSS sample only probes the bright-end of the
quasar population, and to probe the mass and accretion rate of the bulk of
quasars it is necessary to assemble deeper spectroscopic quasar samples
\citep[e.g.,][]{Kollmeier_etal_2006,Gavignaud_etal_2008,Trump_etal_2009,Nobuta_etal_2012}.

\begin{figure}
\includegraphics[width=1.\textwidth]{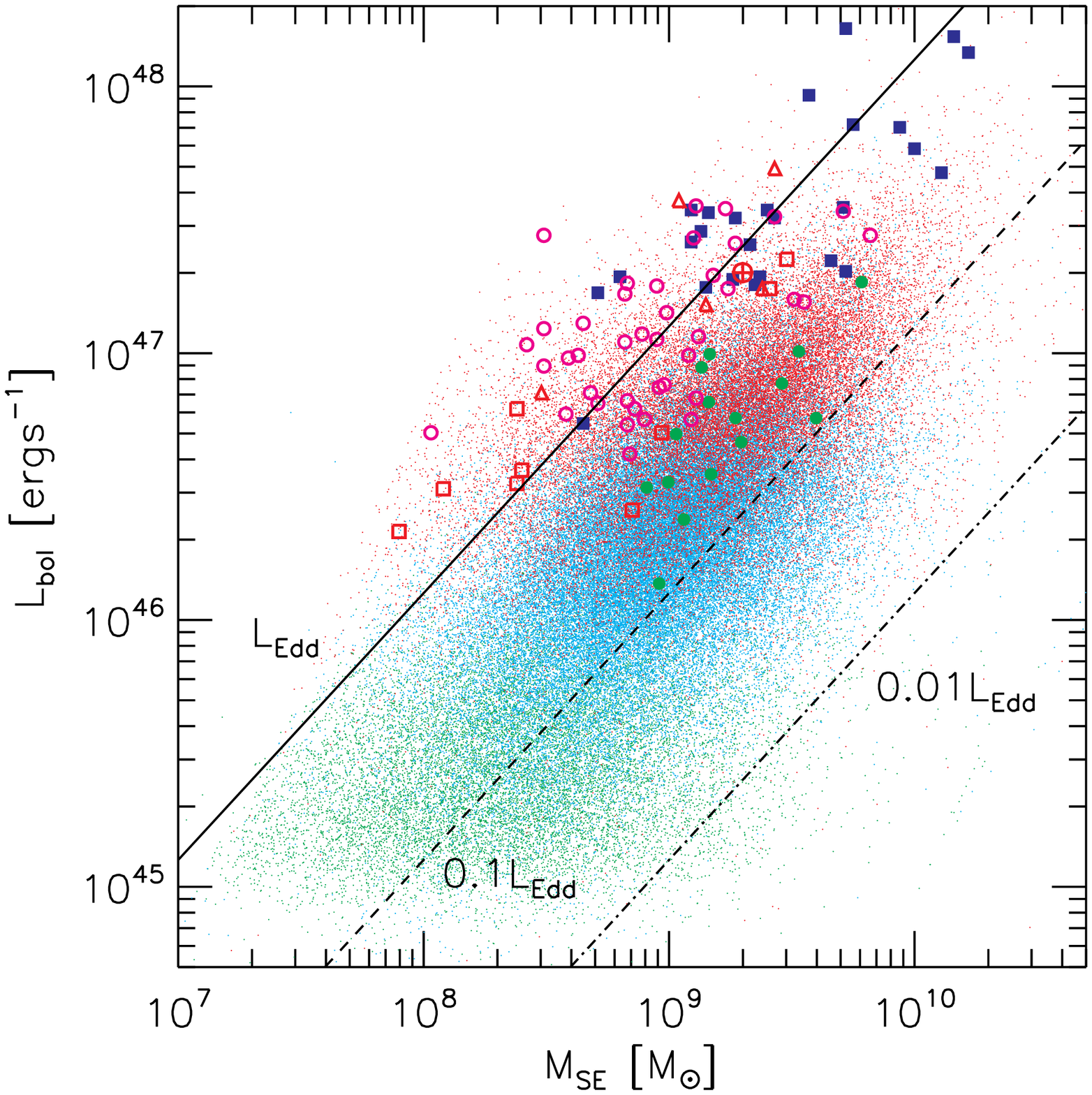}
\caption{The observed quasar mass-luminosity plane based on SE masses for quasars in a wide range of redshifts ($0<z\lesssim7$)
from the samples compiled in Fig.\ \ref{fig:mvir_evo}. The dots are from the SDSS sample in \citet{Shen_etal_2011}, for
quasars at $z<0.7$ (\hbeta-based SE masses; green), $0.7<z<1.9$ (\MgII-based SE masses; cyan), and $1.9<z<5$ (\CIV-based
SE masses; red). The large
symbols are from various $z\gtrsim 2$ samples using \hbeta\ or \MgII-based SE masses. I have used slightly different bolometric
corrections for these non-SDSS samples from those used in the original papers. The SDSS quasars have Eddington ratios $0.01\lesssim
\lambda\lesssim 1$ with a mean value of $\langle\lambda\rangle\sim 0.1-0.3$, while the other high-$z$ samples have even higher
Eddington ratios. Given the flux-limited nature of all these samples and the errors in SE masses, the {\em observed} Eddington
ratio distribution is highly biased relative to the intrinsic distribution (see \S\ref{s:app:ML} for details).
}\label{fig:mvir_Lbol}
\end{figure}

Since I have emphasized the distinction between true BH masses and SE mass
estimates, I shall use the term ``observed'' or ``measured'' to refer to
distributions based on SE mass estimates, to distinguish them from ``true''
distributions. Fig.\ \ref{fig:mvir_Lbol} shows such an {\em observed}
mass-luminosity plane from the same collection of quasars as shown in Fig.\
\ref{fig:mvir_evo}. Note that these samples are flux-limited to different
magnitudes, and several high-redshift samples based on \hbeta\ or \MgII\
(i.e., large symbols) have a higher flux-limit than the SDSS. I also used
slightly different values of bolometric corrections to convert continuum
luminosity to bolometric luminosity for those non-SDSS samples. From this
plot we see that the observed distributions of quasars are bounded between
constant Eddington ratios $0.01\lesssim \lambda \lesssim 1$, with median
values of $\langle\lambda\rangle\sim 0.1-0.3$ for SDSS quasars, and somewhat
higher values for the $z\gtrsim 5$ samples. The dispersion in Eddington ratio
in these flux-limited samples is typically $\sim 0.3$ dex. Similar
distributions were observed by, e.g., \citet{Kollmeier_etal_2006}. However,
as demonstrated in, e.g.,
\citet{Shen_etal_2008b,Kelly_etal_2009a,Kelly_etal_2010,Shen_Kelly_2011,Kelly_Shen_2013},
the observed distribution suffers from the sample flux limit such that
low-Eddginton ratio objects have a lower probability being selected into the
sample, and from the uncertainties and statistical biases of SE masses
relative to true masses. The selection effect due to the flux limit and
errors in SE masses dramatically modify the intrinsic distribution in the
mass-luminosity plane, and must be taken into account when interpreting the
observations.

\begin{figure}
\includegraphics[width=0.7\textwidth,angle=90]{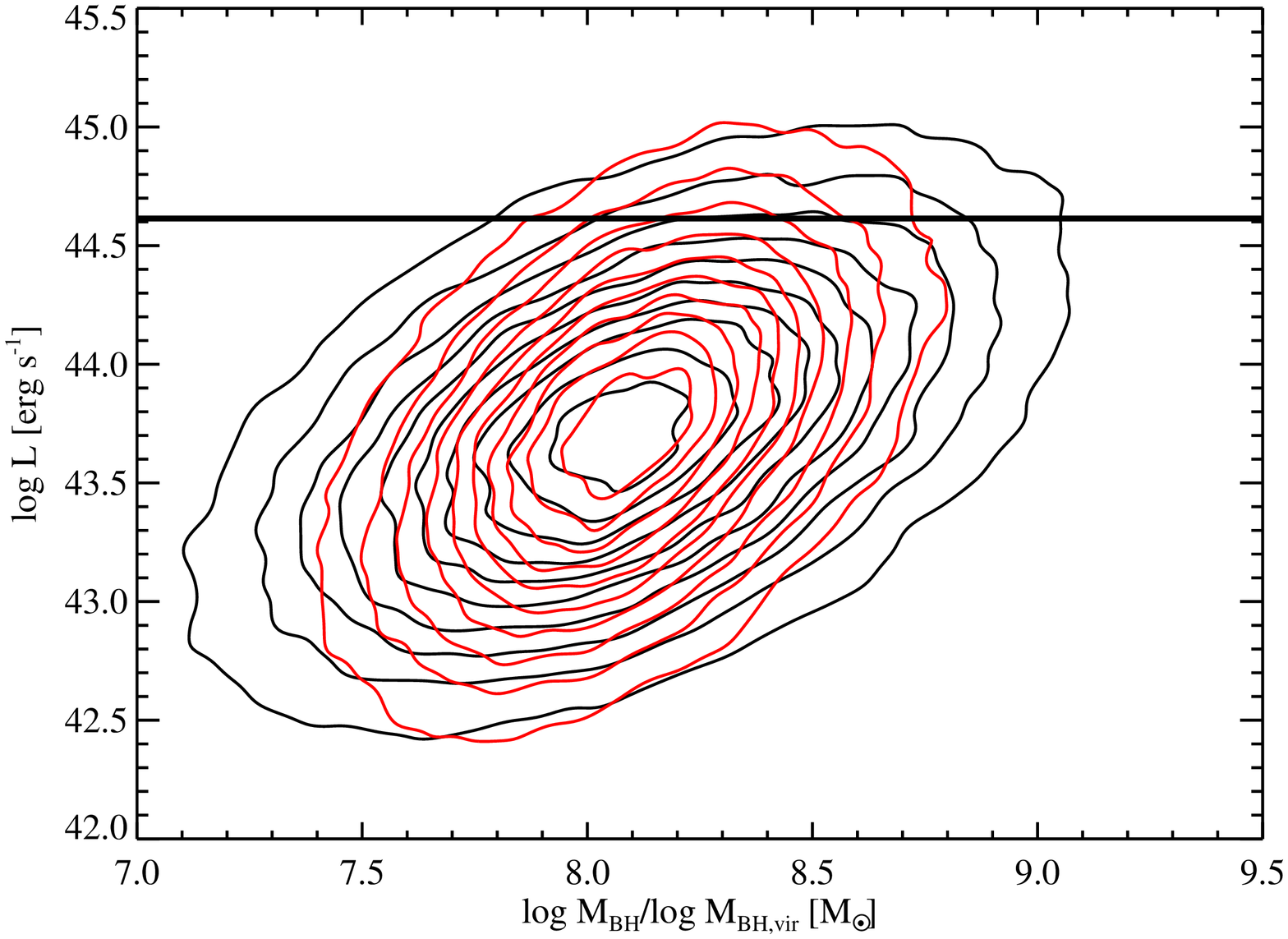}
\caption{The simulated mass-luminosity plane at $z=0.6$ based on the modeling of SDSS quasars in
\citet{Shen_Kelly_2011}, which extends below
the flux limit (the black horizontal line). The $y$-axis plots the restframe $2500$\AA\ monochromatic luminosity,
and the bolometric luminosity is $L_{\rm bol}\sim 5L$. The red contour is the distribution
based on {\em true} BH masses and is determined by the model BHMF and Eddington
ratio distribution in \citet{Shen_Kelly_2011}. The black contour is the distribution based on \hbeta\ SE BH masses. The flux limit only
selects the most luminous objects into the SDSS sample (which are closer to the Eddington limit), and the distribution based on SE virial BH masses
is flatter than the one based on true masses due to both the scatter $\sigma_{ml}$ and a non-zero
$\beta\sim 0.2$ for this redshift bin \citep[see \S\ref{s:bias_l} and][]{Shen_Kelly_2011}.
}\label{fig:ml_plane_z0.6}
\end{figure}

The best approach to tackle these issues is a forward modeling, in which one
specifies an underlying distribution of true masses and luminosities and map
to the observed mass-luminosity plane by imposing the flux limit and
relations between SE virial masses and true masses
\citep[e.g.,][]{Shen_etal_2008b,Kelly_etal_2009a,Kelly_etal_2010}. Then the
comparisons between model and observed distributions constrain the model
parameters with standard Markov Chain Monte Carlo (MCMC) techniques and
Bayesian inference. This is a complicated and model-dependent problem, and
the best efforts so far are the studies by \citet[][]{Shen_Kelly_2011} and
\citet{Kelly_Shen_2013}, building on earlier work by \citet{Shen_etal_2008b},
\citet{Kelly_etal_2009a} and \citet{Kelly_etal_2010}. Alternatively,
\citet{Schulze_Wisotzki_2010} developed a maximum likelihood method (also a
forward modeling method), which accounts for the effect of the flux limit,
but not the errors in SE masses (although the SE errors can be incorporated
in such a framework as well). This maximum likelihood method was subsequently
adopted in \citet{Nobuta_etal_2012} when modeling a faint quasar sample
(again, SE errors not taken into account). Most other quasar mass demographic
studies, however, did not explicitly model either of these effects
\citep[e.g.,][]{Greene_Ho_2007a,Vestergaard_etal_2008,Vestergaard_Osmer_2009}.

\citet{Shen_Kelly_2011} used forward modeling with Bayesian inference to
model the observed distribution in the mass-luminosity plane of SDSS quasars,
taking into account a possible luminosity-dependent bias (i.e., $\beta\neq
0$, see \S\ref{s:bias_l}) to be constrained by the data. The flux limit of
the SDSS sample is taken into account using published selection functions of
SDSS quasars \citep[][]{Richards_etal_2006}. Based on this approach,
\citet{Shen_Kelly_2011} found evidence for a non-zero $\beta$, although the
constraints on $\beta$ are weak and cannot rule out a null value.
\citet{Kelly_Shen_2013} used a more flexible model parametrization to
describe the underlying true distributions (in BH mass and Eddington ratio),
but fixed $\beta=0$ to test how sensitively the results in
\citet{Shen_Kelly_2011} depend on different model parameterizations. They
found that the main conclusions are generally consistent, although the
results in the latter work are less constrained than in the former, due to
the more flexible models. Both studies revealed that based on the SDSS quasar
sample alone, it is difficult to constrain the BHMF to better than a factor
of a few at most redshifts. This is both because the SDSS sample only probes
the tip of the active SMBH population at high-$z$, and to a larger extent,
because the errors of SE masses are poorly understood. However, there are
some solid conclusions from the two studies:

\begin{enumerate}

\item[1.] The {\em observed} distribution in the mass-luminosity plane is
    quite different from the intrinsic distribution, due to the flux
    limit and uncertainties in SE masses. This is demonstrated in Fig.\
    \ref{fig:ml_plane_z0.6}, which shows a quasar $M-L$ plane at $z=0.6$
    based on the modeling of SDSS quasars by \citet{Shen_Kelly_2011}. In
    this plot luminosity $L$ is the restframe $2500$\AA\ monochromatic
    luminosity, and the bolometric luminosity is $L_{\rm bol}\sim 5L$.
    The red contours are the {\em true} distribution of quasars, while
    the black contours are the {\em measured} distribution based on
    \hbeta\ SE virial masses. The black horizontal line indicates the
    flux limit of the sample, hence only objects above this line would be
    selected in the SDSS sample, which form the {\em observed}
    distribution. The flux limit only selects the most luminous objects
    into the SDSS sample, missing the bulk of low Eddington ratio
    objects; even the highest mass bins are incomplete due to the flux
    limit. The distribution based on SE virial BH masses is flatter than
    the one based on true masses due to the scatter and
    luminosity-dependent bias of these SE masses.

    For the {\em observed} distribution based on SE masses, there are
    fewer objects towards larger SE masses and luminosity. This was
    interpreted as the lack of massive black holes accreting at high
    Eddington ratios, or the so-called ``sub-Eddington boundary'' claimed
    by \citet{Steinhardt_Elvis_2010a}. However, such a feature is caused
    by the flux limit and errors in SE masses, and there is no evidence
    that high-mass quasars on average accrete at lower Eddington ratios,
    not for broad-line objects at least\footnote{There could be a
    significant population of high-mass, non-broad-line AGNs accreting at
    low Eddington ratios, possibly via a different accretion mode than
    broad line quasars. }. This conclusion seems to be robust against
    different model parameterizations of the underlying true
    distributions in the forward modeling approach
    \citep{Kelly_Shen_2013}. Of course here I am assuming no systematic
    biases in these FWHM-based SE masses measured in
    \citet{Shen_etal_2011}. It is possible that $\sigma_{\rm line}$-based
    SE masses are more reliable, in which case there would be a
    ``rotation'' in the mass-luminosity plane using $\sigma_{\rm
    line}$-based SE masses, as discussed in \S\ref{s:line_width}. This
    also tends to reduce this ``sub-Eddington boundary'' in the observed
    plane \citep[e.g.,][]{Rafiee_Hall_2011a,Rafiee_Hall_2011b}, but a
    full modeling taking into account both the flux limit and SE mass
    errors is yet to be performed with $\sigma_{\rm line}$-based SE
    masses (i.e., the interpretation by Rafiee \& Hall is still based on
    ``observed'' rather than true distributions).

\item[2.] The intrinsic Eddington ratio distribution at fixed true mass
    is broader ($\sim 0.4$ dex) than the {\em observed} Eddington ratio
    distribution in flux-limited samples ($\lesssim 0.3$ dex), and the
    mean Eddington ratio in the flux-limited samples based on SE masses
    is higher\footnote{These SE masses are on average overestimated due
    to the luminosity-dependent bias discussed in \S\ref{s:bias_l}, which
    tends to underestimated the true Eddington ratios. But the mean
    observed Eddington ratio based on SE masses of the flux-limited
    sample is still higher than the mean value for all quasars extending
    below the flux limit \citep[see fig.\ 19 of][]{Shen_Kelly_2011}. }
    than the mean Eddington ratio for all active SMBHs (most of which are
    not detected). This is consistent with earlier studies by
    \citet{Shen_etal_2008b} and \citet{Kelly_etal_2010}. Some deeper
    spectroscopic surveys indeed start to find these lower Eddington
    ratio objects
    \citep[e.g.,][]{Babic_etal_2007,Gavignaud_etal_2008,Nobuta_etal_2012},
    and are consistent with the model-extrapolated distributions from
    \citet{Shen_Kelly_2011} and \citet{Kelly_Shen_2013}; however, since
    in general these deep data are noisier and the selection function is
    less well quantified than SDSS, care must be paid when inferring the
    dispersion in Eddington ratios for these faint quasars.

\end{enumerate}

The next step to utilize this quasar mass-luminosity plane is to measure the
abundance of quasars in this plane, and study its redshift evolution. This is
a much more powerful way to study the cosmic evolution of quasars than
traditional 1D distribution functions such as the luminosity function (LF)
and the quasar BHMF.

\begin{figure}
\centering
    \includegraphics[height=0.4\textheight]{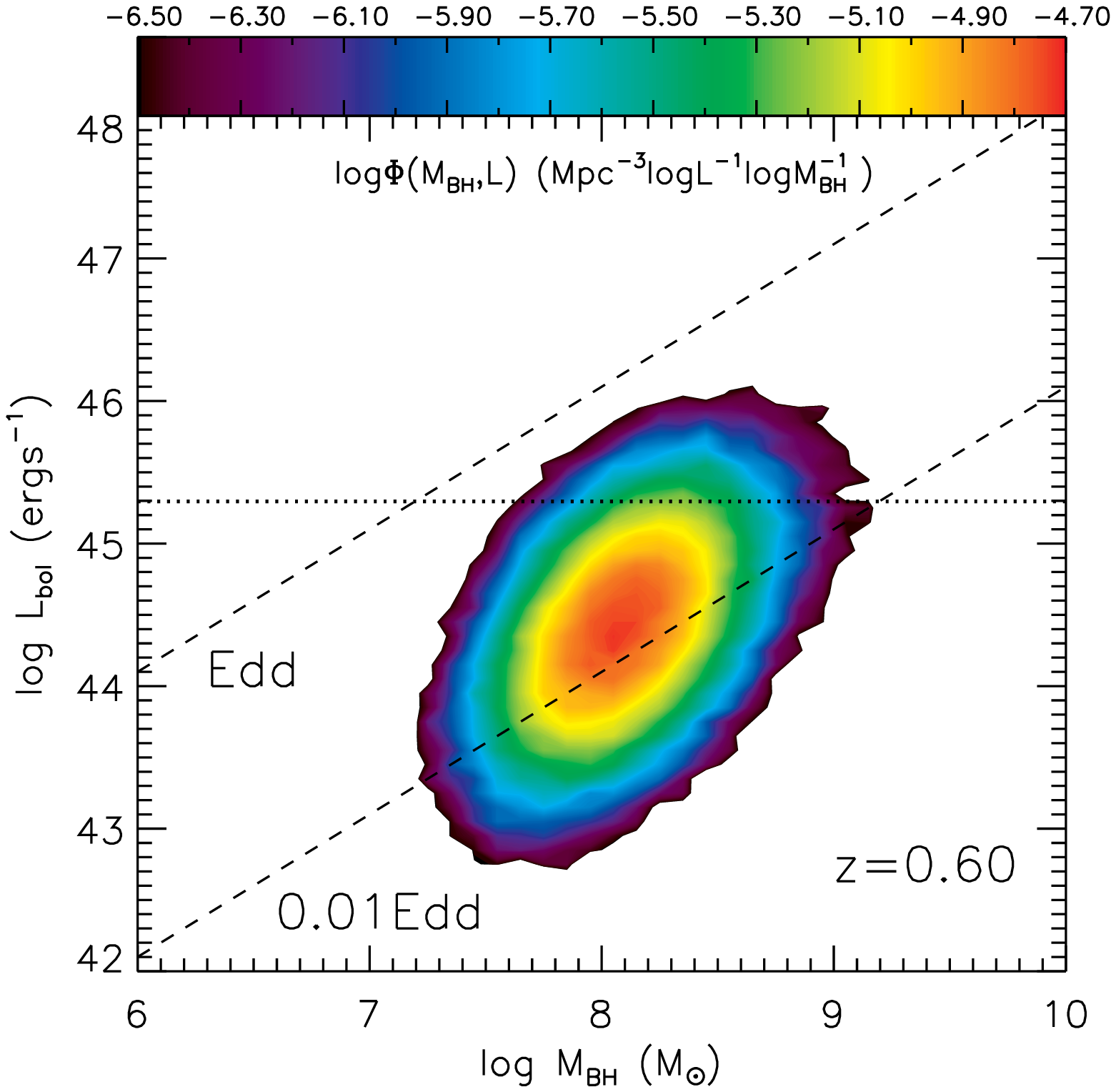}
    \includegraphics[height=0.4\textheight]{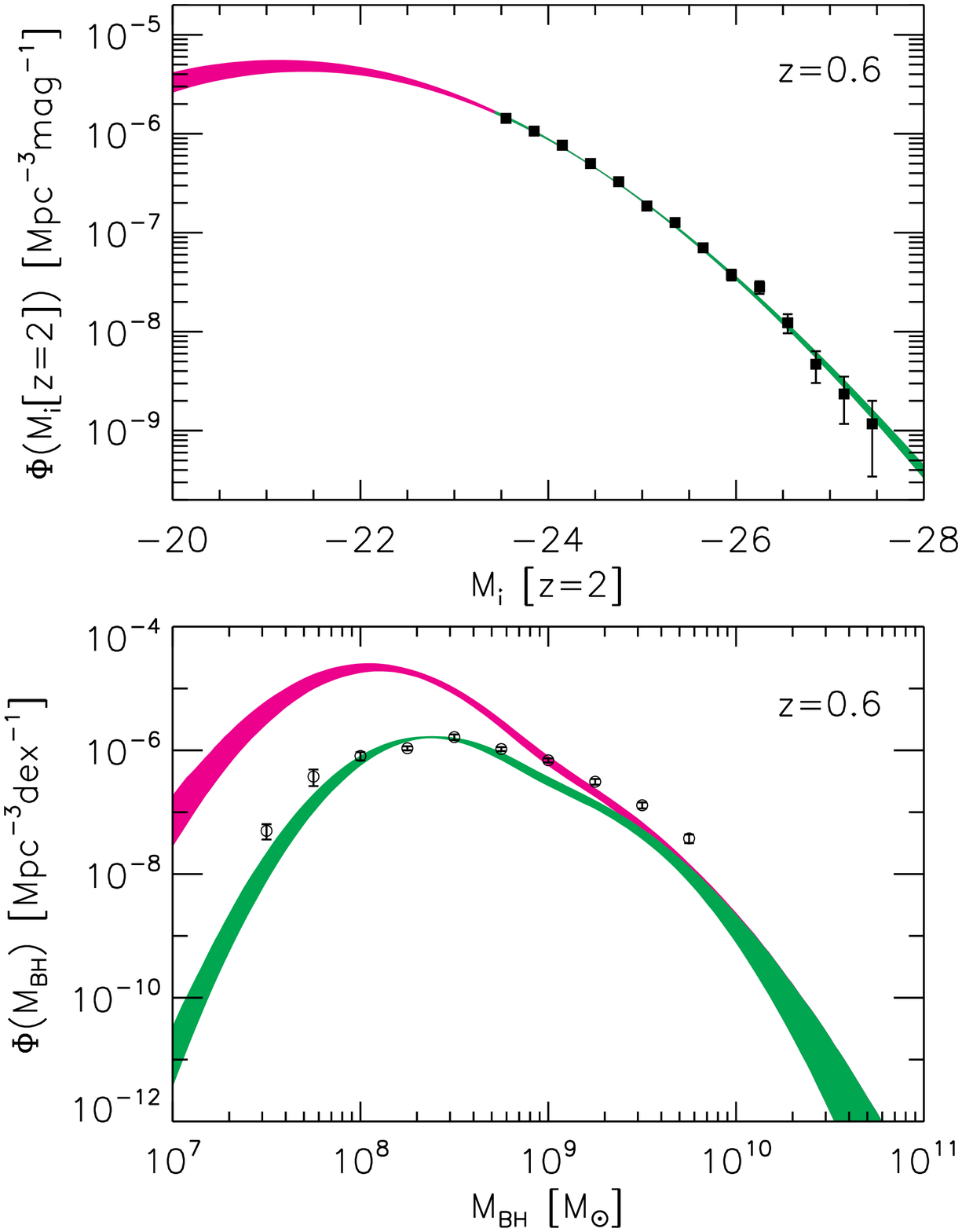}
    \caption[my caption]{ An example of the forward modeling of quasar demographics in
    the mass-luminosity plane by \citet{Shen_Kelly_2011}, modeled at $z=0.6$. {\em Left:} The simulated mass-luminosity plane (with {\em true} BH
    masses), which
    extends below the SDSS flux limit (the black horizontal line). Shown here is the comoving number density map [$\Phi(M_{\rm BH},L)$], where only
    regions
    with $\Phi(M_{\rm BH},L)>10^{-6.5}\,{\rm Mpc}^{-3}\log L^{-1}\log M_{\rm BH}^{-1}$ are shown. The two diagonal lines indicate constant Eddington
    ratios of 1 and 0.01. The flux limit only selects the most luminous objects into
    the SDSS sample. {\em Right:} Projections of
    the 2D distribution onto the luminosity and BH mass axes, i.e., the LF (upper right) and BHMF (bottom right) of quasars at $z=0.6$. The points
    are binned observational data and the lines are best-fit models: the magenta lines show the results for all broad-line quasars (corrected for the
    flux limit) and the
    green lines show those for the flux-limited sample. The thickness of the lines indicates the 1$\sigma$ model uncertainty. I have used the
    $i$-band
    absolute magnitude instead of bolometric luminosity
    in presenting the LF, and I have also taken into account the difference between true BH masses (colored lines) and SE virial BH masses (points).
    This forward-modeling framework
    accounts for the selection incompleteness in BH mass due to the flux limit, and the uncertainties in SE virial BH mass estimates, and can
    constrain
    the
    2D distribution down to $\sim 3$ magnitudes fainter than the flux limit \citep{Shen_Kelly_2011}. However, to make more robust constraints at the
    faint end, deeper
    quasar samples are highly desirable.\\

    \quad The 2D distribution of quasars offers significantly more information on the evolution of quasars than 1D distributions (i.e., LF or BHMF)
    alone. Specifically if we focus on
    snapshots of the mass-luminosity plane in short time intervals (e.g., shorter than typical quasar lifetime), then objects will move along certain
    evolutionary tracks in the plane, determined by the evolution of Eddington ratio (possibly also coupled with the change of the radiative
    efficiency
    as accretion rate evolves).
    This evolution in the mass-luminosity plane can be modeled in great detail, and compared with either analytical models or numerical
    simulations of SMBH growth. This is somewhat similar to the simple arguments in the quasar LF literature (such as the ``pure luminosity
    evolution''
    model), but the
    diagnostic capabilities of the mass-luminosity plane would be far more powerful and self-consistent in terms of SMBH growth and light curve
    evolution. }
    \label{fig:zbin2_ml_plane_all}
\end{figure}

I demonstrate the power of the mass-luminosity plane in quasar demographic
studies in Fig.\ \ref{fig:zbin2_ml_plane_all}. This is the same simulated,
{\em true} quasar $M-L$ plane at $z=0.6$ as in Fig.\ \ref{fig:ml_plane_z0.6},
using the models in \citet{Shen_Kelly_2011} constrained using SDSS quasars.
The 2D density (i.e., abundance) of quasars in this plane is shown in the
color-coded contours. The traditional LF and BHMF (shown in the right panels)
are then just the projection onto each axis. In the right panels I also
demonstrate the differences between using true BH masses and SE virial
masses, as well as the effect of the sample flux limit. It is clear from this
demonstration that the 1D distribution functions lose information by
collapsing on one dimension, and a better way to study the demography of
quasars is to measure their abundance in 2D, since the mass and luminosity of
a quasar are physically connected by the Eddington ratio. The ultimate goal
is to study the evolution of the quasar density in the $M-L$ plane as a
function of time. Recent studies have started to work in this direction
\citep[e.g.,][]{Shen_Kelly_2011,Kelly_Shen_2013}, although deeper quasar
samples and a better understanding of SE mass errors are needed to utilize
the full power of the $M-L$ plane.

To summarize, the quasar mass-luminosity plane has great potential in
studying quasar evolution, and efforts have been underway to investigate this
plane in detail
\citep[e.g.,][]{Steinhardt_Elvis_2010a,Steinhardt_Elvis_2010b,
Steinhardt_etal_2011a,Steinhardt_Elvis_2011,Shen_Kelly_2011,Kelly_Shen_2013}.
However, it should always be kept in mind that the ``observed'' distribution
in the $M-L$ plane is {\em not} the true distribution. I strongly discourage
direct interpretations of the observed distributions based on SE masses and
flux-limited data, which can easily lead to superficial or even spurious
results.

\subsection{Evolution of BH-Bulge Scaling Relations}\label{s:app:evo}

Another important application of SE virial mass estimators is to study the
$M_{\rm BH}-$host scaling relations in broad line AGNs, and to probe the
evolution of these relations at high redshift. Measuring the $M_{\rm
BH}$-host relations in low redshift quasars and AGNs has been done using both
RM masses and SE masses
\citep[e.g.,][]{Laor_1998,Greene_Ho_2006,Bentz_etal_2009c,Xiao_etal_2011}.
Assuming some virial coefficient $\langle f\rangle$, these studies were able
to add active objects in these scaling relations and extend the dynamic range
in BH mass.

In the past a few years, there have been a huge amount of effort to quantify
the evolution of these scaling relations up to $z\sim 6$, by measuring host
properties in broad-line quasars. Some studies directly measure the galaxy
properties by decomposing the quasar and galaxy light in either imaging or
spectroscopic data
\citep[e.g.,][]{Treu_etal_2004,Peng_etal_2006a,Peng_etal_2006b,Woo_etal_2006,Treu_etal_2007,
Woo_etal_2008,Shen_etal_2008x,Jahnke_etal_2009,
McLeod_Bechtold_2009,Decarli_etal_2010,Merloni_etal_2010,Bennert_etal_2010,Cisternas_etal_2011,Targett_etal_2012};
other studies use indirect methods to infer galaxy properties, such as using
the narrow emission line width to infer bulge velocity dispersion
\citep[e.g.,][]{Shields_etal_2003,Shields_etal_2006,Salviander_etal_2007,Salviander_Shields_2012}.
Molecular gas (using CO tracers) has also been detected in the hosts of
$z\sim 6$ quasars, allowing rough estimates on the host dynamical mass of
these highest redshift quasars \citep[e.g.,][and references
therein]{Walter_etal_2004,Wang_etal_2010}. In all cases the BH masses were
estimated using the SE methods based on different broad emission lines. With
a few exceptions, most of these studies claim an excess in BH mass relative
to bulge properties either in the $M_{\rm BH}-\sigma_*$ relation or in the
$M_{\rm BH}-M_{\rm bulge}/L_{\rm bulge}$ relation, and advocate a scenario
where BH growth precedes spheroid assembly.

It is worth noting that measuring host galaxy properties of Type 1 AGNs could
be challenging, and systematic biases may arise when measuring the stellar
velocity dispersion from spectra \citep[e.g.,][]{Bennert_etal_2011}, or host
luminosities from image decomposing
\citep[e.g.,][]{Kim_etal_2008,Simmons_Urry_2008}. Conversions from
measurables (such as host luminosity) to derived quantities (such as stellar
mass) are also likely subject to systematics, especially for low-quality
data. Thus careful treatments are required to derive unbiased host
measurements.

On the other hand, it is also worrisome that the errors in SE BH mass
estimates may affect the observed offset in the BH scaling relations at
high-redshift. As discussed extensively in \S\ref{s:bias}, there are both
physical and practical concerns that the applications of locally-calibrated
SE estimators to high-redshift quasars may cause systematic biases. Even if
the extrapolations are valid, the luminosity-dependent bias discussed in
\S\ref{s:bias_l} may still lead to an average overestimation of quasar BH
masses in flux-limited surveys. \citet{Shen_Kelly_2010} studied the impact of
the luminosity-dependent bias on flux-limited quasar samples, and found an
``observed'' BH mass excess of $\sim 0.2-0.3$ dex for $L_{\rm bol}\gtrsim
10^{46}\,{\rm erg\,s^{-1}}$ with a reasonable value of $\sigma^\prime_l=0.4$
dex (see \S\ref{s:bias_l} for details). This sample bias using SE mass
estimates becomes larger (smaller) at higher (lower) threshold quasar
luminosities.

Another statistical bias was pointed out by \citet{Lauer_etal_2007b}, which
is at work even if there is no error in BH mass estimates. The basic idea is
that since there is an intrinsic scatter between BH mass and bulge properties
($\sim 0.3$ dex for the local sample), and since the distribution functions
in BH mass and galaxy properties are expected to be bottom-heavy, a
statistical excess (bias) in the average BH mass relative to bulge properties
arises when the sample is selected based on BH mass (or based on quasar
luminosity, assuming the Eddington ratio is constant). This is similar to the
Malmquist-type bias discussed in \S\ref{s:bias_m}. One can work out
\citep[e.g.,][]{Lauer_etal_2007b,Shen_Kelly_2010} that the BH mass offset
introduced by this bias depends on the slope of the galaxy distribution
function, as well as the scatter in the BH-host scaling relations. For simple
power-law models of the galaxy distribution function on property $S$ with a
slope $\gamma_s$, and lognormal scatter $\sigma_\mu$ at fixed galaxy property
$S$, this bias takes a similar form as the Malmquist-type bias in
\S\ref{s:bias_m}:
\begin{equation}
\Delta\log M_{\rm BH,Lauer}=-\ln(10)\gamma_s\sigma^2_{\mu}/C\ ,
\end{equation}
where $C$ is the coefficient of the mean BH-host property ($S$) scaling
relation $\log M_{\rm BH}=C\log S + C^\prime$. Thus if the intrinsic scatter
in the BH-host scaling relations increases with redshift, then this
statistical bias alone can contribute a significant amount to the observed BH
mass offset in the high-redshift samples \citep[e.g.,][]{Merloni_etal_2010}.
A larger intrinsic scatter in these scaling relations at high redshift is
expected, if the tightness of the local BH-host scaling relations is mainly
established via the hierarchical merging of less-correlated BH-host systems
at higher redshift
\citep[e.g.,][]{Peng_2007,Hirschmann_etal_2010,Jahnke_Maccio_2011}. The real
situation is of course more complicated, and one must consider a realistic
Eddington ratio distribution at fixed BH mass and the effect of the flux
limit. There could also be other factors that may complicate the usage of
AGNs to probe the evolution of these BH-host scaling relations, as discussed
in detail in \citet{Schulze_Wisotzki_2011}. But overall a BH mass excess due
to the Lauer et~al.\ bias is expected when select on quasar luminosity. An
interesting corollary is that a deficit in BH mass is expected if the sample
is selected based on galaxy properties. This may explain the findings that
high-redshift submillimeter galaxies (SMGs) tend to have on average smaller
BHs relative to expectations from local BH-host scaling relations
\citep[e.g.,][]{Alexander_etal_2008}.

The Lauer et~al.\ bias caused by the intrinsic scatter in BH-host scaling
relations works independently with the luminosity-dependent bias caused by
errors in SE masses, so together they can contribute a substantial (or even
full) amount of the observed BH mass excess at high redshift
\citep[e.g.,][]{Shen_Kelly_2010}. Both biases are generally worse for samples
with a higher luminosity threshold given the curvature in the underlying
distribution function\footnote{This is not always the case. If the scatter
($\sigma_\mu$, $\sigma^\prime_l$) increases at the low-mass end, then both
biases could be worse at the low-mass/luminosity end. }, thus higher-$z$
samples with higher intrinsic AGN luminosities will have larger BH mass
biases, leading to an apparent evolution. There are several samples that are
probing similar luminosities as the local RM AGN sample
\citep[e.g.,][]{Woo_etal_2006}. Since the SE mass estimators were calibrated
on the local $M_{\rm BH}-\sigma_*$ relation using the RM AGN sample, one
argument often made is that both biases should be calibrated away for the
high-$z$ sample with similar AGN luminosities. This argument is flawed,
however, because the local RM AGN sample is heterogeneous and is not sampling
uniformly from the underlying BH/galaxy distribution functions, while the
high-$z$ sample usually is sampling uniformly from the underlying
distributions -- this is exactly why both biases will arise for the high-$z$
samples. The only exception that might work is to compare two quasar samples
at two different redshifts with the same luminosity threshold, where the
predicted BH mass biases should be of the same amount, and see if there is
evolution in the average host properties. But even in this case it requires
that the underlying distributions (slope and scatter) and measurement
systematics are the same for both the low-$z$ and high-$z$ samples. Proper
simulations that take into account the measurement systematics (in both BH
mass and host properties) and underlying distributions should be performed to
verify the interpretations upon the observations.


To summarize, there might be true evolutions in the BH-host scaling
relations\footnote{If the mass and velocity dispersion of galaxies bear any
resemblance to the virial mass ($M_{\rm h,vir}$) and virial velocity ($V_{\rm
h,vir}$) of their host dark matter halos, then one of the two relations,
$M_{\rm BH}-\sigma_*$ and $M_{\rm BH}-M_{\rm bulge}$, must evolve since the
$M_{\rm h,vir}-V_{\rm h,vir}$ relation is redshift-dependent. }, but the
current observations are inconclusive, due to the unknown systematics in the
BH mass and host galaxy measurements. Better understandings of these
systematics, the selection effects, as well as theoretical priors are all
needed to probe the evolution of these scaling relations, and such efforts
have been underway
\citep[e.g.,][]{Croton_2006,Robertson_etal_2006,Lauer_etal_2007b,Hopkins_etal_2007,Di_Matteo_etal_2008,Booth_Schaye_2009,Shankar_etal_2009,Shen_Kelly_2010,Hirschmann_etal_2010,
Schulze_Wisotzki_2011,Jahnke_Maccio_2011,Portinari_etal_2012,Salviander_Shields_2012,Zhang_etal_2012}.

\section{Summary and Future Perspectives}\label{s:future}

To conclude, there have been considerable progress over the past several
decades on the development of BH weighing methods for quasars. We now have a
working technique based on reverberation mapping of broad line AGNs that can
measure active virial BH masses with an accuracy of a factor of a few ($\sim
0.5$ dex). Rooted in the RM technique, efficient SE virial mass estimators
have been developed to measure BH mass for large statistical samples of broad
line quasars based on single-epoch spectroscopy. These methods greatly
facilitate quasar studies in the era of modern, large-scale spectroscopic
surveys.

However, there are outstanding issues regarding the reliability of these
virial (RM or SE) mass estimators, and consequences of their significant
uncertainties, which is the focus of this review. Specifically I have the
following highlighting remarks.

\begin{enumerate}

\item[$\bullet$] There are genuine concerns that the current RM sample
    does not represent the whole quasar/AGN population, and the limited
    sample size and luminosity/redshift ranges of the RM sample, as well
    as the poorly understood BLR structure and dynamics, may impact the
    applicability of these locally-calibrated SE relations to high-$z$
    and/or high-luminosity quasars (\S\ref{s:phy_concern}). Due to
    limitations of the current RM sample and the uncertainty in the
    average virial coefficient $\langle f\rangle$, systematic biases on
    the order of a factor of a few are likely present due to these
    physical caveats.

\item[$\bullet$] Even when the extrapolation of these SE estimators to
    high-$z$ quasars is justified, rigorous statistical biases will arise
    from the uncertainties (scatter) in these SE masses
    (\S\ref{s:conse}). In particular I demonstrated the conceptual
    difference between errors in SE masses and the distribution of SE
    masses within restricted luminosity ranges. Since luminosity is used
    in the estimation of SE masses, the variance in SE mass is reduced
    when luminosity is constrained. I also derived the
    luminosity-dependent bias
    \citep[e.g.,][]{Shen_etal_2008b,Shen_Kelly_2010,Shen_Kelly_2011} that
    at fixed true mass, the SE masses are over(under)-estimated in the
    mean when luminosity is higher (lower) than the mean luminosity at
    this fixed true mass, due to the stochastic variations between
    luminosity and line width that contribute to the uncertainty of SE
    masses. Simple simulations were performed to demonstrate these
    effects, and suggest that sample biases on the order of a factor of a
    few are present in flux-limited bright quasar sample. Thus these
    error-induced biases are {\em as significant as} the unknown
    systematic biases in SE masses, and cannot be ignored. More
    importantly, even when we eliminate all systematic biases (zero-point
    uncertainty) of these SE estimators in the future, these
    error-induced statistical biases will largely remain given the
    imperfect nature of these estimators. The formalism in
    \S\ref{s:bias_l} provides guidance on how to quantify these
    error-induced sample biases with simulations.

\item[$\bullet$] Properly accounting for the selection effect due to the
    sample flux limit, and statistical biases arising from errors in SE
    masses, are crucial to interpreting the observed distributions of
    quasars in statistical samples (\S\ref{s:app}). I discussed how the
    ``observed'' distributions of BH mass and Eddington ratio for
    threshold data and with SE masses differ from the ``true''
    distributions (\S\ref{s:app:ML}), and cautioned on some recent claims
    based directly on the observed distributions. I further commented on
    the impact of the error-induced biases in SE masses on studies of the
    evolution of the BH-host scaling relations (\S\ref{s:app:evo}), and
    concluded that the current observations are inconclusive for the
    claimed evolution.

\end{enumerate}

Looking forward, there is an urgent need to expand the current RM sample with
lag measurements, and, to acquire exquisite (velocity-resolved) RM data to
utilize the full power of this technique. Only with a substantially larger RM
sample that properly probes the AGN parameter space, and with a much better
understanding of the BLR geometry and dynamics (for different lines) based on
these RM data and ancillary data, can we improve these virial BH mass
estimators further. Since resource-wise, RM is a consuming exercise, it would
be interesting to explore the possibilities of more efficient RM with
wide-field multi-object imaging and spectroscopy, as well as dedicated
facilities for single-object-mode monitoring.

There are a few recent innovative proposals regarding RM that are worth
mentioning. \citet{Zu_etal_2011} proposed an alternative method to measure RM
lags, by fitting the observed continuum and emission line light curves with
recently-developed statistical models to describe quasar variability
\citep[e.g., the ``damped random walk'' model developed
by][]{Kelly_etal_2009b,Kozlowski_etal_2010}. Compared with the traditional
cross-correlation method in measuring RM lags
\citep[e.g.,][]{Gaskell_Peterson_1987}, this new method can improve lag
measurements by simultaneously fitting multiple lines and quantifying error
correlations. In addition, there have been efforts to build dynamical BLR
models to directly model the RM light curves in the time domain
\citep[e.g.,][]{Brewer_etal_2011,Pancoast_etal_2012}; such forward modeling
\citep[as discussed in, e.g.,][]{Netzer_Peterson_1997} can in principle
provide direct constraints on the geometry of the BLR and the mass of the BH,
and so it is worthwhile to explore its potential further. Finally,
alternative strategies in RM experiments with no or few spectroscopic data
\citep[e.g.,][]{Haas_etal_2011,Chelouche_Daniel_2012,Fine_etal_2012}, while
not as good and reliable as traditional spectroscopic RM, may speed up the
process of probing the diversity of AGN parameters in the context of RM.

\section*{Acknowledgements}

I thank all my collaborators and colleagues for many stimulating and
enlightening discussions on this subject over the past a few years. In
writing this review I benefited a lot from inspiring conversations with
and/or feedback from Brandon Kelly, Luis Ho, Brad Peterson, Gordon Richards,
Roberto Assef, Aaron Barth, Neil Brandt, Kelly Denney, Jenny Greene, Pat
Hall, Zo{\l}tan Haiman, Chris Kochanek, Juna Kollmeier, Tod Lauer, Xin Liu,
Youjun Lu, Chien Peng, Alireza Rafiee, Andreas Schulze, Charles Steinhardt,
Michael Strauss, Benny Trakhtenbrot, Scott Tremaine, Marianne Vestergaard,
David Weinberg, Jong-Hak Woo, and Qingjuan Yu. I acknowledge support from the
Carnegie Observatories through a Hubble Fellowship from Space Telescope
Science Institute. Support for Program number HST-HF-51314.01-A was provided
by NASA through a Hubble Fellowship grant from the Space Telescope Science
Institute, which is operated by the Association of Universities for Research
in Astronomy, Incorporated, under NASA contract NAS5-26555.

\label{lastpage}
\end{document}